\documentclass[pdflatex,sn-apa]{sn-jnl}

\usepackage{graphicx}%
\usepackage{multirow}%
\usepackage{amsmath,amssymb,amsfonts}%
\usepackage{amsthm}%
\usepackage{mathrsfs}%
\usepackage[title]{appendix}%
\usepackage{xcolor}%
\usepackage{textcomp}%
\usepackage{manyfoot}%
\usepackage{booktabs}%
\usepackage{algorithm}%
\usepackage{algorithmicx}%
\usepackage{algpseudocode}%
\usepackage{listings}%

\usepackage{soul}
\colorlet{usercolorname}{yellow!0}
\sethlcolor{usercolorname}

\usepackage{subcaption}
\usepackage{multirow}
\usepackage{makecell}

\usepackage{CJKutf8}

\usepackage{ccicons}

\begin{document}
\begin{CJK}{UTF8}{ipxm}












\received{3 June 2025}
\revised{2 May 2026}
\accepted{20 July 2026}






\title[Japanese Consumer Experiences with Dark Commercial Patterns]{Unease, Ambivalence, and Endured Disloyalty in Japanese Consumer Experiences with Dark Commercial Patterns}


\author*[1,2]{\fnm{Katie} \sur{Seaborn}}\email{katie.seaborn@cst.cam.ac.uk}

\author[1]{\fnm{Jo} \sur{Yukami}}\email{yukami.j.aa@m.titech.ac.jp}

\author[1]{\fnm{Tatsuya} \sur{Itagaki}}\email{itagaki.t.ad@m.titech.ac.jp}

\author[1]{\fnm{Mizuki} \sur{Watanabe}}\email{watanabe.m.ca@m.titech.ac.jp}

\author[1]{\fnm{Yijia} \sur{Wang}}\email{wang.y.cf@m.titech.ac.jp}

\author[1]{\fnm{Ping} \sur{Geng}}\email{geng.p.aa@m.titech.ac.jp}

\author[1]{\fnm{Takao} \sur{Fujii}}\email{fujii.t.av@m.titech.ac.jp}

\author[1]{\fnm{Yuto} \sur{Mandai}}\email{mandai.y.aa@m.titech.ac.jp}

\author[1]{\fnm{Miu} \sur{Kojima}}\email{kojima.m.ap@m.titech.ac.jp}

\author[1]{\fnm{Suzuka} \sur{Yoshida}}\email{yoshida.s.av@m.titech.ac.jp}

\affil*[1]{\orgdiv{}, \orgname{Institute of Science Tokyo}, \orgaddress{\city{Tokyo}, \country{Japan}}}

\affil[2]{\orgdiv{}, \orgname{University of Cambridge}, \orgaddress{\city{Cambridge}, \country{UK}}}

\abstract{Dark commercial patterns and deceptive user interface (UI) designs (or DPs) trick consumers into actions that benefit the shareholders. The legal and ethical implications of DPs are shaped by the sociocultural context. Special types of DPs exist in Japan, but the impact of these DPs on consumer attitudes and behaviour remains underexplored. 
We report on the first comparative mixed methods user study with Japanese consumers ($N=84$), half of whom ($n=40$) experienced a range of DPs---including the Japanese varieties---in a simulated e-commerce website. 
We discovered that the Japanese DPs were among the least noticeable and caused the highest simulated financial harm, with Untranslation perceived as highly disruptive and Alphabet Soup highly deceptive.
Comparative analyses with a group that experienced the DP-free version of the online store ($N=44$) revealed a sharp negative difference in positive emotions and acceptance. Qualitative analyses surfaced cultural norms in consumer--business relationships, notably unease, ambivalence, and endured disloyalty (不誠実 or fuseijitsu).
No evidence of sampling biases was found for the participants involved in a publicly broadcast programme on the study ($n=10$), indicating true deception and unacceptability. 
Our findings suggest that while reactions toward and ability to recognize a given DP may vary across individuals, the mere presence of DPs tends to have negative effects on most Japanese consumers.\\

\textbf{Highlights}\\
・Japan-based dark commercial patterns were among the most deceptive and harmful.\\
・Reduced positive emotions and acceptance were found compared to a deception-free version.\\
・Cultural norms like ambivalence and enduring disloyalty shaped consumer responses.\\

\textbf{Published version}\\
Seaborn, K., Yukami, J., Itagaki, T., Watanabe, M., Wang, Y., Geng, P., et al, 2026. Unease, ambivalence, and endured disloyalty in Japanese consumer experiences with dark commercial patterns. Int. J. Human–Computer Studies. 1–19. doi: https://doi.org/10.1016/j.ijhcs.2026.103903\\

\textbf{License}\\
This work is licensed under CC-BY-NC-ND \ccbyncnd\ 4.0.
}


\maketitle

\section{Introduction}\label{isect1}

Dark commercial patterns and deceptive user interface (UI) designs (hereafter DPs)\footnote{Our choice of the abbreviation ``DP'' is guided by context and critique. We recognize that the word ``dark'' is racially charged~\citep{network2025technoskepticism,benjamin2019race}, a point that publisher ACM has raised (\href{https://www.acm.org/diversity-inclusion/words-matter}{https:/\allowbreak{}/\allowbreak{}www.\allowbreak{}acm.\allowbreak{}org/\allowbreak{}diversity-\allowbreak{}inclusion/\allowbreak{}words-\allowbreak{}matter}). Also, in design contexts, ``dark'' is vague and typically associated with colour palette rather than hidden elements. At the same time, early adopters and the Japanese media introduced the term ``dark pattern'' or {ダークパターン} to the Japanese public, and it has stuck. The 2022 OECD report~\citep{oecd2022}, entitled ``dark commercial patterns,'' continues to be influential in the Japanese media, regulatory, and public spheres, as well. Even Brignull's recent book ``Deception Patterns''~\citep{Brignull2023} is entitled ``Dark Patterns'' in Japanese. We compromise by collapsing both terms under the heading of ``DP.''} cover a range of digital products---manipulative UI~\citep{Narayanan2020}, disloyal patterns~\citep{richards2021duty}, and detrimental persuasive technologies~\citep{Fogg2002}---purposefully designed to alter user actions and choices in ways that are advantageous to the shareholders and often disadvantageous to the user~\citep{gray2023dpsysreview,Brignull2023}. DPs are often deployed by businesses in pursuit of commercial objectives~\citep{fansher2018hashtag,Brignull2023}. Businesses have power over consumers, who use the DP-laden digital services and products they supply. As \citet[p. 67]{Narayanan2020} starkly write, DPs represent ``abuse of the tremendous power that designers hold in their hands.''
Recognizing this, professionals from all walks of life have contributed to fundamental work on DPs. One trajectory has focused on the nature of DPs~\citep{gray2018darkside} and developing legal arguments against their use~\citep{gray2021legal}. Another body of work has sought to identify what types of DPs exist based on observation and analysis~\citep{digeronimo2020,hidaka2023linguistic,mathur2019atscale,yada2022dark} and subsequently develop ontologies and taxonomies of actual DPs~\citep{gray2024ontology}. Other groups involving academics and authorities have sought a means of regulating DPs~\citep{oecd2022,oecd2023}, establishing a lively and interdisciplinary community.

An ongoing challenge is how to evaluate the \emph{user side} of DPs in everyday consumer contexts~\citep{borberg2022so,nazarov2022clustering,BongardBlanchy2021}. Methodological gaps and potentialities remain open, e.g., what measures to use~\citep{mathur2021whatdark} and in relation to what theory~\citep{chang2024theory}. Crucially, DPs are a sensitive topic for academia and industry. Ethical restrictions at institutions can prevent the study of real digital products or limit sharing of pertinent information~\citep{borberg2022so,hidaka2023linguistic,nazarov2022clustering}. The various terms used to describe professional design choices and commercial UIs are negative and contentious~\citep{fansher2018hashtag,mathur2021whatdark,gray2018darkside,gray2021legal,feng2023analysis,kyi2023gdpr}. While working with companies and shareholders is ideal, the framing and potential legal implications are barriers. For instance, Japanese public broadcaster NHK\footnote{NHK is the romanized initialism for the Japan Broadcasting Corporation.} and DP expert Atsushi Hasegawa collaboratively reached out to 30 major Japanese companies colloquially known to employ DPs, but of the 16 (53\%) that responded, only three admitted to doing so.\footnote{\url{https://www.nhk.or.jp/minplus/0016/topic062.html} (note: in Japanese).}

Given such barriers, researchers create and employ a variety of \emph{simulations}. Simulated DPs can be injected into real websites and apps through browser plugins and other means~\citep{roffarello2022steal,utz20219gdprconsent,berens2022cookie,bermejo2021cookie}. Non-interactive mock screen shots and mockups of complete systems, inspired by real examples, can be used as research material to gather user perceptions~\citep{bongardblanchy2023,schafer2023countermeasures}. A few research teams have developed full or partial interactive systems from scratch---systems that aim to represent actual user experiences (UX) based on real equivalents~\citep{voigt2021dark,van2022shopping,cranor2022cookie,grassl2021dark,habib2022cookie,Seaborn2024lbw}. Understandably, most simulations target \emph{specific contexts or DPs in isolation}, even though most systems bear multiple DPs~\citep{digeronimo2020,hidaka2023linguistic,gunawan2021webmobile}. Simulations have, for example, been streamlined to isolate app permission requests~\citep{bongardblanchy2023}, constrained to screens and user flows involving cookie consent banners~\citep{cranor2022cookie,bermejo2021cookie,berens2022cookie,habib2022cookie}, and designed to include only certain types of DPs 
even when other DPs might typically be present in the same context~\citep{van2022shopping}. 
Many researchers also rely on \emph{non-interactive simulations}---mockups, blueprints, fake screen shots, and videos~\citep{bongardblanchy2023,schafer2023countermeasures,digeronimo2020}. 
Interactive simulations that present a realistic environment and user flow amplify ecological validity in comparison to static or narrow presentations of DP stimuli, especially when DPs work by way of user interaction~\citep{carter2008exiting} and over time~\citep{Gray2025time}. While always limited, representative and interactive simulations are among the best tools at our disposal. Our work offers a new platform with experimental components (\nobreak{}versions with and without DPs) contextualized to a typical context: online stores.

Another challenge is the focus on Western, English-speaking consumer populations and user groups~\citep{seabornanother2024}, an ongoing ``pattern'' in HCI research~\citep{Linxen2021}. Yet, the socio-cultural context can be crucial. Law-wise, one example is the case of India, where the Department of Consumer Affairs, through the 2019 Consumer Protection Act, has provided contextual regulations and reforms.\footnote{\url{https://consumeraffairs.nic.in/theconsumerprotection/guidelines-prevention-and-regulation-dark-patterns-2023}.} Recent work on the UX side has highlighted the Japanese context~\citep{hidaka2023linguistic,Seaborn2024lbw}. In 2022, novel varieties of DPs called ``Linguistic Dead-Ends'' were found in Japan~\citep{hidaka2023linguistic}, with initial work hinting at their duplicity~\citep{Seaborn2024lbw}. DPs rose to the fore of public consciousness in Japan during the COVID-19 pandemic, where about 53\% of households went online to shop in 2022 (up 34.3\% from 2018).\footnote{\url{https://www.statista.com/statistics/1182675/japan-online-shopping-penetration-households/}.} 
In 2022, 93.5\% of the top 200 Google Play Store apps were found to contain about 3.9 DPs on average~\citep{hidaka2023linguistic}. 
Regulatory bodies in Japan began to take action~\citep{oecd2022}. Notably, the 2023 OECD report was led by the Consumer Affairs Agency of the Government of Japan, with a concerted focus on consumer vulnerability and policy~\citep{oecd2023}.
Even so, the degree to which the average Japanese consumer is aware of DPs, how they feel about such designs, and the implications for typical use remain underexplored.

In response, we conducted a comparative study ($N=84$) on DPs, notably the Japan-based Linguistic Dead-Ends~\citep{hidaka2023linguistic}, with a diverse sample of Japanese consumers carrying out typical online shopping tasks. Addressing the above gaps, we leveraged two guiding lenses~\citep{mathur2021whatdark} for evaluating the impact of DPs on Japanese consumers. We therefore asked two research questions (RQs). First, we asked \textbf{\emph{RQ1: To what extent can the average Japanese consumer identify DPs? [Individual Autonomy]}} For this, we captured objective measures of noticeability, deceptiveness, and task completion, as well as analyzed qualitative reports on the experience. Second, we asked \textbf{\emph{RQ2: How does the average Japanese consumer feel about each form of DP? [Individual Welfare]}} We used subjective self-reports of affective state (before and after the online shopping experience) and acceptability, supplemented by qualitative insights, along with an objective measure of simulated financial loss. Due to ethical restrictions, we developed a simulation (refer to \hyperref[sec:website]{Section~\ref{sec:website}}) containing a range of DPs based on those typically found in real e-commerce websites (refer to \autoref{fig:userflow} and \autoref{sec:tasks} for details). 
We contribute:
\begin{itemize}
    \item empirical findings on DPs from an observational study with Japanese consumers in an interactive platform;
    \item empirical confirmation of the duplicity of DPs so far only found in Japan, called Linguistic Dead-Ends~\citep{hidaka2023linguistic}, notably the relative \emph{disruptive} nature of subtype Untranslation and \emph{deceptive} nature of subtype Alphabet Soup;
    \item mixed methods findings on attitudes, behaviour, and deceptiveness to the range of DPs found in everyday Japanese e-commerce platforms, notably on the overlooked measure of financial loss (simulated here) and revealing unease, ambivalence, and culturally sensitive judgments relating to trust, i.e., disloyalty or {不誠実} (fuseijitsu);
    \item a comparative control with Japanese consumers experiencing a DP-free version of the website, confirming a sharp negative difference in positive feelings and acceptance after use of the DP version; \item initial evidence that potential sampling biases in relation to publicly broadcasting the research had no relevant effect; and
    \item methodology for conducting interactive user studies with a simulated website translatable to in-situ studies.
\end{itemize}

We offer the first empirical evidence, using mixed methods and a comparator, that the DPs Japanese consumers experience every day are unwelcome. We hope to inspire further interactive research on DPs in Japan and beyond.

\section{Background}\label{isect2}

We begin by positioning our research 
within the larger area of human participant and UX work on the effects of DPs employed in e-commerce and similar applications. We then ground our objectives within the Japanese context.

\subsection{Consumer reactions to UI deception}\label{isect3}

Approaches to evaluating the human factor in consumer experiences with DPs are as diverse as the interdisciplinary group of researchers working in the field---covering HCI and design, as well as psychology, legal scholarship, and beyond.

\subsubsection{Evaluating the impact of DPs on consumers}\label{isect4}

Measuring potential harm 
is multifaceted and under-theorized~\citep{mathur2019atscale,chang2024theory}. \citet{chang2024theory} discovered that only 46.8\% of DP papers ($N=51$) used a concrete theory, and of those that did, most ($n=28$) referenced the general nudge theory~\citep{thaler2021nudge} as a baseline mechanism for understanding DPs. Still, \citet{mathur2021whatdark} offer four guiding ``lenses''---two at the individual level---when evaluating the impact of DPs: \emph{individual welfare}, i.e., personal harm, \emph{collective welfare}, i.e., harm to the market, \emph{regulatory objectives}, i.e. undermining the law and legislation by regulatory organizations, and \emph{individual autonomy}, i.e., personal loss of decision-making ability. A common track that falls under the ``individual'' lenses has considered how people \emph{perform} in the face of DPs. Since our work involves individuals in a simulated environment, we focused on the \emph{individual} lenses.

\emph{Individual autonomy} may be at the heart of DPs when viewed as mechanisms that affect choice architecture. Researchers have addressed a common set of operationalizations that include subjective and objective measures. Mere user awareness of DPs remains a common measure~\citep{bhoot2021enduser,bongardblanchy2023,borberg2022so,digeronimo2020,Nimkoompai2022}.
\citet{bhoot2021enduser}, for instance, asked participants to identify instances of deception in a series of screenshots presented in a questionnaire form. 
User acceptance of DPs is also widely assessed~\citep{berens2022cookie,borberg2022so,luguri2021shining,Owens2022,utz20219gdprconsent,voigt2021dark}. \citet{berens2022cookie}, for example, varied the visual appearance of accept and reject options for cookie banners and comparatively evaluated user acceptance of each option. 
Perceived control---the degree to which users feel in control of their experience within an interface or interaction---is also commonly investigated~\citep{grassl2021dark,Hogan2022}.
Some researchers have also evaluated behaviour change as a result of DPs~\citep{bermejo2021cookie,grassl2021dark,Hogan2022,luguri2021shining,utz20219gdprconsent,van2022shopping,Koh2023yu}. \citet{Koh2023yu} manipulated the presence of certain DPs---low-stock messages, activity messages, a countdown timer, and limited-time messages---to assess changes in user product selection. A more specific form of behaviour change pursued in DP research is decision-making ability~\citep{tokuhara2023choicedelay,borberg2022so,habib2022cookie,Hogan2022,utz20219gdprconsent}.
For example, \citet{tokuhara2023choicedelay} used task time to measure selection of buttons on an interface that used timed reveal of the options plus a delay. The pattern across the literature is use of subjective and objective measures. Following this, we employed observational methods to objectively track where and when participants were (and were not) able to detect DPs (refer to \hyperref[sec:checklist]{Section~\ref{sec:checklist}}). We also used qualitative think-aloud protocols and interview methods to identify whether DPs were caught or missed (refer to \hyperref[sec:procedure]{Section~\ref{sec:procedure}}) and how perceived autonomy was affected, if at all.

One aspect of \emph{individual welfare} relevant to e-commerce contexts is \emph{financial loss}. This factor is one of the lenses proposed by \citet{mathur2021whatdark} and offered as a key individual harm in \citet{Brignull2023}. Examples of financial harm include unintended purchases, last-minute fees that the consumer feels compelled to accept, hidden costs, and even legal fees~\citep{Brignull2023,mathur2021whatdark}. Yet, despite how DPs are characterized, we could find no work that directly measures financial loss in simulations or for real consumers. Here, we calculated the simulated financial losses experienced by participants as fake users and the gains to the fake online store.

Another aspect of \emph{individual welfare} is the \emph{affective dimension}. \citet{Hogan2022} found that participants conned into signing up for opt-ins experienced deeply negative emotions. \citet{bhoot2021enduser} found that DPs pushing paid security features heightened user anxiety. \citet{chaudhary2022videostream} found that viewer mood was negatively affected after using streaming platforms with DPs. Similar to \citet{chaudhary2022videostream}, we used a classic pre-post evaluation of mood with standardized measures common in UI design and user studies (refer to \hyperref[sec:q]{Section~\ref{sec:q}}).

\subsubsection{Expanding the scope of user studies
}\label{isect5}

Isolating the impact of a \emph{specific} DP is common. 
For example, many have researched cookie consent banners~\citep{habib2022cookie,berens2022cookie,cranor2022cookie,bermejo2021cookie} in the wake of the new European General Data Protection Regulation (GDPR) act. 
In Japan, \citet{sakamotoinvestigation2020} discovered that about 65\% of $\sim$180,000 websites had deceptive consent banners. 
But focusing on one pattern alone obscures the full experience of deception on a given platform. For instance, US~\citep{digeronimo2020,gunawan2021webmobile} and Japanese~\citep{hidaka2023linguistic} apps have an average of 3.9--7.4 DPs.
These numbers may be higher, given limitations in recording times (about $\sim$5--10 minutes). We based our simulation on a full-length consumer experience. 
We crafted 25 cases made up of 51 low-level DPs categorized within the seven OECD high-level types~\citep{oecd2022} (refer to \hyperref[sec:system]{Section~\ref{sec:system}}). We compromised between coverage of all DP types and actual prevalence within a single platform---to the extent known, which is still an open question.

\begin{figure*}
    \centering
    \includegraphics[width=\textwidth]{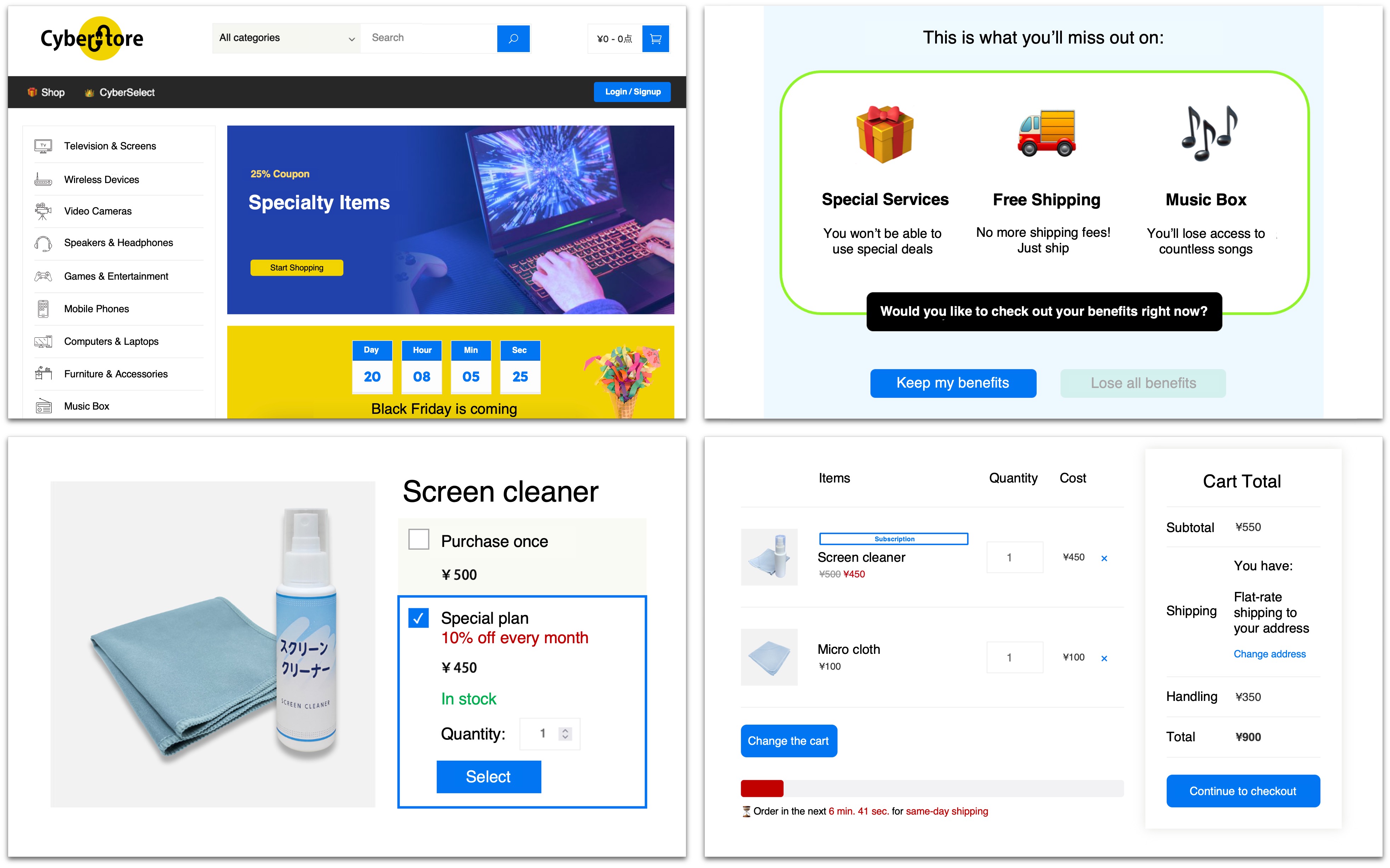}
    \caption{Screen shots of the ``CyberStore'' website: the homepage (top left), a cancellation page (top right), a product page (bottom left), and the check-out page (bottom right). Screen cleaner photos by NHK. Note: Translated to English from Japanese.}
    \label{fig:website}
\end{figure*}

Another limitation in existing work is the \emph{mode of presentation}.
Most studies have relied on \emph{non-interactive} materials, i.e., screen shots and videos. Context matters when evaluating the degree of impact. In \citet{vanderHam2015}, for instance, people walking in reality performed better during follow-up pointing and map-drawing tasks, likely because they had full access to the whole environment, while those in the virtual condition were limited to the context prescribed for the experiment. Screen shots similarly lack the interactive factor essential to experiencing many DPs, such as the procedural Roach Motel~\citep{mildner2023}. Videos prevent the participant from taking action as a user. The presentation of the stimuli may also bias responses: having a stimulus placed next to a set of response options indicates that something is there. In this work, we present a fully interactive e-commerce experience, developed to professional quality (refer to \autoref{fig:website}). We also took care to avoid mentioning DPs in our recruitment and procedures until the end of the study, thus controlling for bias around finding certain patterns or having certain experiences.

\subsection{Cultural sensitivity and UI deception: The case of Japan}\label{isect6}

\subsubsection{New varieties of DPs in Japan}\label{isect7}

In 2022, \citet{hidaka2023linguistic} found two new types of DPs in Japan, which they termed ``Linguistic Dead-Ends.'' One was ``Alphabet Soup,'' where the local syllabary is (mis)used in confusing and misleading ways. The Japanese language features two syllabaries: kana, comprising katakana and hiragana, and kanji, which were derived from Chinese characters. Katakana is often used for novel terms and foreign words. \citet{hidaka2023linguistic} found cases where katakana was used to represent English words in place of existing Japanese words, limiting user comprehension and working in sync with other DPs, such as Preselection. This DP is on the rise in Japan; in early 2023, the government raised \nobreak{}awareness of purposeful confusion between the Chinese yuan and Japanese yen, which share the same symbol ({￥}), in digital markets.\footnote{\url{https://www.kokusen.go.jp/news/data/n-20230419_2.html} (note: in Japanese).} The second subtype is ``Untranslation,'' where access to essential information is blocked through sudden use of another language. For example, information about data sharing and security is suddenly presented in English, while the rest of the app is in Japanese. In 2024, \citet{gray2024ontology} adopted these DPs into their ontology under the meso-level ``Language Inaccessibility'' category, with Alphabet Soup representing a new low-level category, thus extending the theoretical basis of DPs. While these DPs could transcend the Japanese context, this has not been explored yet, likely because this work is still new. Moreover, only preliminary work has explored Japanese consumer perceptions of these patterns~\citep{Seaborn2024lbw}. Our study aimed to address this gap.

\subsubsection{Cultural mores and the Japanese consumer}\label{isect8}

Cultural sensitivity is needed to understand the reactions of Japanese consumers to DPs against larger patterns related to trust and decision-making. 
Japanese consumers have been characterized as lacking knowledge of norms and standards in the commercial sphere~\citep{miura2021norms}. This creates mental flexibility in decision-making about purchasing goods, but is premised on low awareness, possibly leading to \nobreak{}consumer susceptibility. Underlying this state of affairs is Japanese \nobreak{}collectivism~\citep{Takano1997}. One factor is the ``conformity orientation'' or tendency to go along with others~\citep{Takano1997}. An assimilation effect~\citep{fujimura1999} can occur even when the consumer has a poor experience, if the other party compensates in some way. Still, if the experience is too poor, the collective spirit can turn sour, leading to rejection of the experience, i.e., a contrast effect~\citep{fujihara1981} or criticism resulting from high expectations not met, i.e., a dependency orientation~\citep{fujihara1981}. Japanese consumers expect a certain quality of service and care that reflects {誠実} (seijitsu), or sincerity as a facet of customer loyalty~\citep{takashi2010seijitsu}. What remained to be explored was whether and how these values and expectations play out in the face of DPs. This guided our thematic analysis (in \hyperref[sec:thematic]{Section~\ref{sec:thematic}}).

\section{System design and user flow}\label{sec:system}

The first author led the design and development of a simulated Japanese e-commerce website, featuring the range of DPs 
found in real commercial offerings (\autoref{fig:teaser}). The aim was to strike a balance between realism and coverage of the DPs that Japanese consumers experience every day. Great care was taken in selecting the DPs and constructing a typical user flow (refer to \autoref{sec:dps}). The first author was an industry developer and HCI researcher with deep expertise in DPs. They ensured a professional simulation, true-to-life UX, and reduced confounds such as poor usability (refer to \autoref{sec:website}).

\begin{figure*}[!ht]
    \begin{subfigure}{.5\textwidth}
        \centering
        \includegraphics[width=1\textwidth]{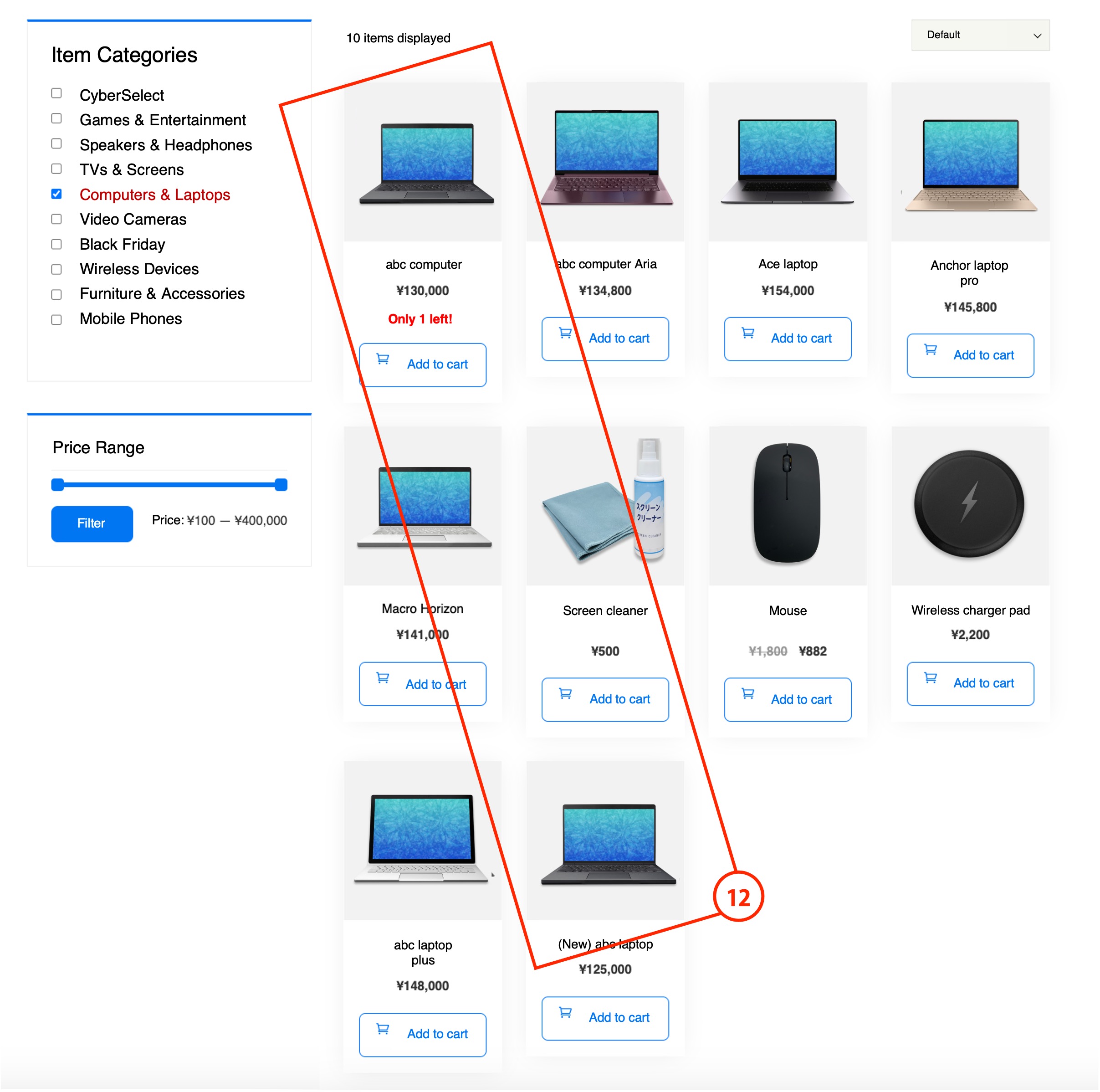}
        \caption{DP Case \#12: The computers and supplies on offer.}
        \label{fig:dpexample1}
    \end{subfigure}%
    \begin{subfigure}{.5\textwidth}
        \centering
        \includegraphics[width=1\textwidth]{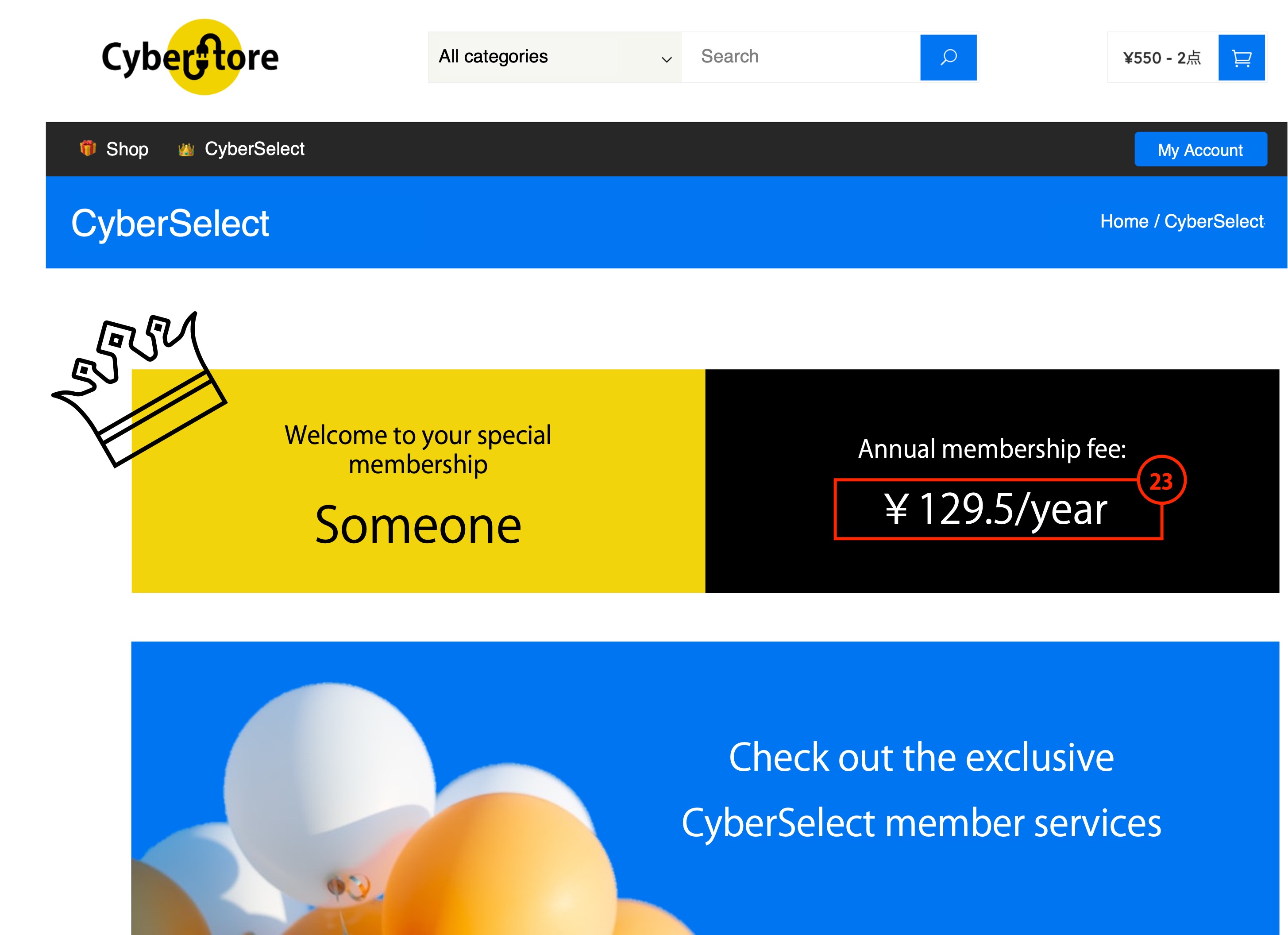}
        \caption{DP Case \#23: The membership fee.}
        \label{fig:dpexample2}
    \end{subfigure}
    \caption{Two examples of typical DPs embedded in the simulated e-commerce website. Note: Translated into English from Japanese.}
\end{figure*}

In describing our selection of DPs, we use the following labels:
\begin{itemize}
    \item DP \textbf{categories}, representing the seven generally established highest-level types of DPs~\citep{oecd2022} under which DP classes are placed.     \item DP \textbf{classes}, representing more specific types of DPs, e.g., Alphabet Soup is a class of the Linguistic Dead-Ends category.
\item DP \textbf{cases}, offering a specific implementation of one or more DP classes, e.g., the ambiguous use of the Japanese yen symbol in the yearly membership fee found on the user account screen (\autoref{fig:dpexample2}).\end{itemize}

\begin{figure*}[!ht]
    \centering
    \includegraphics[width=1\textwidth]{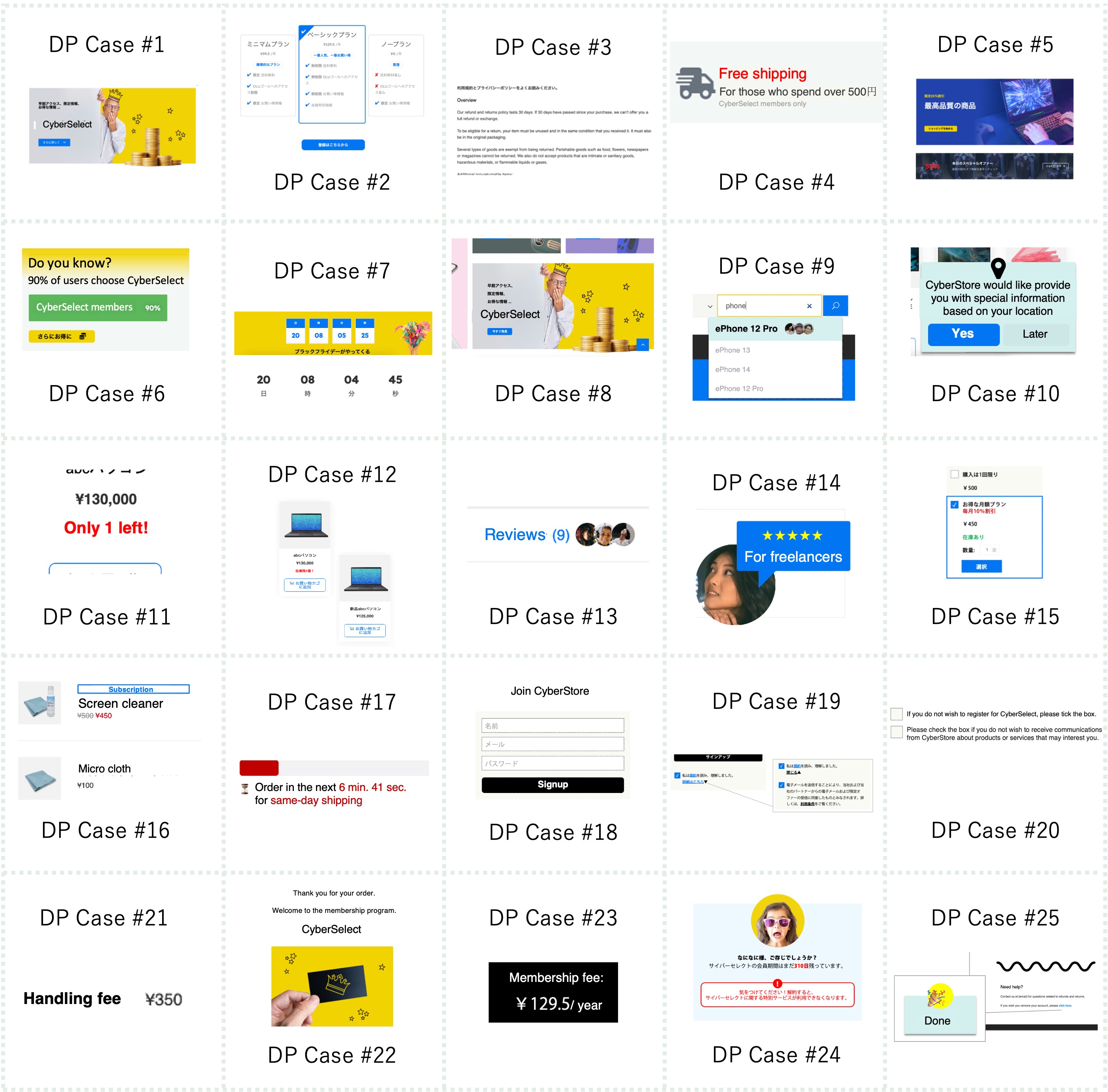}
    \caption{Cheat sheet of the 25 DP cases. Certain cases were translated from Japanese to English when needed for understanding the context and/or DP.}
    \label{fig:cheatsheet}
\end{figure*}

\subsection{Website simulation}\label{sec:website}

\begin{figure}
    \includegraphics[width=1\textwidth]{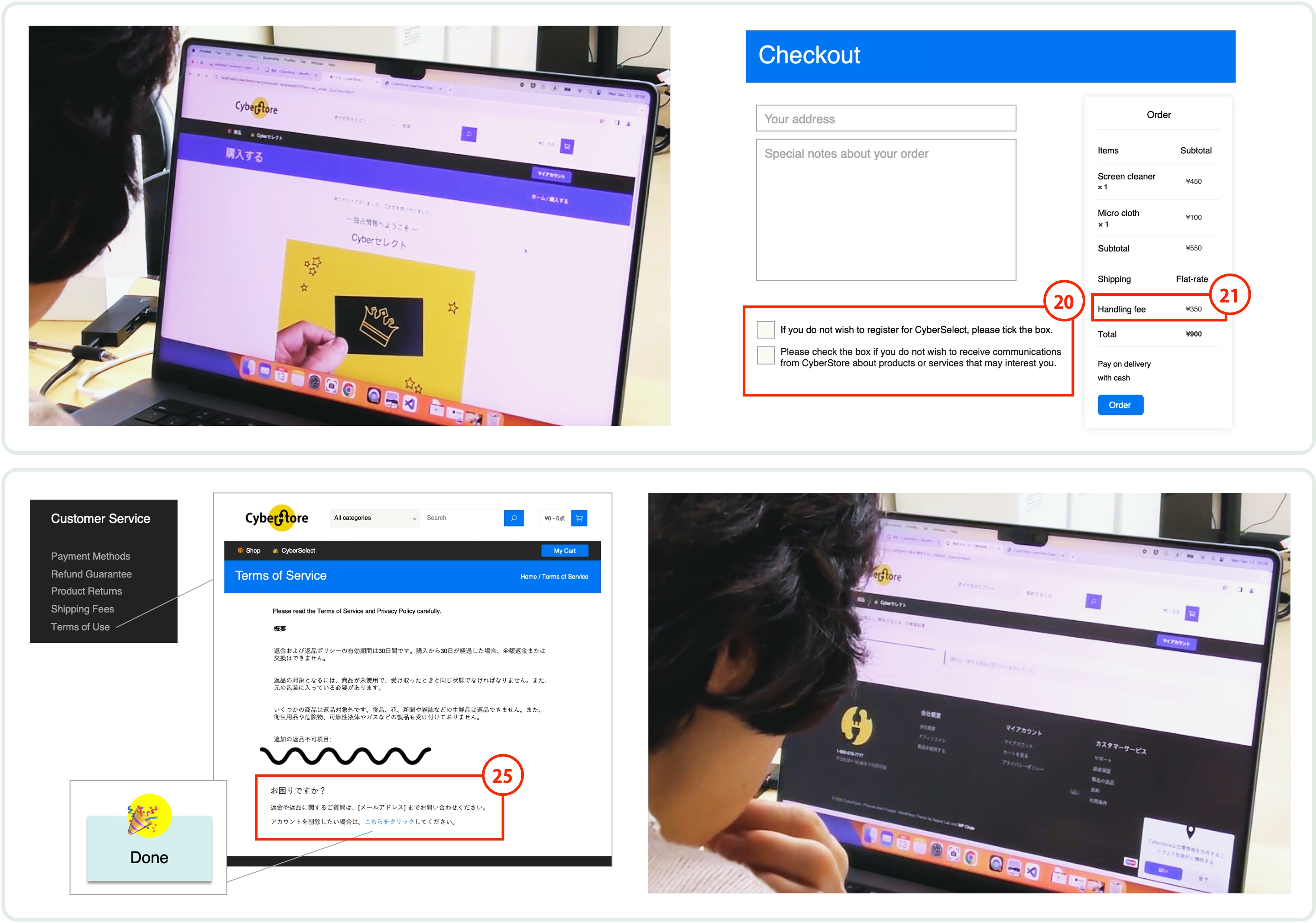}
    \caption{A participant encounters DPs in the simulated e-commerce website. After making a purchase, they are unintentionally signed up for the premium membership because of DP Case \#20, which includes the Trick Wording DP (top). They also did not notice DP Case \#21, Drip Pricing, which added an extra fee to their bill. At the end of the experience, they were unable to close their account because the necessary information was in English: the Untranslation DP (bottom). Note: We translated the close-ups of the screens from Japanese to English for optimal understanding; in kind, we have also flipped the translation for the Untranslation DP.}
    \label{fig:teaser}
\end{figure}

A simulated e-commerce website\enlargethispage{-20pt} was created for a fake electronics retailer (\autoref{fig:teaser}). The design of ``CyberStore'' was directly inspired by real websites used by Japanese people, such as Amazon Japan, Rakuten, Dospara, PC Koubou, Kakaku PC, Softmap, Tsukumo, and Mouse Computer. A similar aesthetic 
was used, in terms of palette and placement of basic elements, such as menus. The user flow represented typical everyday shopping experiences (refer to~\autoref{sec:tasks}). The first author created the website using a WordPress WooCommerce theme\footnote{Powered by the \href{https://wordpress.org/plugins/woocommerce/}{WooCommerce} \href{https://wordpress.org/}{WordPress} plugin with a customized \href{https://wpcirqle.com/}{WP Cirqle} theme as the base.} and then modified the theme by creating and embedding DPs according to the user flow (refer to \autoref{sec:dps} and \autoref{fig:userflow}). Two supporting developers created one DP each.

The website was developed in an iterative fashion, relying on a rapid experimentation procedure~\citep{soni2010rapid} comprised of short Agile-like~\citep{martin2003agile} cycles of rapid prototyping and pilot testing in-lab and with collaborators. Rather than assessing deceptibility at this stage, we focused on ensuring good usability and UX to avoid potential confounds related to poor usability and UX. This way, we could distinguish any negative perceptions about DPs from the baseline design.

\begin{figure*}
    \centering
    \includegraphics[width=\textwidth]{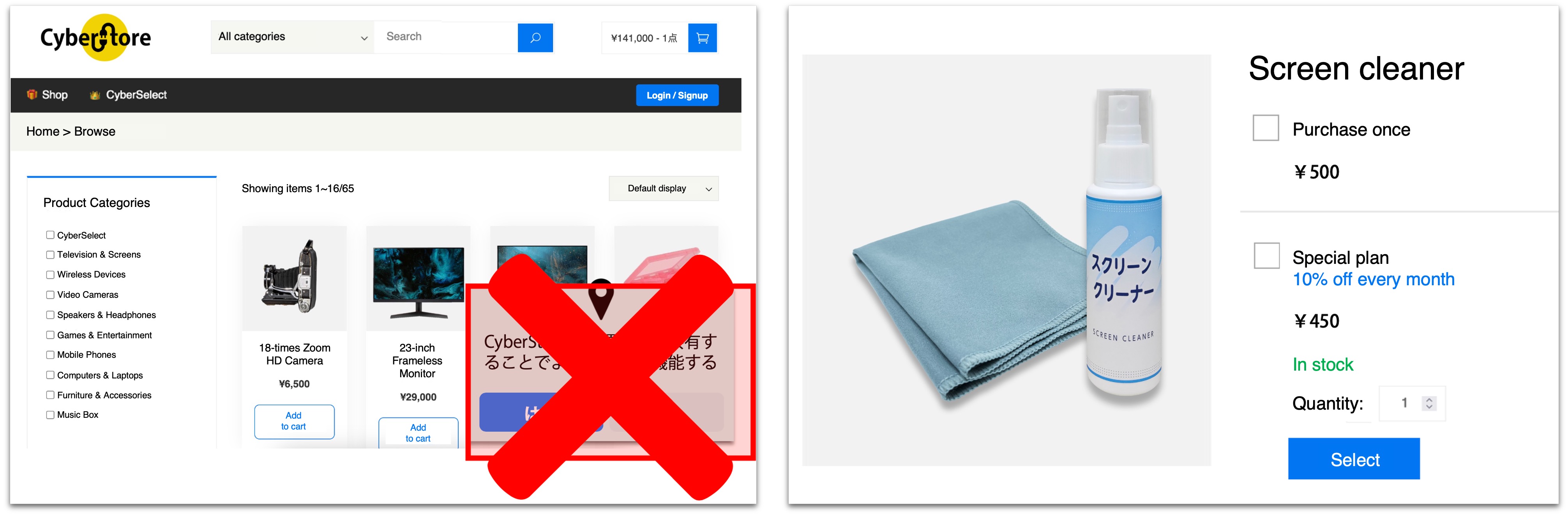}
    \caption{Screen shots of the ``CyberStore'' website without DPs. On the left is the Browse page with the location pop-up nagging DP crossed out. On the right is a DP-free version of the product page.}
    \label{fig:nodp}
\end{figure*}

The DP-free version of the website used as an experimental control was essentially the same website minus the DPs. For example, \autoref{fig:nodp} demonstrates how the location pop-up (Nagging DP) was removed.

\subsection{Selection and design of DPs}\label{sec:dps}

We used the seven \textbf{categories} of DPs compiled in 2022 by the Organisation for Economic Co-operation and Development~\citep[OECD;][]{oecd2022}, specifically Forced Action, Interface Interference, Nagging, Obstruction, Sneaking, Social Proof, Urgency, and Linguistic Dead-Ends. This selection was based on the taxonomical work of \citet{conti2010malicious}, \citet{bosch2016tales}, \citet{gray2018darkside}, \citet{mathur2019atscale}, and \citet{luguri2021shining}. In Japan, the OECD categorizations of DPs are currently the best known and discussed in the media. For example, public broadcaster NHK covered these categories in-depth for popular television show Close-Up Gendai on April 3rd, 2024.\footnote{\url{https://www.nhk.or.jp/minplus/0016/topic062.html} (note: in Japanese).} We also included the Japan-based DPs found in 2023 by \citet{hidaka2023linguistic}. This study was the first to evaluate user perceptions of these DPs, which have so far only been found within Japan.

\begin{figure}[!ht]
    \includegraphics[width=1\textwidth]{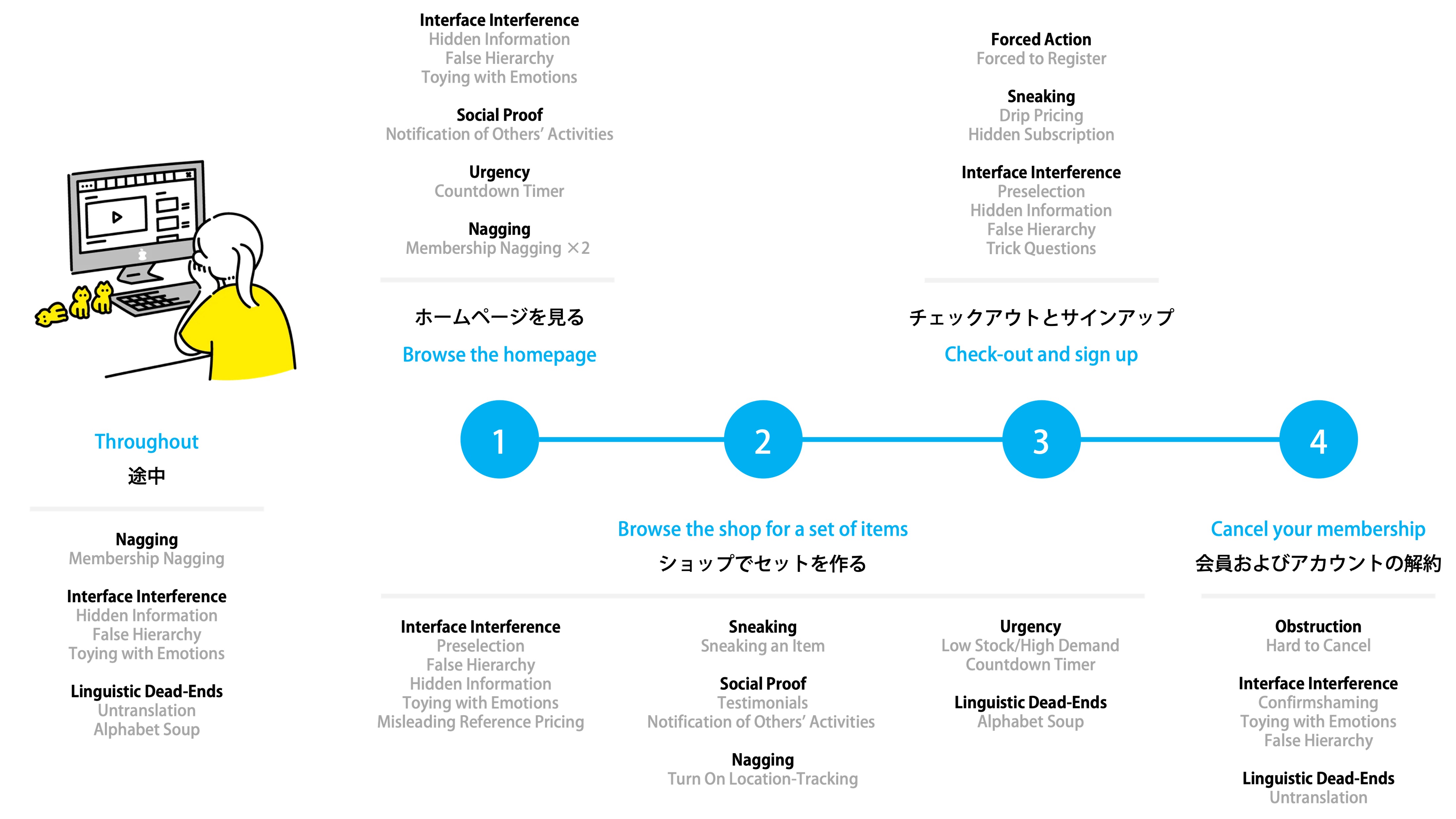}
    \caption{User flow of the website representing a typical online shopping experience involving all DPs. Illustration © \href{https://www.shigureni.com}{shigureni}.}
    \label{fig:userflow}
\end{figure}

For each category of DP, the first author selected a range of DP \textbf{classes}, or subtypes of DPs placed within these larger categories. They next created DP \textbf{cases}, or instances of an interaction featuring at least one DP class (and hence at least one DP category) and typically several as a gestalt~\citep{digeronimo2020}. These cases were drawn from common examples used in e-commerce websites in Japan and abroad, with many sourced from Brignull's ``Hall of Shame,''\footnote{\url{https://www.deceptive.design/hall-of-shame}.} a listing of real DPs that have received social, legal, and academic attention. The first author discussed the selection with the team and collaborator NHK, deciding on 25 DP cases made up of 51 instances of DP classes that covered all DP \nobreak{}categories. Each DP case was strategically placed within the user flow (refer to \autoref{fig:userflow}).

All DP categories, classes, and cases can be found in \autoref{fig:cheatsheet} and the visual guide on OSF.\footnote{\url{https://osf.io/65wzr}.}
A textual list of the cases is included in Appendix C (Supplementary Materials). Here, we cover a few examples to give the reader a sense of the simulation and its realism.

\begin{itemize}
    \item \textbf{Example~1: Same Computer, Different Story (DP Case \#12)}. A range of computers is available in the store. Two, however, are the same model. Even so, they have different names and price points: a {￥}5000 (USD $\sim${\$}32) difference. The more expensive one is placed first, ``above the fold''~\citep{bocchi2016above} or within the viewport at the top of the screen on page load. The less expensive one is placed ``below the fold,'' requiring scrolling and searching over other products. This DP is called Misleading Reference Pricing~\citep{gray2024ontology}. The False Hierarchy DP~\citep{gray2024ontology} creates a false sense of difference in importance through a visual hierarchy where one is hidden below the other. The Toying with Emotions DP~\citep{gray2024ontology} reinforces the Low Stock/High Demand~\citep{gray2024ontology} indicator with red, bold text and an exclamation mark, suggesting that the more expensive option is about to run out of stock. These DPs work in concert to exploit our tendency to choose the first item presented to us: the immediacy effect~\citep{loewenstein1991negative}. Consumers who do not fully browse or compare items may choose the first item that they come across, especially if a scarcity bias is in effect~\citep{Mittone2009}.

    \item \textbf{Example~2: Alphabet Soup, a Japanese DP (DP Case \#23)}. This DP, as yet only found in Japan~\citep{hidaka2023linguistic}, makes use of the local syllabary to mislead and confuse. In DP Case \#23, a member has signed up for a plan set to ``{￥}129.5'' a year. However, the meaning of this number is unclear, because the yen is a non-decimal currency, so 
    ``.5'' yen ({￥}) does not exist. The user may not notice, assume that the amount will be rounded up, or decide that the decimal point should be a comma and is also in the wrong place. The reality may be that this is Chinese yuan, which uses the same symbol as the Japanese yen (i.e., {￥}), or some other mislabeled currency that uses decimals, such as USD, which may also be a more appropriate yearly fee here (e.g., assuming that the price should be rounded up to {￥}130~in yen, this would be $\sim$USD 80 cents, compared to USD {\$}129.50).
\end{itemize}

\section{Methods}\label{isect9}

We conducted a two-phase between-subjects comparative study comprised of Phase 1, an in-person observational user study with the DP-filled version of the simulated online shop~\citep{stanton2017human}, and Phase 2, an online follow-up study with a non-DP version of the shop. In first phase, we used mixed methods, including a concurrent think-aloud protocol~\citep{van2003retrospective,alshammari2015ask}, pre- and post-task questionnaires to assess effects on baseline measures, and a semi-structured interview to identify whether, when, and how participants noticed any deception~\citep{blandford2016qualitative}. In the follow-up comparison study, we only used the questionnaires, because there were no DPs to observe. We received institutional ethics board (IRB) approval from Tokyo Institute of Technology\footnote{Now Institute of Science Tokyo.} (\#2023265). The first phase was conducted in December 2023 and the second in February 2025.

\subsection{Participants}\label{isect10}

We recruited 84 Japanese people who identified as having experience with online stores as consumers in two phases. The second phase recruitment was carried out based on reviewers' requests for a controlled comparison. Our goal was to gather a diverse sample of Japanese consumers in terms of age, educational background, and gender. 

For the first phase, wherein we conducted the observational study, we recruited 40 people through NHK, a public broadcaster and third-party recruiter with access to a range of participant pools. We had two cohorts: one anonymous group ($n=30$) and a smaller group ($n=10$) of non-anonymous participants for the public broadcast. The recruiter used several different participant pools, and each had a different rate of compensation. Participants were compensated in accordance with each pool. This ranged from $\sim$5000 to 8000 yen (USD $\sim${\$}30 to 50).

For the second phase---a follow-up study with the DP-free control---we recruited 44 people through Yahoo! Crowdsourcing, an online service with high reach across Japan that is linked to unique smartphones, nearly guaranteeing individuality and demographics~\citep{seaborn2025ycs}. Compensation accorded with institutional ethics at 600 yen (USD $\sim$4) for 30 minutes.

In the first phase anonymous group ($n=30$), there were fourteen women and sixteen men (none of another gender). Two were aged 18--24, three 25--34, two 35--44, eight 45--54, nine 55--64, four 65--74, and two 75+. Most had gone to university but had not completed their studies ($n=11$). Eight held a bachelor's degree, two a graduate degree, one a degree from a preparatory school, and seven the equivalent of a high school diploma. One had not completed high school yet. Most did not know about DPs (only four of 28; for two, this data was missing).

The first phase non-anonymous participants ($n=10$) included five women and five men (none of another gender). Two were aged 25--34, two 35--44, three 45--54, two 55--64, and one 65--74. Four held a high school diploma or equivalent, three had attended university but had no degree, and three held bachelor's degrees. Essentially all had not heard of DPs (one vaguely knew about the concept). The number of recruits was capped at ten by the recruiter.

In the second phase, anonymous participants ($n=44$) included 38 men,	six women, and	one who preferred not to say. Most ($n=18$) were aged
45--54,	13 were aged
35--44,	six were aged
55--64, four were aged 18--24, and three were aged 25--34.
Twenty-seven held bachelor's degrees, 
two went to university but did not have a degree,
two had not completed high school,
three had completed college, 
two were active students,
and eight had high school equivalent degrees.

First-phase participants consented in advance to use of their anonymous data and screen recordings in the public broadcast and research. The non-anonymous participants consented to use of their session and interview videos for the broadcast. The participants in the follow-up study were not included in the programme.

\subsection{Procedure}\label{sec:procedure}

All participants went through the same general procedure. We discuss one exception for the non-anonymous cohort. We also describe exceptions and special procedures for the follow-up study, given the online format and use as a control without DPs.

To avoid priming effects~\citep{head1988priming,bermejo2021cookie}, the true nature of the study was kept hidden until the end of the session. During recruitment, participants were told that they would evaluate the interactive features of a new e-commerce website. 

To start, participants were informed about the study procedure and then gave their consent. In the first phase, participants were asked to sit in front of the research laptop. In the second phase, online participants were asked to make themselves comfortable in front of their own computer. They filled out the pre-questionnaire (refer to \hyperref[sec:q]{Section~\ref{sec:q}}).

In the first phase, participants watched a training video about the concurrent think-aloud protocol before starting the main task. This method, long used within HCI, requires participants to voice their thoughts, feelings, and locus of attention out loud while they evaluate a system~\citep{van2003retrospective,alshammari2015ask}. This requires some cognitive effort and training with an example. For this, we created a short video wherein a researcher demonstrated the method while evaluating a cook book. Participants were asked to practice with the same cook book for $\sim$2 minutes. We note that the second phase cohort did not follow this procedure. While we acknowledge this as a potential threat to internal validity, there were several practical reasons at play. Since it was an online study, the procedural addition of a think-aloud protocol would be difficult. The study was also carried out via Yahoo! Crowdsourcing, which does not allow non-anonymous engagement, so video conferencing for think-aloud observation would not be permitted. Given that there were no DPs in the second phase, and no participant in first phase labelled other elements of the interface as deceptive, we would not expect the phase two cohort to perceive the clean version as deceptive.

In the main activity, participants in both phases went through the shopping website task-by-task using a written list mapped to the user flow (refer to \autoref{fig:userflow} and \autoref{sec:tasks}). Notably, participants were able to freely carry out the tasks in any way they wished, in any order. This maximized participant autonomy and created a more natural flow to the experience. At the same time, not all participants were able to encounter all DPs because some skipped certain pages or were unable to complete certain tasks (refer to \hyperref[sec:findings]{Section~\ref{sec:findings}}). In the first phase, a researcher filled out the observation checklist (refer to \hyperref[sec:checklist]{Section~\ref{sec:checklist}}) and took notes. When finished, participants completed the post-task questionnaire (\nobreak{}refer to \hyperref[sec:q]{Section~\ref{sec:q}}). Second phase participants then ended the study by returning to the recruitment platform and inputting a code for compensation.

In the first phase, a researcher interviewed the participant about their experience (refer to \hyperref[sec:semi]{Section~\ref{sec:semi}}). This represented the first step towards disclosure about the study's true aims. At this point, the concept of a DP was explained in relation to gathering deeper insights into their experience.
Participants were shown a PDF containing screen shots of each page in the website and were asked to point to and describe any notable parts. The observing researcher made a note of whether the participant noticed any DPs (refer to \hyperref[sec:checklist]{Section~\ref{sec:checklist}}). We considered this a primed ``second chance'' to notice DPs.

In the last step of the first phase, the host researcher revealed the focus of the study. Participants were asked if they wished to add to their comments in light of this knowledge. They were allowed to revisit the interactive website or review the PDF again, if desired. They were also asked to keep the true focus of the study on DPs a secret.

The non-anonymous participants in the first phase also participated in an extended interview for the public broadcast. They were asked to summarize, revisit, and add on to their earlier comments, if needed.

The first phase sessions took $\sim$50--60 minutes plus $\sim$15--30 minutes for the non-anonymous participants (for the broadcast). The second phase took $\sim$30 minutes.

\subsection{Tasks}\label{sec:tasks}

The user flow (\autoref{fig:userflow}) was designed to represent a typical series of tasks found in Japanese e-commerce websites. These were defined and ordered to maximize the likelihood of participants encountering all DPs embedded in the website without explicit direction (refer to \hyperref[sec:dps]{Section~\ref{sec:dps}}). The tasks were pilot tested with Japanese lab members ($n=8$)
and individual collaborators from NHK ($n=2$) who had experience with online shopping in Japan. The tasks were:
\begin{itemize}
    \item Browse the homepage freely
    \item Browse the shop for a specific set of items: a computer, a bag, and cleaning supplies
    \item Check out and account sign up (with preset and fake data)
    \item Close the account
\end{itemize}

Participants were given a written list of these tasks. However, they were not forced to complete each task nor were they prevented from exploring the website. 
Even so, they were given a 3-minute time limit for the last task, which proved difficult to achieve in pilot tests because of the severity of the DPs involved, specifically Untranslation.

\subsection{Data collection}\label{isect11}

We aimed to understand the effect of DPs on \textbf{individual welfare}, i.e., personal harm, and \textbf{individual autonomy}, i.e., loss of autonomy, two key lenses when evaluating DPs~\citep{mathur2021whatdark}. Here, we detail our instruments, measures, and data collection procedures in pursuit of viewing the experience through these ``lenses''~\citep{mathur2021whatdark}. We also wished to confirm the degree to which our sample was diverse and representative of the Japanese consumer population by gathering demographic information in a non-interfering way. Note that, in the second phase, we only captured the measures in \hyperref[sec:q]{Section~\ref{sec:q}}, since there were no DPs for participants to experience. The second phase was carried out to elucidate whether the pre/post measures related to the DPs or the online shopping experience itself.

\subsubsection{Questionnaire instruments and measures (both phases)}\label{sec:q}

We captured demographics and experience measures in two questionnaires, one pre-task and one immediately following completion of the task. Measures included:
\begin{itemize}
    \item \textbf{Affective State [Individual Welfare] (Pre/Post)}: We used valence and arousal to assess affective state before (as a baseline) and after the task. These variables are common subjective measures of emotion, and arousal can be particularly illuminating~\citep{Mauss2009}. To capture affective state, we used the valence and arousal dimensions from the face-based tactile version~\citep{iturregui2020towards} of the Self-Assessment Manikin (SAM)~\citep{bradley1994measuring}. The SAM is language-free and uses a 9-point semantic differential scale, where 1 is low valence/arousal and 9 is high valence/arousal.
    \item \textbf{Acceptability [Individual Welfare] (Post)}: We used the Japanese translation~\citep{yamano2015jpsus} of the 10-item System Usability Scale (SUS)~\citep{lewis2018system} to assess interface acceptability~\citep{bangor2008empirical} with a 5-point Likert scale. While nominally about ``usability,'' SUS scores in effect represent user acceptance~\citep{bangor2008empirical} and have been correlated with acceptability measures~\citep{NikAhmad2021}. We interpreted the SUS scores according to the ``acceptability ranges'' in \citet{bangor2008empirical} (refer to \hyperref[sec:quant]{Section~\ref{sec:quant}}).
    \item \textbf{Demographics [Representativeness] (Post)}: We captured gender (man, woman, non-binary/X-gender, another gender), age (categorical in five-year increments), and education (from high school onwards). We included these items in the post-task questionnaire to avoid order biases and stereotype threats~\citep{Kalton1982}.
\end{itemize}

\subsubsection{Observation checklist and measures (Phase 1)}\label{sec:checklist}

We provided a spreadsheet checklist to observers who oversaw the session. The checklist was divided into the sections and pages in the user flow (refer to \autoref{fig:userflow}). Observers were tasked with recording three measures for each DP experienced by participants, as appropriate:
\begin{itemize}
    \item \textbf{Noticeability [Individual Autonomy]}: This refers to whether and when the DP was \emph{noticed} by each participant, if at all. This measure, similar to one in \citet{utz20219gdprconsent}, addresses the extent to which the DP prevents user autonomy through concealment or lack of awareness of actions and choices. The options were: \emph{On their own} (during the na\"{i}ve interaction); \emph{After prompting} (via the think-aloud protocol); \emph{During the reveal} (in the interview); \emph{Did not notice} (at any stage); \emph{Missed chance} (did not encounter during the interactive part); and \emph{Did not notice during reveal}.
    \item \textbf{Deceptibility [Individual Autonomy]}: We captured in what way each DP was \emph{deceptive}, if noticed, as a measure of subjective agreement about \emph{how} the DP affected participant autonomy. The options were: \emph{Deceitful,} felt manipulated or without a choice in a negative way, as in \citet{Machuletz2020}; \emph{Disruptive,} felt disturbed or interrupted and wanted to avoid the DP; \emph{Indifferent}, accepted it without concern; and \emph{Unaffected}, i.e., tricked.

    \item \textbf{Task Completion [Individual Autonomy]}: We generated counts and percentages based on each participant's ability to complete certain tasks that modified the path through the experience based on the user flow. Specifically, these were: avoiding the ``CyberSelect'' premium membership set as the default on the purchase screen (DP Case \#22: Hidden Subscription) and account cancellation (DP Case \#25: Untranslation~+~Hard to Cancel).

    \item \textbf{Simulated financial Loss [Individual Welfare]}: We created metrics (counts and percentages of those affected) on simulated \emph{financial loss}, a measure of harm for individual consumers~\citep{mathur2021whatdark}, associated with DPs and tasks, where possible. These included: buying the more expensive computer (DP Case \#12: Misleading Reference Pricing); subscribing to the cleaner (DP Case \#15: Preselection~+~False Hierarchy~+~Hidden Subscription); not noticing the sneaked-in cloth (DP Case \#16: Sneaking in an Item); accidentally signing up for the premium account (DP Case \#20: Preselection~+~Trick Questions~+~Membership Nagging~+~Hidden Subscription); and not noticing the extra fee at checkout (DP Case \#21: Drip Pricing).
\end{itemize}

To account for priming~\citep{head1988priming}, disclosure of our focus on DPs was staged. Observers used the checklist to distinguish between three stages: (i) first experiencing the website na\"{i}vely; (ii) after being prompted to comment on the page, as part of the concurrent think-aloud protocol; and (iii) during the semi-structured interview, wherein the focus on DPs was revealed (refer to \hyperref[sec:semi]{Section~\ref{sec:semi}}). Subsequently, the ``best performing'' DPs were those that remained unnoticed.

\subsubsection{Semi-structured interview (Phase 1)}\label{sec:semi}

We used a semi-structured interview approach~\citep{blandford2016qualitative}. We aimed to avoid priming by way of a funnel approach to the questions~\citep{wilson2013interview}, which also allowed us to assess level of familiarity with DPs. We started with broad, open questions, then questions about prior experiences, then questions about the current experience, and lastly an open question about general impressions. 
The guiding questions were: \emph{Have you heard of DPs before? Have you experienced DPs before? Were there any aspects of the design that you felt were deceptive during operation? What other feelings or impressions do you have about the design?}
Refer to Appendix B for the Japanese versions of the questions. During the interview, we provided a PDF of all pages on the website (refer to Appendix A). We asked participants to point out relevant areas in the PDF that aligned with their perceptions and feelings.

\subsection{Data analysis}\label{isect12}

We used Google Sheets and the online R calculators provided by Statistics Kingdom\footnote{\url{https://www.statskingdom.com/}.} for most data analyses. As per the IRB decree (discussed in \hyperref[sec:cohorts]{Section~\ref{sec:cohorts}}), the data from the two Phase 1 cohorts (anonymous $n=30$, non-anonymous $n=10$) were kept separate, but presented together and compared, where possible (refer to \hyperref[sec:comparison]{Section~\ref{sec:comparison}}). The Phase 2 data ($n=44$) were used as a control condition and treated as a between-subjects group, i.e., a ``No DPs'' group compared to the first phase ``DPs'' group.

\subsubsection{Quantitative data}\label{sec:quant}

We calculated a range of descriptive statistics for all quantitative data, notably counts and percentages, mean (M), median (MD), standard deviation (SD), and interquartile range (IQR), where appropriate. For the observational data, we created descriptives by participant, DP class, DP case, and DP category. 
The questionnaire measures were analyzed according to the instrument. For the SUS, we used the acceptability interpretation~\citep{bangor2008empirical}, where 51.6 was unacceptable, 51.7--71 was marginal, and 71.1--100 was acceptable.

Most measures were treated as single point data. We also carried out inferential statistics tests. These included within-subject evaluations of a change in affective state before and after the experience and between-subject evaluations of differences between the anonymous and non-anonymous cohorts in Phase 1, as well as differences between Phase 1 and Phase 2. 
We also explored relationships between the other variables.
Parametric (e.g., paired t-tests and Welch's t-tests) and non-parametric tests (e.g., Wilcoxon signed rank tests) were used based on type of data and the results from normality tests, notably the Shapiro-Wilk test. One anonymous participant did not supply ``before'' affect scores and another did not fill out the SUS; their data were excluded from analyses using these measures. All other data were retained.

\subsubsection{Qualitative data}\label{isect13}

We used a hybrid thematic analysis approach to analyze the qualitative data from the interviews. Hybrid thematic analysis combines deductive and inductive theme development procedures~\citep{Swain2018}, where both the data itself and applied theories drive the development of codes and themes. This approach acknowledges the constructive nature of knowledge production and the non-neutrality of researchers, as well as allows for cyclical development between novel data and existing frameworks. As per \citet{Swain2018}, our hybrid thematic analysis was carried out in three phases: (i) table preparation with a priori code creation and initial data engagement; (ii) continued a priori coding alongside a posteriori coding with deeper engagement in the data; and (iii) combining codes into meaningful themes.

Two researchers---one a native Japanese speaker with some knowledge of English and one a native English speaker with advanced Japanese---carried out concurrent coding with the aim of developing knowledge about the Japanese context. The Japanese researcher led the process, developing most initial codes. Then, the two discussed the codes, working together to complete Phase III and craft a thematic framework. Disagreements were discussed until consensus was reached.

\begin{table*}[ht!]
\caption{Noticeability by location and DP case, divided by cohort. A: Anonymous ($n=30$). N-A: Non-Anonymous ($n=10$).}
\label{tab:notice}
\small
\begin{tabular}{l|rr|rr|rr|rr|rr|rr|rr}
\toprule
DP Case \#& \multicolumn{2}{l|}{First} & \multicolumn{2}{l|}{Second} & \multicolumn{2}{l|}{Third} & \multicolumn{2}{l}{Total Noticed} & \multicolumn{2}{l|}{\%} & \multicolumn{2}{l}{Missed} & \multicolumn{2}{l}{\%} \\
 & \multicolumn{1}{r}{A} & \multicolumn{1}{r|}{N-A} & \multicolumn{1}{r}{A} & \multicolumn{1}{r|}{N-A} & \multicolumn{1}{r}{A} & \multicolumn{1}{r|}{N-A} & \multicolumn{1}{r}{A} & \multicolumn{1}{r|}{N-A} & \multicolumn{1}{r}{A} & \multicolumn{1}{r|}{N-A} & \multicolumn{1}{r}{A} & \multicolumn{1}{r|}{N-A} & \multicolumn{1}{r}{A} & \multicolumn{1}{r}{N-A} \\
\midrule
\multicolumn{15}{l}{Throughout} \\
1 & 9 & 1 & 2 & 0 & 3 & 2 & 14 & 3 & 47\% & 30\% & 16 & 7 & 53\% & 70\% \\
2 & 9 & 5 & 5 & 0 & 11 & 9 & 5 & 2.8 & 17\% & 28\% & 25 & 7.2 & 83\% & 72\% \\
3 & 12 & 4 & 4 & 0 & 5 & 3 & 21 & 7 & 70\% & 70\% & 9 & 3 & 30\% & 30\% \\
\midrule
\multicolumn{15}{l}{Homepage} \\
4 & 12 & 5 & 1 & 0 & 2 & 1 & 7.5 & 3 & 25\% & 30\% & 22.5 & 7 & 75\% & 70\% \\
5 & 18 & 8 & 4 & 0 & 3 & 1 & 25 & 9 & 83\% & 90\% & 5 & 1 & 17\% & 10\% \\
6 & 5 & 0 & 1 & 1 & 3 & 1 & 9 & 2 & 30\% & 20\% & 21 & 8 & 70\% & 80\% \\
7 & 6 & 4 & 1 & 0 & 1 & 1 & 8 & 5 & 27\% & 50\% & 22 & 5 & 73\% & 50\% \\
8 & 5 & 1 & 2 & 0 & 2 & 0 & 9 & 1 & 30\% & 10\% & 21 & 9 & 70\% & 90\% \\
\midrule
\multicolumn{15}{l}{Shopping} \\
9 & 5 & 0 & 2 & 0 & 3 & 1 & 3.3 & 0.3 & 11\% & 3\% & 26.7 & 9.7 & 89\% & 97\% \\
10 & 21 & 9 & 1 & 0 & 10 & 8 & 10.7 & 5.7 & 36\% & 57\% & 19.3 & 4.3 & 64\% & 43\% \\
11 & 21 & 6 & 1 & 0 & 2 & 1 & 24 & 7 & 80\% & 70\% & 6 & 3 & 20\% & 30\% \\
12 & 45 & 14 & 4 & 0 & 1 & 4 & 12.5 & 4.5 & 42\% & 45\% & 17.5 & 5.5 & 58\% & 55\% \\
13 & 22 & 5 & 0 & 0 & 1 & 1 & 23 & 6 & 77\% & 60\% & 7 & 4 & 23\% & 40\% \\
14 & 17 & 4 & 0 & 0 & 3 & 0 & 20 & 4 & 67\% & 40\% & 10 & 6 & 33\% & 60\% \\
15 & 32 & 5 & 3 & 0 & 10 & 4 & 15 & 3 & 50\% & 30\% & 15 & 7 & 50\% & 70\% \\
16 & 22 & 8 & 1 & 1 & 0 & 0 & 23 & 9 & 77\% & 90\% & 7 & 1 & 23\% & 10\% \\
17 & 4 & 2 & 0 & 1 & 2 & 1 & 6 & 4 & 20\% & 40\% & 24 & 6 & 80\% & 60\% \\
\midrule
\multicolumn{15}{l}{Checkout} \\
18 & 7 & 5 & 0 & 0 & 1 & 0 & 8 & 5 & 27\% & 50\% & 22 & 5 & 73\% & 50\% \\
19 & 40 & 18 & 0 & 0 & 2 & 5 & 8.4 & 4.6 & 28\% & 46\% & 21.6 & 5.4 & 72\% & 54\% \\
20 & 73 & 13 & 8 & 0 & 6 & 5 & 17.4 & 3.6 & 58\% & 36\% & 12.6 & 6.4 & 42\% & 64\% \\
21 & 8 & 4 & 2 & 0 & 2 & 0 & 12 & 4 & 40\% & 40\% & 18 & 6 & 60\% & 60\% \\
22 & 7 & 2 & 2 & 0 & 4 & 2 & 6.5 & 2 & 22\% & 20\% & 23.5 & 8 & 78\% & 80\% \\
23 & 0 & 4 & 3 & 0 & 0 & 0 & 3 & 4 & 10\% & 40\% & 27 & 6 & 90\% & 60\% \\
\midrule
\multicolumn{15}{l}{Account} \\
24 & 12 & 14 & 0 & 0 & 8 & 6 & 6.7 & 6.7 & 22\% & 67\% & 23.3 & 3.3 & 78\% & 33\% \\
25 & 30 & 8 & 6 & 0 & 3 & 3 & 19.5 & 5.5 & 65\% & 55\% & 10.5 & 4.5 & 35\% & 45\% \\
\bottomrule
\end{tabular}
\end{table*}

\subsubsection{Comparison of cohorts}\label{sec:cohorts}

Institutional ethics restrictions prevented us from combining the data sets of the anonymous ($n=30$) and non-anonymous ($n=10$) participants. The risk of potential sampling biases and resulting performance or behaviour changes influenced this decision. The non-anonymous participants responded to a special call for participation in the public broadcast. This could have resulted in self-selection and especially voluntary response biases~\citep{Abeler2014}, where participants chose to participate or not based on the chance to be on television. \citet{Tripepi2010} point out that this can negatively affect the ``internal validity of the study if it is related to the exposure'' (p.c98). In our case, the recruiter 
was a public entity. 
Still, the non-anonymous advertisement may have attracted people who wished to be on television, including actors and celebrities. At the same time, we had no reason to believe that this would affect internal validity. The subject of the study---DPs---was kept confidential. Also, DPs can be so subtle and insidious that special measures are needed even for experts during heuristic and content evaluations, i.e., DP ``blindness'' or ``evasiveness''~\citep{digeronimo2020,hidaka2023linguistic}.

As per the IRB decree, we separated the results by cohort, presenting the descriptive statistics for each side-by-side. We also \nobreak{}compared the cohorts statistically to determine whether the above issues played a role. As per the IRB ruling, we separated these statistics from the main results in \hyperref[sec:comparison]{Section~\ref{sec:comparison}}. Given the uneven groups, we used non-parametric statistical tests, specifically the Chi-squared test~\citep{greenwood1996guide} for thematic frequencies and the Mann-Whitney \textit{U} test for ordinal or ratio data, i.e., SAM and SUS scores~\citep{Akritas1997mannwhit}.

After finding that there were no significant differences, we then grouped the data from both cohorts as a single Phase 1 dataset ($n=40$) and compared these data to the Phase 2 dataset ($n=44$).


\section{Findings}\label{sec:findings}

We report on our mixed methods findings by RQ, dividing but comparing by cohort, where appropriate. 
We recommend referring to the visual cheat sheet for the DP cases in \autoref{fig:cheatsheet} when reading the text results.

\subsection{Who is deceived? Noticeability, deceptibility, task completion, and simulated financial loss (Phase 1, RQ1)}\label{isect14}

We considered whether and at which stage the DPs were noticeable, how deceptive they were, whether task completion was achieved, and what potential simulated financial loss for participants (or gain for CyberStore) could occur.

\subsubsection{Noticeability}\label{isect15}

Full descriptive statistics are presented in \autoref{tab:notice}. On average, over half of DP cases went unnoticed by the anon. (58\%, M~=~17.3, SD~=~7, MD~=~19.3, IQR~=~12) and non-anon. participants (55\%, M~=~5.5, SD~=~2.2, MD~=~6.0, IQR~=~2.7).

The most noticeable DP cases were the homepage call-outs (anon.: $n=25$, 83\%; non-anon.: $n=9$, 90\%), the low stock indicators (anon.: $n=24$, 80\%; non-anon.: $n=7$, 70\%), the testimonials on the product pages (anon.: $n=23$, 77\%; non-anon.: $n=6$, 60\%), and Sneaking an Item (anon.: $n=23$, 77\%, non-anon.: $n=9$, 90\%). For the anon., the least noticeable DP cases were the CyberSelect membership fee featuring Alphabet Soup and the Social Proof on the search dropdown (three each). For the non-anon., the least noticeable were the search dropdown Social Proof  and the CyberSelect membership call-outs (one each).

Descriptive statistics by DP category and class are presented in \autoref{tab:noticebycat}. The most noticeable category for the anon. participants was Obstruction (67\%) and Forced Action for the non-anon. participants (73\%). For classes, Sneaking an Item (Sneaking) was most noticeable for the anon. (77\%) and non-anon. (90\%) participants, and Testimonials (Social Proof) was also most noticeable for the anon. participants (77\%). For anon. and non-anon. participants, Alphabet Soup (Linguistic Dead-End; 90\% and 75\%, respectively) and Hidden Information (Interface Interference; 70\% each) were least noticeable. Misleading Reference Pricing (Interface Interference) was also least noticeable for anon. participants (90\%). Notification of Others' Activities (Social Proof) was least noticeable for non-anon. participants (80\%).

\begin{table*}[ht!]
\caption{Noticeability by DP category and class across cohorts. A: Anonymous ($n=30$). N-A: Non-Anonymous ($n=10$).}
\label{tab:noticebycat}
\small
\begin{tabular}{lrrrr}
\toprule
DP Category and Class & \multicolumn{2}{l}{Noticed \%} & \multicolumn{2}{l}{Missed \%} \\
 & \multicolumn{1}{l}{A} & \multicolumn{1}{l}{N-A} & \multicolumn{1}{l}{A} & \multicolumn{1}{l}{N-A} \\
\midrule
\textbf{Nagging:} & \textbf{41\%} & 32\% & \textbf{59\%} & 68\% \\
Membership Nagging & 38\% & 26\% & 62\% & 74\% \\
Turn On Location-Tracking & 53\% & 60\% & 47\% & 40\% \\
\midrule
\textbf{Interface Interference:} & \textbf{33\%} & 42\% & \textbf{68\%} & 58\% \\
Hidden Information & 20\% & 20\% & 80\% & 80\% \\
False Hierarchy & 29\% & 36\% & 71\% & 64\% \\
Toying with Emotions & 39\% & 62\% & 61\% & 38\% \\
Preselection & 39\% & 46\% & 61\% & 54\% \\
Misleading Reference Pricing & 10\% & 30\% & 90\% & 70\% \\
Trick Questions & 73\% & 40\% & 27\% & 60\% \\
Confirmshaming & 27\% & 70\% & 73\% & 30\% \\
\midrule
\textbf{Linguistic Dead-Ends:} & \textbf{38\%} & 40\% & \textbf{62\%} & 60\% \\
Alphabet Soup & 10\% & 25\% & 90\% & 75\% \\
Untranslation & 67\% & 55\% & 33\% & 45\% \\
\midrule
\textbf{Social Proof:} & \textbf{48\%} & 30\% & \textbf{53\%} & 70\% \\
Notification of Others' Activities & 38\% & 20\% & 62\% & 80\% \\
Testimonials & 77\% & 60\% & 23\% & 40\% \\
\midrule
\textbf{Urgency:} & \textbf{48\%} & 58\% & \textbf{52\%} & 43\% \\
Countdown Timer & 23\% & 45\% & 77\% & 55\% \\
Low Stock/High Demand & 73\% & 70\% & 27\% & 30\% \\
\midrule
\textbf{Sneaking:} & \textbf{42\%} & 40\% & \textbf{58\%} & 60\% \\
Sneaking an Item & 77\% & 90\% & 23\% & 10\% \\
Hidden Subscription & 35\% & 30\% & 65\% & 70\% \\
Drip Pricing & 40\% & 40\% & 60\% & 60\% \\
\midrule
\textbf{Forced Action:} & \textbf{27\%} & 50\% & \textbf{73\%} & 50\% \\
Forced to Register & 27\% & 50\% & 73\% & 50\% \\
\midrule
\textbf{Obstruction:} & \textbf{67\%} & 70\% & \textbf{33\%} & 30\% \\
Hard to Cancel & 67\% & 70\% & 33\% & 30\%\\
\bottomrule
\end{tabular}
\end{table*}

\subsubsection{Deceptibility}\label{isect16}

Reactions and the relative lack thereof are presented in \autoref{tab:deceptability}. Percentages for the reactions when not tricked, i.e., when DPs were noticed at one of the three stages, were calculated to indicate the \emph{relative spread of the \textbf{types} of reactions}---Deceitful, Disruptive, Indifferent---to any given, noticed DP.

Most DP cases (anon.: 58\%; non-anon.: 55\%) went unnoticed: no reaction or apparent affect (anon.: M~=~17.3, SD~=~7, MD~=~19.3, IQR~=~12; non-anon.: M~=~5.5, SD~=~2.2, MD~=~6, IQR~=~2.7).
When noticed, anon. participants reacted with indifference to 47\% of cases (M~=~6.4, SD~=~6, MD~=~5, IQR~=~3.5), while non-anon. participants reacted the same way to 52\% of cases (M~=~2.3, SD~=~2, MD~=~2, IQR~=~3). 29\% were deemed deceitful by anon. participants (M~=~3.7, SD~=~4.9, MD~=~2, IQR~=~2) and 25\% were deemed so by non-anon. participants (M~=~1.3, SD~=~1.6, MD~=~1, IQR~=~1). Anon. participants deemed 24\% disruptive (M~=~2.6, SD~=~2.9, MD~=~2, IQR~=~2), while non-anon. participants deemed 23\% disruptive (M~=~0.9, SD~=~1, MD~=~1, IQR~=~1.3).

The most \emph{deceitful} DP cases to the anon. participants were the sneaked-in cloth (18 or 78\% of 23 reactions; DP Case \#16) and the English terms of service (17 or 81\% of 20 reactions; DP Case \#3). The English terms of service was also deemed highly deceitful by non-anon. participants (6 or 86\% of 7 reactions; DP Case \#3), who also found the account cancellation procedure highly deceitful (4.5 or 82\% of 5.5 reactions; DP Case \#25).
The most \emph{disruptive} DP case to the anon. participants were the trick wording (13 or 75\% of 17.4 reactions; DP Case \#20) and the membership options ad (3.6 or 72\% of 5 reactions; DP Case \#2). For non-anon. participants, the most disruptive DP case was the membership banner ads (everyone; DP Case \#8).

Anon. participants were generally \emph{indifferent} to low stock indicators (22 or 92\% of 24 reactions; DP Case \#11), the same computer priced differently (11 or 88\% of 12.5 reactions; DP Case \#12), social media profile icons (19 or 83\% of 22 reactions; DP Case \#13), and testimonials (17 or 85\% of 20 reactions; DP Case \#14). The non-anon. participants were indifferent to the social media profile icons in the search dropdown (everyone; DP Case \#9) and testimonials (everyone; DP Case \#14).

\begin{table*}[ht!]
\caption{Deceptibility by location and DP case, divided by cohort. Note: The percentages for Deceitful, Disruptive, and Indifferent are relative, i.e., the reaction when the DP was noticed. A: Anonymous ($n=30$). N-A: Non-Anonymous ($n=10$).}
\label{tab:deceptability}
\footnotesize
\begin{tabular}{p{.8cm}|rr|rr|rr|rr|rr|rr|rr|rr}
\toprule
\multirow{2}*{\makecell{DP \\Case \#}} & \multicolumn{2}{l}{Deceitful} & \multicolumn{2}{l|}{\%} & \multicolumn{2}{l}{Disruptive} & \multicolumn{2}{l|}{\%} & \multicolumn{2}{l}{Indifferent} & \multicolumn{2}{l|}{\%} & \multicolumn{2}{l}{Unaffected} & \multicolumn{2}{l}{\%} \\
 & A & N-A & A & N-A & A & N-A & A & N-A & A & N-A & A & N-A & A & N-A & A & N-A \\
\midrule
\multicolumn{17}{l}{Throughout} \\
1 & 2 & 0 & 14\% & 0\% & 5 & 1 & 36\% & 33\% & 7 & 2 & 50\% & 67\% & 16 & 7 & 53\% & 70\% \\
2 & 0.8 & 0.2 & 16\% & 7\% & 3.6 & 1.4 & 72\% & 50\% & 0.6 & 1.2 & 12\% & 43\% & 25 & 7.2 & 83\% & 72\% \\
3 & 17 & 6 & 81\% & 86\% & 1 & 1 & 5\% & 14\% & 3 & 0 & 14\% & 0\% & 9 & 3 & 30\% & 30\% \\
\midrule
\multicolumn{17}{l}{Homepage} \\
4 & 1 & 0 & 13\% & 0\% & 1 & 0.5 & 13\% & 17\% & 5.5 & 2.5 & 73\% & 83\% & 22.5 & 7 & 75\% & 70\% \\
5 & 5 & 0 & 20\% & 0\% & 3 & 1 & 12\% & 11\% & 17 & 8 & 68\% & 89\% & 5 & 1 & 17\% & 10\% \\
6 & 2 & 1 & 22\% & 50\% & 1 & 0 & 11\% & 0\% & 6 & 1 & 67\% & 50\% & 21 & 8 & 70\% & 80\% \\
7 & \multicolumn{1}{l}{} & 1 & 0\% & 20\% & 2 & 0 & 25\% & 0\% & 6 & 4 & 75\% & 80\% & 22 & 5 & 73\% & 50\% \\
8 & 1 & 0 & 11\% & 0\% & 3 & 1 & 33\% & 100\% & 5 & 0 & 56\% & 0\% & 21 & 9 & 70\% & 90\% \\
\midrule
\multicolumn{17}{l}{Shopping} \\
9 & 1.3 & 0 & 40\% & 0\% & 0.3 & 0 & 10\% & 0\% & 1.7 & 0.3 & 50\% & 100\% & 26.7 & 9.7 & 89\% & 97\% \\
10 & 1.7 & 1.3 & 16\% & 24\% & 7.3 & 4.3 & 69\% & 76\% & 1.7 & 0 & 16\% & 0\% & 19.3 & 4.3 & 64\% & 43\% \\
11 & 0 & 1 & 0\% & 14\% & 2 & 2 & 8\% & 29\% & 22 & 4 & 92\% & 57\% & 6 & 3 & 20\% & 30\% \\
12 & 0.5 & 0.25 & 4\% & 6\% & 1 & 1.3 & 8\% & 28\% & 11 & 3 & 88\% & 67\% & 17.5 & 5.5 & 58\% & 55\% \\
13 & 3 & 1 & 13\% & 17\% & 1 & 0 & 4\% & 0\% & 19 & 5 & 83\% & 83\% & 7 & 4 & 23\% & 40\% \\
14 & 2 & 0 & 10\% & 0\% & 1 & 0 & 5\% & 0\% & 17 & 4 & 85\% & 100\% & 10 & 6 & 33\% & 60\% \\
15 & 4 & 1 & 27\% & 33\% & 5 & 0.7 & 33\% & 22\% & 6 & 1.3 & 40\% & 44\% & 15 & 7 & 50\% & 70\% \\
16 & 18 & 4 & 78\% & 44\% & 0 & 0 & 0\% & 0\% & 5 & 5 & 22\% & 56\% & 7 & 1 & 23\% & 10\% \\
17 & 3 & 1 & 50\% & 25\% & 0 & 0 & 0\% & 0\% & 3 & 3 & 50\% & 75\% & 24 & 6 & 80\% & 60\% \\
\midrule
\multicolumn{17}{l}{Checkout} \\
18 & 3 & 0 & 38\% & 0\% & 0 & 1 & 0\% & 20\% & 5 & 4 & 63\% & 80\% & 22 & 5 & 73\% & 50\% \\
19 & 0.4 & 1 & 5\% & 22\% & 5 & 1.6 & 60\% & 35\% & 3 & 2 & 36\% & 43\% & 21.6 & 5.4 & 72\% & 54\% \\
20 & 1 & 1 & 6\% & 28\% & 13 & 2.2 & 75\% & 61\% & 3.4 & 0.4 & 20\% & 11\% & 12.6 & 6.4 & 42\% & 64\% \\
21 & 6 & 3 & 50\% & 75\% & 0 & 0 & 0\% & 0\% & 6 & 1 & 50\% & 25\% & 18 & 6 & 60\% & 60\% \\
22 & 1.5 & 1 & 23\% & 50\% & 3 & 0.5 & 46\% & 25\% & 2 & 0.5 & 31\% & 25\% & 23.5 & 8 & 78\% & 80\% \\
23 & 2 & 0 & 67\% & 0\% & 1 & 1 & 33\% & 25\% & 0 & 3 & 0\% & 75\% & 27 & 6 & 90\% & 60\% \\
\midrule
\multicolumn{17}{l}{Account} \\
24 & 3 & 3.7 & 45\% & 55\% & 2.3 & 1.7 & 35\% & 25\% & 1.3 & 1.3 & 20\% & 20\% & 23.3 & 3.3 & 78\% & 33\% \\
25 & 14 & 4.5 & 72\% & 82\% & 3 & 0 & 15\% & 0\% & 2.5 & 1 & 13\% & 18\% & 10.5 & 4.5 & 35\% & 45\%\\
\bottomrule
\end{tabular}
\end{table*}

We now consider the shares of types of reactions by deceptibility (\autoref{tab:decepticat}). Both cohorts found Obstruction DPs the most deceitful (anon.: 53\%; non-anon.: 60\%) and Nagging DPs the most disruptive (anon.: 24\%; non-anon.: 18\%). Both cohorts were most indifferent to Urgency DPs (anon.: 41\%; non-anon.: 40\%), but non-anon. participants were also indifferent to Forced Action DPs (40\%). Anon. participants found Sneaking an Item (Sneaking; 60\%), Hard to Cancel (Obstruction; 53\%), and Untranslation (Linguistic Dead-End; 48\%) most deceitful. Non-anon. participants found Hard to Cancel (Obstruction; 60\%), Untranslation (Linguistic Dead-End; 45\%), Sneaking an Item (Sneaking; 40\%), and Confirmshaming (Interface Interference; 40\%) most deceitful. Trick Questions (Interface Interference) were most disruptive for anon. participants (53\%), while Turn On Location-Tracking (Nagging) was most disruptive for the non-anon. cohort (50\%). Anon. participants were indifferent to Low Stock/High Demand (Urgency; 67\%) and Testimonials (Social Proof; 63\%). For non-anon., these were Testimonials (Social Proof; 50\%) and Sneaking an Item (Sneaking; 50\%).

\begin{table}[ht!]
\caption{Deceptibility by DP category and class across cohorts. A: Anonymous ($n=30$). N-A: Non-Anonymous ($n=10$).}
\label{tab:decepticat}
\begin{tabular}{lrrrrrrrr}
\toprule
DP Category and Class & \multicolumn{2}{l}{Unaffected \%} & \multicolumn{2}{l}{Indifferent \%} & \multicolumn{2}{l}{Deceitful \%} & \multicolumn{2}{l}{Disruptive \%} \\
 & A & N-A & A & N-A & A & N-A & A & N-A \\
\midrule
\textbf{Nagging:} & \textbf{59\%} & 68\% & \textbf{11\%} & 10\% & \textbf{5\%} & 3\% & \textbf{24\%} & 18\% \\
Membership Nagging & 62\% & 74\% & 11\% & 12\% & 4\% & 2\% & 23\% & 12\% \\
Turn On Location-Tracking & 47\% & 40\% & 10\% & 0\% & 10\% & 10\% & 33\% & 50\% \\
\midrule
\textbf{Interface Interference:} & \textbf{68\%} & 58\% & \textbf{15\%} & 17\% & \textbf{6\%} & 12\% & \textbf{12\%} & 13\% \\
Hidden Information & 80\% & 80\% & 7\% & 8\% & 7\% & 5\% & 7\% & 8\% \\
False Hierarchy & 71\% & 64\% & 13\% & 13\% & 5\% & 10\% & 10\% & 13\% \\
Toying with Emotions & 61\% & 38\% & 23\% & 30\% & 7\% & 12\% & 9\% & 20\% \\
Preselection & 61\% & 54\% & 18\% & 20\% & 5\% & 16\% & 16\% & 10\% \\
Misleading Reference Pricing & 90\% & 70\% & 7\% & 20\% & 0\% & 0\% & 3\% & 10\% \\
Trick Questions & 27\% & 60\% & 10\% & 10\% & 10\% & 10\% & 53\% & 20\% \\
Confirmshaming & 73\% & 30\% & 7\% & 10\% & 13\% & 40\% & 7\% & 20\% \\
\midrule
\textbf{Linguistic Dead-Ends:} & \textbf{62\%} & 60\% & \textbf{7\%} & 13\% & \textbf{26\%} & 23\% & \textbf{6\%} & 5\% \\
Alphabet Soup & 90\% & 75\% & 2\% & 20\% & 3\% & 0\% & 5\% & 5\% \\
Untranslation & 33\% & 45\% & 12\% & 5\% & 48\% & 45\% & 7\% & 5\% \\
\midrule
\textbf{Social Proof:} & \textbf{53\%} & 70\% & \textbf{38\%} & 25\% & \textbf{8\%} & 5\% & \textbf{3\%} & 0\% \\
Notification of Others' Activities & 62\% & 80\% & 29\% & 17\% & 7\% & 3\% & 2\% & 0\% \\
Testimonials & 23\% & 40\% & 63\% & 50\% & 10\% & 10\% & 3\% & 0\% \\
\midrule
\textbf{Urgency:} & \textbf{52\%} & 43\% & \textbf{41\%} & 40\% & \textbf{3\%} & 8\% & \textbf{5\%} & 10\% \\
Countdown Timer & 77\% & 55\% & 15\% & 35\% & 5\% & 10\% & 3\% & 0\% \\
Low Stock/High Demand & 27\% & 30\% & 67\% & 45\% & 0\% & 5\% & 7\% & 20\% \\
\midrule
\textbf{Sneaking:} & \textbf{58\%} & 60\% & \textbf{10\%} & 13\% & \textbf{12\%} & 16\% & \textbf{20\%} & 11\% \\
Sneaking an Item & 23\% & 10\% & 17\% & 50\% & 60\% & 40\% & 0\% & 0\% \\
Hidden Subscription & 65\% & 70\% & 6\% & 6\% & 1\% & 8\% & 28\% & 16\% \\
Drip Pricing & 60\% & 60\% & 20\% & 10\% & 20\% & 30\% & 0\% & 0\% \\
\midrule
\textbf{Forced Action:} & \textbf{73\%} & 50\% & \textbf{17\%} & 40\% & \textbf{10\%} & 0\% & \textbf{0\%} & 10\% \\
Forced to Register & 73\% & 50\% & 17\% & 40\% & 10\% & 0\% & 0\% & 10\% \\
\midrule
\textbf{Obstruction:} & \textbf{33\%} & 30\% & \textbf{3\%} & 10\% & \textbf{53\%} & 60\% & \textbf{10\%} & 0\% \\
Hard to Cancel & 33\% & 30\% & 3\% & 10\% & 53\% & 60\% & 10\% & 0\%\\
\bottomrule
\end{tabular}
\end{table}

\subsubsection{Task completion}\label{isect17}

About half of the anon. participants (12 of 30, or 40\%) \emph{unintentionally signed up for ``CyberSelect''} (there were also four anon. participants who intentionally signed up). Of the non-anon. participants, six (of 10, or 60\%) unintentionally signed up. In total, 18 of 40 (nearly half, or 45\%) fell prey to Trick Wording assisted by Pre-``un''-selection in the form, fuelling the Hidden Subscription DP. This indicates an amplification effect when DPs were combined. 

Most anon. participants (27 of 30, or 90\%) and all ten non-anon. participants were unable to cancel their accounts due to the combination of the Untranslation (with all essential information presented in English) and Hard to Cancel (on top of Untranslation, the page was hard to find and the link to take action even harder to find) DPs. Once again, the combination of DPs seems to enhance susceptibility.

\subsubsection{Simulated financial loss}\label{isect18}

Of the anon. participants, 22 (of 30, or 73\%) \emph{bought one of the ABC computers, priced and placed differently but the same product}. Of these, 17 participants (77\%) bought the expensive one without realizing it, thus falling prey to the trick. 
Data for two of the non-anon. participants were missing. Of the remaining eight, three (38\%) bought one of the ABC computers. Of these, two bought the expensive one, also falling prey to the deception. This resulted in a simulated personal loss of {￥}5000 ($\sim$USD {\$}31) each and a simulated net gain of {￥}95,000 ($\sim$USD {\$}590) for CyberStore.

Most anon. participants (23 of 29, or 79\%) noticed and five accepted the \emph{sneaked-in cloth}.
Nine (90\%) non-anon. participants noticed and half accepted it. This indicates that acceptance varies even when an intended trick is noticed. This resulted in a simulated personal loss of {￥}100 each ($\sim$USD {\$}.60) and a simulated net gain of {￥}700 ($\sim$USD {\$}4.30) for CyberStore.

Of the anon. participants, 12 (of 30 or 40\%) \emph{accidentally signed up for the CyberSelect premium account}. Six (60\%) of the non-anon. participants also signed up accidentally. Let us assume that participants would not close their accounts or cancel the membership. 
Let us also assume that the yearly membership fee presented on the Member page, set to ``{￥}129.5,'' in fact means Chinese yuan rather than Japanese yen (as a result of the Alphabet Soup DP). This would result in a simulated personal loss of $\sim${￥}2900 per year each ($\sim$USD {\$}18) and a simulated net gain of {￥}52,200 ($\sim$USD {\$}324) per year for CyberStore.

Nearly two-thirds of the anon. participants (18 of 30 or 60\%) and non-anon. participants (6 of 10 or 60\%) did not notice the extra fee at checkout. This resulted in a simulated personal loss of $\sim${￥}350 each ($\sim$USD {\$}2.15) and a simulated net gain of {￥}8400 ($\sim$USD {\$}52) for CyberStore.

Taken together, with these DPs alone, a sample of forty Japanese consumers could experience a net average simulated personal loss of $\sim${￥}3907 each ($\sim$USD {\$}24.25) and enable a simulated net gain of {￥}156,300 ($\sim$USD {\$}970) for CyberStore.

\subsection{How did it feel? Affective state, acceptability, and attitudes (Both phases, RQ2)}\label{isect19}

We explored how participants felt about their experience in general and specific DPs through a combination of pre/post quantitative \hbox{self-reported} measures and patterns in the qualitative interview data (Phase 1). We included a comparative control with the non-DP version of the website (Phase 2).

\subsubsection{Affective state (valence, arousal)}\label{isect20}

\begin{figure*}[!ht]
    \centering
    \includegraphics[width=.9\textwidth]{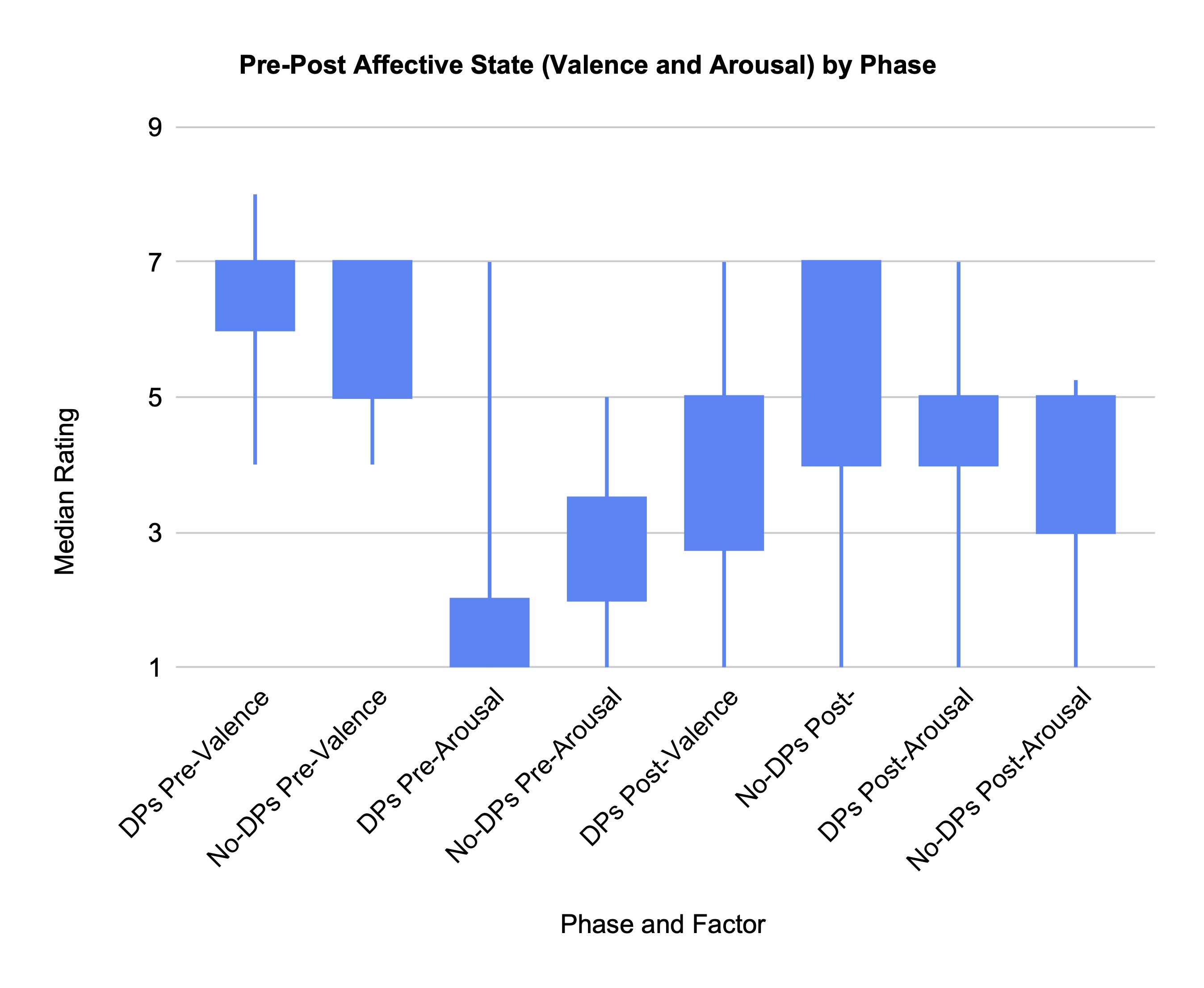}
    \caption{Affective state (valence and arousal) pre/post and by phase.}
    \label{fig:valence}
\end{figure*}

Changes to affective state were found pre/post experience in both phases (\autoref{fig:valence}).
In Phase 1, with the DP-filled version of the website, \emph{valence} tended to greatly decrease, i.e., worsen, following the experience.
Paired t tests found a statistically significant difference in valence before (anon.: M~=~6.9, SD~=~1.4; non-anon: M~=~7, SD~=~1.4) and after (anon.: M~=~4.6, SD~=~2.5; non-anon.: M~=~4.4, SD~=~2.1); for anon., \emph{t}(28) = {\textminus}4.97, \emph{p} $<$.001, \emph{d} =.92 (95\% CI [{\textminus}2.05, 2.05]); for non-anon., \emph{t}(9) = {\textminus}2.9, \emph{p} =.018, \emph{d} =.92 (95\% CI [{\textminus}2.26, 2.26]). Shapiro-Wilk tests indicated that the data were not normally distributed, so follow-up Wilcoxon Signed-Rank tests were run. The results were the same when comparing before (anon.: MD~=~7, IQR~=~2; non-anon.: MD~=~7.5, IQR~=~2) and after (anon.: MD~=~5, IQR~=~4.8; non-anon.: MD~=~4, IQR~=~2.8); for anon., \emph{Z} = {\textminus}3.7, \emph{p} $<$.001, \emph{r} =.78 (95\% CI [{\textminus}1.96, 1.96]); for non-anon., \emph{Z} = {\textminus}2.2, \emph{p} =.028, \emph{r} =.73 (95\% CI [{\textminus}1.96, 1.96]). In Phase 2, a paired \textit{t}-test indicated no statistically significant difference between pre ($M=6.4,
SD=7,
MD=7,
IQR=2$) and post ($M=6.0,
SD=7,
MD=7,
IQR=3$) valence, $p=.146$.

\emph{Arousal} tended to increase after the experience in both phases andespecially in Phase 1. For Phase 1, paired-t tests indicated a statistically significant difference in arousal before (anon.: M~=~2.9, SD~=~2; non-anon.: M~=~3.3, SD~=~1.8) and after (anon.: M~=~5.5, SD~=~2.4; non-anon.: M~=~5.6, SD~=~2.0); for anon., \emph{t}(28) = 5.5, \emph{p} $<$.001, \emph{d} = 1.02 (95\% CI [{\textminus}2.05, 2.05]); for non-anon., \emph{t}(9) = 2.8, \emph{p} =.02, \emph{d} =.89 (95\% CI [{\textminus}2.26, 2.26]). A similar but less pronounced increase was found in Phase 2, comparing pre ($M=4,
SD=3.5,
MD=3.5,
IQR=3$) and post (M=$4.5,
SD=MD=5,
IQR=2.3$) arousal scores, \emph{t}(43) = 2.2, \emph{p} =.03, \emph{d} =.34 (95\% CI [{\textminus}2.02, 2.02]).

For the cross-phase between-subjects comparisons, we used Welch's \textit{t}-tests due to the uneven sample sizes and potential variance issues~\citep{Ruxton2006}. There was no significant difference between the no-DP and DP groups for pre-experience valence, $p=.09$, meaning both cohorts felt the same at the start. However, the no-DP group was more aroused, \emph{t}(79.14) = 2.35, \emph{p} =.02, \emph{d} =.53 (95\% CI [{\textminus}1.99, 1.99]), which may be attributed to the online environment. The post-valence scores for the no-DP group were also higher than those of the DP group, \emph{t}(76.60) = 2.80, \emph{p} =.006, \emph{d} =.62 (95\% CI [{\textminus}1.99, 1.99]), confirming a worse experience for the DP group. There was no difference by group for post-arousal scores, $p=0.06$.

The valence and arousal results across phases suggest that participants encountered \emph{negative and intense} feelings after the experience with the DP-filled version of the website, such as \emph{annoyance, frustration, anger, distress, anxiety, and agitation}~\citep{Hanjalic2005}. While arousal heightened in both phases, this can be attributed to carrying out an engaging task, i.e., the shopping experience. We must also exercise caution when inferring larger emotional patterns from the intersection of valence and arousal~\citep{Kuppens2013}. We supplemented these quantitative analyses with qualitative analyses, presented in \hyperref[sec:thematic]{Section~\ref{sec:thematic}}.

\subsubsection{Acceptability (SUS)}\label{isect21}

\begin{figure*}[!ht]
    \centering
    \includegraphics[width=.9\textwidth]{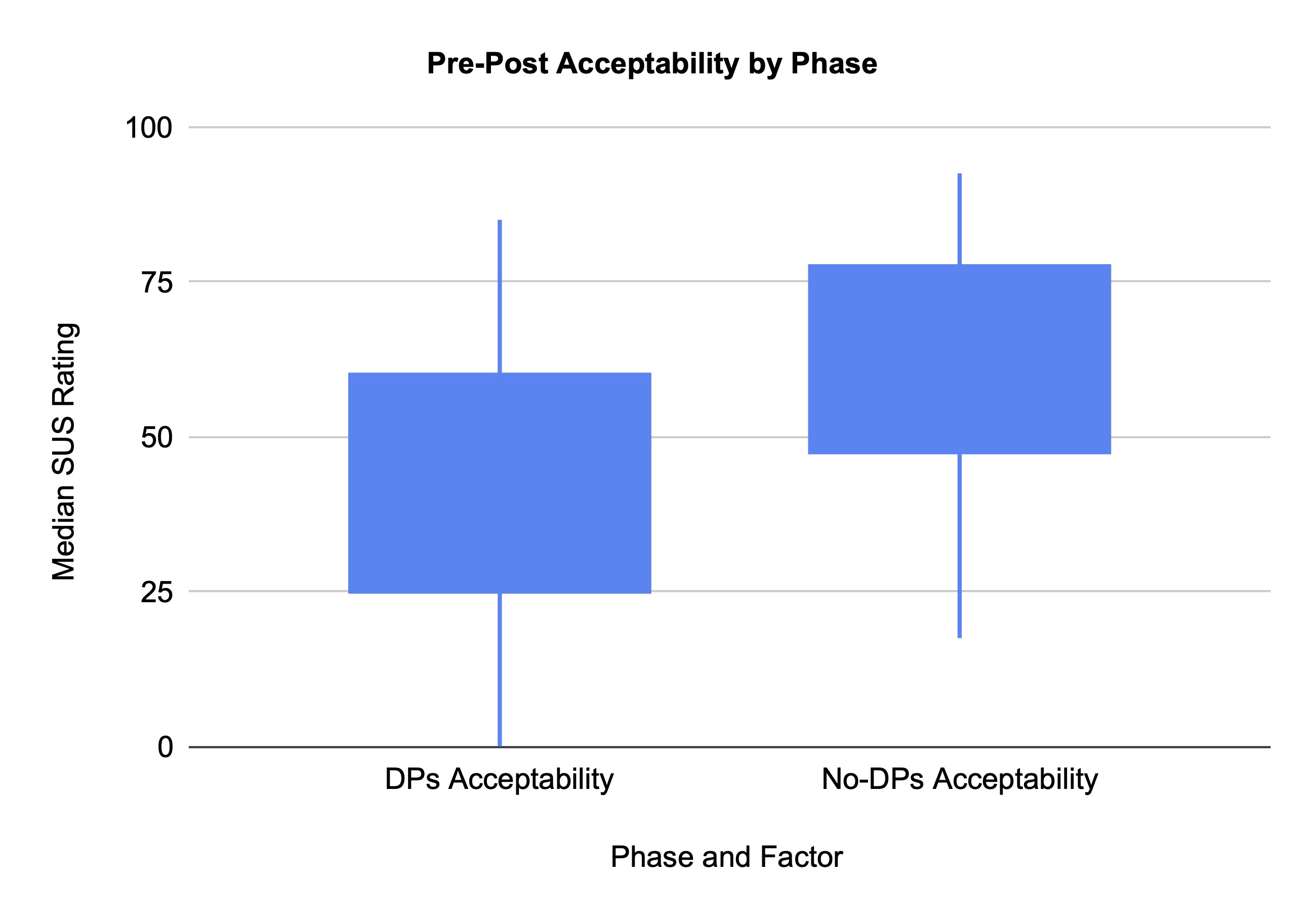}
    \caption{Acceptability by phase.}
    \label{fig:acceptability}
\end{figure*}

In Phase 1, post-experience SUS scores (\autoref{fig:acceptability}) indicated that anonymous participants generally found the website unacceptable ($M=35.5, SD=22.4, MD=32.5, IQR=32.5, hi=85, lo=0$), while the non-anonymous participant scores indicated marginal acceptance ($M=54, SD=20, MD=54, IQR=34, hi=83, lo=30$). Still, the scores were polarized. Altogether, 69\% of anonymous ($n=20$) and 70\% of non-anonymous ($n=7$) participants deemed the website unacceptable. Three participants in both cohorts accepted the website. Six anonymous participants provided marginal scores. This contrasted with Phase 2. A Welch's \textit{t}-test indicated that acceptability was significantly lower in Phase 1 with the DP version of the website compared to Phase 2 with the non-DP version ($M=62.2,
SD=65,
MD=65,
IQR=30,
hi=92.5,
lo=17.5$), $t(74.36) = -4.67, p <.001, d = 1.04$ (95\% CI [{\textminus}1.99, 1.99]). 
In summary, the DP-filled version of the website was deemed far less acceptable than the DP-free one.

\subsubsection{Attitudes (thematic analysis of qualitative accounts)}\label{sec:thematic}

The thematic framework is presented in \autoref{tab:themes}. Participants reacted in a variety of ways. 
Many were simply bamboozled (18, 11\%) by the design or information~\citep{Herbig1994}, while others were not aware of norms about business practices and therefore did not react strongly~\citep{miura2021norms}. Nearly half had a mixture of positive and negative feelings, or ambivalence (42, 25\%). Many conformed~\citep{Takano1997} or were accustomed to DPs, or let it go if the benefits were perceived to be greater than the detriments~\citep{fujimura1999}. Others experienced cognitive dissonance, denying that they were deceived while acknowledging instances of deception. 

Some attitudes and behaviours related to consumer-centrism (38, 23\%), a facet of Japanese consumer relations~\citep{fujihara1981,fujimura1999}. Notable in its smaller representation was disloyalty ({不誠実} or fuseijitsu). Loyalty, or {誠実} (seijitsu), is key to Japanese consumer relations~\citep{takashi2010seijitsu}. Still, many themes broadly point to an overarching sense of disloyalty; we coded direct references. Also, the online environment obscures the ``personal'' connection driving 
impressions of loyalty and sincerity.

Overall, there was great unease (69, 41\%), especially about the interaction design (28, 17\%). Notably, the Linguistic Dead-Ends~\citep{hidaka2023linguistic} and various price-related DPs aroused suspicion. Still, little was mentioned about the fake reviews, again highlighting apparent norms around social proofs in Japan.

\begin{table}[!ht]
\caption{Thematic framework of consumer attitudes (165 comments from $N=40$). Rep.: Representation by counts and percentages.}
\label{tab:themes}
\footnotesize
\begin{tabular}{p{0.1\linewidth}p{0.19\linewidth}p{0.27\linewidth}p{0.11\linewidth}p{0.27\linewidth}p{0.04\linewidth}}
\toprule
\textbf{Theme} & \textbf{Sub-Theme} & \textbf{Definition} & \textbf{Source} & \textbf{Example} & \textbf{Rep.} \\
\midrule
Bamboozled & \makecell{Knowledge of \\Norms Missing\\ (規範意識の無さ)} & Not aware of norms and standards for products and services, and made assumptions about the available options.
& \citet{miura2021norms} & "Business as usual" (A1) & 4 (2\%) \\
 & \makecell{Information Overload\\ (情報過負荷)} & Amount of information prevented effective decision-making. & \citet{Herbig1994} & "I don't always read the [ToS] ... I skipped it right away" (Y5) & 5 (3\%) \\
 & \makecell{Confusion\\ (混乱)} & Could not parse what was going on or what just happened to them. & \textit{Inductive} & "I didn't realize I became a paying member." (Y8) & 9 (5\%) \\
 \midrule
Ambivalence & \makecell{Conformity \\Orientation\\ (同調志向)} & Tended to accept and conform to the other party. & \citet{Takano1997} & "That damned 100 yen cloth ... ah well, it's okay." (A16) & 9 (5\%) \\
 & \makecell{Normalization\\ (規範化)} & Accepted the experience as normal, expressing that they did not mind or were not deceived. & \textit{Inductive} & "I don't feel that bad, but I do feel rushed when the countdown is displayed." (A14) & 14 (8\%) \\
 & \makecell{Assimilation Effect\\ (同化効果)} & Tended to evaluate the experience as meeting expectations if the difference between expectation and outcome was small or perceived as equitable. & \citet{fujimura1999} & "I bought it like I didn't care because it was cheap." (A21) & 7 (4\%) \\
 & \makecell{Cognitive Dissonance\\ (不協和)} & Had positive and negative thoughts or feelings about the experience, at the same time. & \textit{Inductive} & "Basically {[}no deception{]}. {[}But{]} the screen that solicited me was worrying." (A2) & 12 (7\%) \\
 \midrule
Consumer-Centrism & \makecell{Dependency\\ Orientation\\ (甘え志向)} & As per collectivist culture, 
had high expectations for companies, leading to harsh judgments when expectations were not met. & \citet{fujihara1981} & "The 50\% discount rate feels too cheap." (B1) & 15 (9\%) \\
 & \makecell{Contrast Effect\\ (対比効果)} & Tended to evaluate the experience as not meeting expectations if the difference between expectation and outcome was too large. & \citet{fujimura1999} & "I was disgusted when I bought the cleaner and it came with a cloth." (A26) & 11 (7\%) \\
 & \makecell{Disloyalty\\ (不誠実)} & Felt that the company was being dishonest towards them: a break in customer loyalty or 誠実. & \textit{Inductive} & "I might contact the Consumer Affairs Agency to let them know that I've fallen for a dodgy website." (A22) & 12 (7\%) \\
 \midrule
Unease & \makecell{About the Interaction \\Design\\ (IxDに対する不安)} & Unease resulting from the interactive experience and user interface. & \textit{Inductive} & "Different colours ... I wouldn't say it's a scam, but I think it's scary." (Y3) & 28 (17\%) \\
 & \makecell{About Foreign\\ Services\\ (海外のサービス\\に対する不安)} & Unease when there was a lack of Japanese support. & \textit{Inductive} & "... the smell of being made abroad. If it's English, I don't feel like reading it ... I can't read it." (A18) & 17 (10\%) \\
 & \makecell{About Information\\ (情報不足\\に対する不安)} & Unsure about the information, or lack thereof, when making decisions. & \textit{Inductive} & "I don't know if the price is really lower because I don't know the original price." (A6) & 16 (10\%) \\
 & \makecell{About User Reviews\\ (口コミの違和感)} & Discomfort with the reviews on the website. & \textit{Inductive} & "Some products had five stars even though the reviews were not favourable." (A28) & 8 (5\%)\\
 \bottomrule
\end{tabular}
\end{table}

\subsection{Comparison of Phase 1 cohorts}\label{sec:comparison}

The quantitative and qualitative results for both first-phase cohorts---anonymous and non-anonymous---indicated similar trends and patterns, with one caveat: acceptability. Statistically, the results were equivalent for valence and arousal. However, a Mann-Whitney \textit{U} test indicated that acceptability differed, \emph{U} = 213, \emph{z} = 2.17, \emph{p} =.030, \emph{r} =.35, (95\% CI [84.15, 205.85]). This may suggest a social acceptability bias given the nature of the items on the SUS (e.g., ``I thought the system was easy to use'') and the subsequent interview for public broadcast. It could also be a function of the non-anonymous cohort's small sample size.

Given the 
uneven number of participants, we could not conduct inferential statistics on the thematic data in most cases. The descriptive statistics suggested no differences. A Chi-squared test on the sub-theme with the most representation, \emph{Unease About the Interaction Design} (28, 17\%), was not statistically significant, $p=.610$.

Altogether, this indicates that any potential sampling biases or performance changes related to non-anonymous participation in the broadcast did not affect attitudes or behaviours before, during, or after the experience.

\section{Discussion}\label{isect22}

The dark commercial patterns and deceptive designs in CyberStore, our simulated e-commerce website, evoked a range of reactions and feelings from Japanese consumers. This was not for show. More than half of DPs went unnoticed, while nearly two-thirds of DPs elicited no reaction (RQ1). These results align with but also supersede non-interactive survey-based work. For instance, \citet{BongardBlanchy2021} found that 42--64\% of people missed DPs with Hidden Information in screen shots, while 80\% of our cohort missed the Hidden Information DPs in the interactive website. Most participants also did not complete the full experience because they were unable to cancel their accounts (RQ1). This reflects the literature on user inability to identify the presence of DPs~\citep{digeronimo2020,mildner2023}, even when primed~\citep{mildner2023,bongardblanchy2023,BongardBlanchy2021}.
Whether noticed DPs were deceitful or elicited indifference varied (RQ1). For example, Sneaking (despite the name) and Obstruction DPs were among the easiest to notice. This aligns with previous work, even though the noticeability rate differs, e.g., our Sneaking an Item DP varied from 77 to 90\% compared to 40.6\% in \citet{Nimkoompai2022}.
Yet, the overall experience was negative for most people (RQ2), as confirmed in the follow-up study with a new cohort experiencing a DP-free version of the website.

Adopting the proposed normative perspectives of \citet{mathur2021whatdark} was illuminating. For example, three-quarters of participants were tricked by Misleading Reference Pricing, choosing the higher-priced (and higher-placed) version of the same computer; we were able to calculate a simulated personal loss of $\sim$USD {\$}50. Across all DPs, the loss averaged $\sim$USD {\$}24.25 per person (RQ1). Part of this involved half of participants being tricked into signing up for the premium membership. Many, but not all, participants felt unease and confusion, while others had mixed feelings and expressed frustration at the lack of consumer-centrism (RQ2). Facing trickery in the context of positive expectations and Japanese collectivism~\citep{Takano1997}, our participants endured {不誠実} (fuseijitsu) or disloyalty and cognitive dissonance or {不協和} (fuseiwa) (RQ2). We now examine patterns in the experience and consumer culture to better understand these findings.

\subsection{Japan-centred deception: Linguistic Dead-Ends}\label{isect23}

The Japanese DPs were not detected by the majority of our Japanese participants. The Untranslation subtype prevented all but a few from cancelling their accounts. Our thematic analyses indicated that, while obvious, this pattern was unnerving and disloyal. This is partly linked to low English literacy rates in Japan.\footnote{\url{https://www.nippon.com/en/japan-data/h01843/}.} However, use of English was also perceived by participants as not Japanese consumer-centred. Our results highlight great risk in these patterns for the Japanese public.  Other nations with low English fluency rates may also be at risk.

Where Untranslation was clear and disruptive, Alphabet Soup was subtle and deceptive. Almost all participants missed the two membership fee cases. The Alphabet Soup in the annual membership fee resulted in a simulated yearly loss of about {￥}2900 (USD $\sim$ {\$}18) for every person tricked. Altogether, the Alphabet Soup DPs were missed 75--90\% of the time. Our implementation of Alphabet Soup reflects the case covered in a recent warning from the Japanese Consumer Affairs Division about confusing use of the {￥} currency symbol, which signifies both Japanese yen and Chinese yuan.\footnote{\url{https://www.kokusen.go.jp/news/data/n-20230419_2.html} (note: in Japanese).} Legal repercussions are likely to emerge. Companies can avoid use of the ambiguous {￥} symbol, such as by using the unambiguous {円} (yen) symbol instead, or make the currency clear to Japanese consumers with other signifiers, such as currency codes like JPY and CNY.

We offer the first complete set of empirical results on Linguistic Dead-Ends, extending the purely descriptive analysis of \citet{hidaka2023linguistic} and initial work of \citet{Seaborn2024lbw}.  
Our results lend strength to the inclusion of these patterns in formal ontologies like that of \citet{gray2024ontology}. Our findings also distinguish these DPs from other language-based ones. Loaded~\citep{luguri2021shining} or emotional language as ``felt persuasion''~\citep{Gray2021felt} is obvious to most people. Linguistic Dead-Ends, however, can be obvious \emph{and} disruptive, as with Untranslation forcing the user to expend extra effort to translate or give up on certain options, as well as utterly subtle patterns like Alphabet Soup working beneath conscious awareness for most users. 
Future work should explore the prevalence of these DPs outside of Japan and, if found, conduct empirical work on consumer impact. Another trajectory is how these DPs intersect with other DPs, similar to what \citet{soe2020norway} did for consent banners. Linguistic Dead-Ends could be combined with other language-based DPs to explore complementary or conflicting effects. A more holistic understanding of user impact 
could be achieved by translating methods from prior DP work, such as perceived control~\citep{grassl2021dark} when encountering Untranslation DPs. Since these are text-based patterns, machine learning approaches~\citep{mathur2019atscale,yada2022dark,feng2023analysis} could be adopted to reveal prevalence. 
Countermeasures~\citep{schafer2023countermeasures} enabled by browser plugins could follow. Finally, comparing prevalence across device types, such as apps vs. websites~\citep{van2022shopping,gunawan2021webmobile}, could reveal differences between the Linguistic Dead-End subtypes. For example, Untranslation in an app may be harder to handle than in a web browser that may provide automated translation.

\subsection{Unease, ambivalence, and endured disloyalty}\label{isect24}

Reactions to DPs varied by pattern and person, but can be understood by way of the Japanese consumer context and culture-independent models of cognition. CyberStore engendered a sense of unease or worse, similar to the findings of \citet{bhoot2021enduser} regarding deception-borne anxiety. We also confirmed that mood decreased. While we cannot directly compare, given the different measures and contexts of use (e-commerce website vs. video platform), our results echo those of \citet{chaudhary2022videostream}, with large shifts down in valence and up in arousal, indicating post-experience negativity. Still, we recognize that DP density could have been a factor, inflating affective reactions and likewise decreasing acceptance due to the sheer number of DPs encountered. As yet, no research has experimentally considered DP density. Furthermore, most work on prevalence so far has focused on a small slice of the experience, whereas we offered a fully-fledged, longer engagement. Also, our findings map onto lower-density results, as well as suggest that individual factors play a role, i.e., whether or not a given DP matters or has an effect can vary by individual. Future work should clarify the role, if any, of DP density. 

Some participants expressed denial when confronted with the deceptive elements. 
Others blamed themselves, conforming to the business~\citep{takashi2010seijitsu}. This suggests self-serving~\citep{miller1975self} attribution biases~\citep{heider2013psychology}, where participants characterized the problem as being beyond their responsibility. The qualitative themes of consumer-centrism~\citep{fujihara1981}, with harsh judgments stemming from high expectations, help explain these reactions. Still others dismissed the bad and embraced the good, a contrast effect~\citep{fujimura1999} and form of cognitive dissonance. Such reactions may seem positive on the surface, but companies should be cautious, as cognitive dissonance can lead to post-experience dissatisfaction and disengagement~\citep{ShahinSharifi2014}. Still others claimed not to care, but this does not preclude negative effects~\citep{habib2022cookie}, like actual financial losses.

A consumer-centric stance is expected in Japanese business culture~\citep{fujihara1981}. This is notably different from other perspectives on consumer--company relations, such as brand trust~\citep{voigt2021dark}. In Japan, failure to cater to the consumer can be deemed an act of disloyalty. Our findings revealed a sense of discontent linked to {不誠実} (fuseijitsu) or Japanese expectations of consumer loyalty and sincerity. This was expressed with examples and commentary but also in subtle ways through the language participants used. For instance, A3 wrote ``{セール品を買}\textbf{{わせようとしてくる}}'' or ``They \textbf{try to get you to} buy items on sale.'' The emphasized parts suggest felt manipulation rather than acceptable conduct. Even though A3 did not state this explicitly, the wording makes it clear.

Discontent and insincerity also link to trust. Consider what A26 wrote: ``{解約時、脅し文句のように思えた}'' or ``When I cancelled, it seemed like a \textbf{threat}.''  Consumers enter into a relationship built on trust with companies, which can be shaken by DPs like Confirmshaming---here, the threatening language implied by the emphasized parts in A26's quote---and interpreted as insincere conduct. As \citet{Gray2021felt} discovered, perceived (dis)trust and prior experience mediated interpretations of deception. This is culturally sensitive. \citet{Gray2021felt}, for instance, found that English speakers keyed into peripheral cues, like where the website was hosted, and Mandarin Chinese speakers avoided bundleware. However, our thematic analysis showed that our Japanese participants reacted to hints of dishonesty alongside unease based on deficient information and cues to foreignness. B6 interpreted the impetus behind the Untranslation pattern succinctly: ``{利用規約が英語表記、日本人にあきらめさせようとしている}'' or ``Terms and conditions are written in English, trying to get Japanese people to give up.'' Whether obvious or subtle, DPs that fail to achieve {誠実} (seijitsu) elicit distrust and may be rendered ineffective on that basis alone. This provides support for what \citet{nakano2022zadak} recommended: Japanese companies should focus on building long-term trust rather than using DPs for short-term financial gains. 

The affective results, notably the negative affect revealed by comparison to the non-DP control, underscore the low acceptability of the simulated online store. Relying on initial customer confusion, lack of norms awareness~\citep{miura2021norms}, or simple overload~\citep{Herbig1994} will engender poor experiences that may lead to future disengagement. 
Even participants who were normalized or assimilated~\citep{fujimura1999} to annoying and malicious tricks in online stores perceived the simulated version negatively. The potential financial gains may not be worth the risk. ``Good faith'' relationships are long-lasting and positive in affect. Forcing consumers to endure negative, disloyal experiences is against the value of {誠実} (seijitsu). This value could be codified into law under the framing of DPs as ``disloyal patterns''~\citep{richards2021duty} or ``disloyal design''~\citep{gunawan2024dark}. Disloyalty here is based on the idea of wrongful self-dealing, i.e., company-centredness instead of consumer-centredeness~\citep{gunawan2024dark}. Japan could act as a case study for formalizing this legal theory, given the importance of customer-centred conduct represented by the consumer relations value of {誠実} (fuseijitsu or sincerity).

\subsection{Social norms, social acceptability, and showtime performance}\label{isect25}

A subversive finding related to the social elements: the Social Proofs that tapped into biases around social cues and acceptability. In the social media age, people have normalized such cues~\citep{bond201261}. As \citet{luguri2021shining} found, Social Proofs were statistically linked to greater acceptance (140 of 634, or 22.1\%). Yet, the results in our case were more severe. For instance, only a small number of participants questioned the veracity of the reviews (8 of 40 people, or 5\% of 165 comments). This may be understood through the collectivist slant of Japanese culture~\citep{Takano1997} alongside social mobilization effects, where socially relevant cues can influence people in unconscious ways~\citep{bond201261}. Collectivist cultures, like Japan, may be more susceptible to Social Proofs or social validation~\citep{cialdini2001science} effects~\citep{Cialdini1999}. In Japan, where social acceptance of commercial goods directs the market~\citep{Synodinos2001}, cues to social acceptance and normalization may be especially strong. Companies must be sensitive to even these, the most banal forms of manipulation.

The IRB required us to separate the anonymous and non-anonymous cohorts in Phase 1, in case participation in a live broadcast would affect results~\citep{Tripepi2010,Abeler2014}. However, there was almost no difference between these groups on any metric. This showcases the power of deceptive design: everyone is susceptible and reacts in similar ways, regardless of motive for participation or being in the public eye.

Altogether, most of the Japanese consumers who participated in our study did not want to experience deception.
Designers, developers, and decision-makers will need to reflect and engage in critical discussions on DP practices~\citep{gray2021legal}, before regulatory shifts arrive.

\subsection{Limitations}\label{isect26}

We acknowledge that the use of a simulated website, while common~\citep{chang2024theory,gray2023dpsysreview}, limits the ecological validity of our results. Our simulated financial loss measure was similarly limited; the results should be checked in ecologically valid settings. Our simulation also employed more DP cases than the current known average of 3.9~in Japanese apps~\citep{hidaka2023linguistic}. The effect, if any, of DP density should be examined in future work, such as by comparing no DPs to low-density (e.g., 1 or 2), average density (e.g., 4), and high density (e.g., 20--40) conditions. This would be enlightening for assessing the validity of results from simulation work, but also cover the spread of densities known via descriptive work~\citep{gunawan2021webmobile,digeronimo2020,hidaka2023linguistic}.
Future simulation work may aim for higher ecological validity by reducing the number of DPs. Future work should also prioritize collaborations with companies and conduct A/B testing, although we recognize that establishing such collaborations on a sensitive topic like DPs is challenging.

Our recruitment was also limited in a few ways. In Phase 1, we used a third-party recruiter known to participants (NHK). This enabled us to recruit a diverse and random sample, but participants may have been influenced by the recruiter, especially the $n=10$ who appeared on the public broadcast. We were also not able to recruit equal numbers of participants for the anonymous, non-anonymous, and follow-up cohorts. We accounted for this by dividing our reporting on descriptive statistics per group, conducting non-parametric statistical tests robust to uneven groups, and presenting these comparisons separately (in \hyperref[sec:comparison]{Section~\ref{sec:comparison}}). We recognize that our results should be confirmed with larger samples. 
We also recognize that Phase 2 differed in participant composition (notably in terms of gender distribution, with far more men in Phase 2), recruitment method, and task context (fully anonymous). While we are not sure whether and what differences could result based on gender, the anonymous format may have allowed more honest or ``uninhibited'' responses, as indicated by a meta-analysis of anonymous and non-anonymous study formats~\citep{ClarkGordon2019}. Those in Phase 1, participating face-to-face in front of the researchers, may have reacted differently. As such, we offer these results as suggestive rather than definitive.
Finally, we did not capture demographic factors that could have affected our results, such as daily Internet usage, having a technical degree or career, and knowledge of English for the Untranslation DP. 

Methodological rigour was potentially limited. We did not register our protocol in advance. We also changed the protocol. First, based on the pilot test, we used a retrospective rather than concurrent think-aloud method. Second, after a peer review cycle, we added the follow-up study as a clean control. This study may have internal validity issues because we did not include the think-aloud protocol (as explained in \hyperref[sec:procedure]{Section~\ref{sec:procedure}}). Since the procedures varied by phase, the phases were not fully matched. While we believe our reasoning for this choice is sound, we admit that whether this difference affected the results is unknown---and that is a limitation. We also did not measure affective reactions to specific DPs. 
We encourage follow-up preregistered studies and precise measurement of DP effects. For additional rigour and as a precaution, we would recommend having the main and control groups carry out the same procedure to clarify the potential threats to internal validity.

\subsection{Action items for stakeholders}\label{isect27}

We offer a list of action items for those invested in promoting awareness about and taking measures against DPs, contextualized for Japan but broadly applicable.

\begin{itemize}
    \item Public awareness of DPs in general and the specific varieties found in Japan, i.e., Linguistic Dead-Ends, is needed. Public organizations, such as the Japanese Consumer Affairs Agency, can run campaigns and produce materials for the everyday public about these culturally sensitive digital threats.
    \item Computer literacy programmes should include a module on DPs. In Japan, the explicit inclusion of Linguistic Dead-Ends in these materials will be vital. In other nations, these DPs can act as a case study on the role of cultural context, and perhaps spark a critical search for novel DPs in those contexts.
    \item Social media influencers, especially those involved in consumer affairs, digital marketing, public education, and design practice, can issue calls for the regulation and removal of DPs, perhaps under a shared hashtag~\citep{fansher2018hashtag}, raising awareness among Japanese consumers and professionals alike.
    \item Companies can create or revise guidelines about design practice to specifically include DPs, especially the Japanese varieties of DPs. \citet[p. 81--84]{Narayanan2020} provide specific guidance. Professional seminars or even lunchtime tutorials could raise general awareness across internal stakeholders.
\end{itemize}

\section{Conclusion}\label{isect28}

Consumer perceptions of deceptive patterns were elusive for high-tech Japan. We have now offered a better understanding of how the average Japanese consumer responds to DPs commonly deployed in e-commerce websites. We have demonstrated variability in perceptions and deceptions while highlighting an overall trend of deceptibility for most Japanese participants and DPs. Notably, we offer empirical evidence of the negative effects of the Japan-based DPs, Linguistic Dead-Ends, on consumers. We have contextualized our results within the Japanese consumer context, finding patterns related to the company--consumer relationship. The next step is large-scale experimental work, ideally A/B testing in real commercial contexts. At this juncture, we can safely suggest that companies operating within the Japanese e-commerce market avoid using DPs so as to facilitate {誠実} (seijitsu) and long-term economic success.

\newpage

\backmatter

\section*{CRediT authorship contribution statement}

\textbf{Katie Seaborn:} Writing -- original draft, Visualization, Validation, Supervision, Software, Resources, Project administration, Methodology, Investigation, Funding acquisition, Formal analysis, Data curation, Conceptualization.
\textbf{Jo Yukami:} Methodology, Formal analysis, Data curation.
\textbf{Tatsuya Itagaki:} Investigation, Data curation.
\textbf{Mizuki Watanabe:} Investigation, Data curation.
\textbf{Yijia Wang:} Investigation, Data curation.
\textbf{Ping Geng:} Investigation, Data curation.
\textbf{Takao Fujii:} Investigation, Data curation.
\textbf{Yuto Mandai:} Investigation, Data curation.
\textbf{Miu Kojima:} Investigation, Data curation.
\textbf{Suzuka Yoshida:} Project administration, Investigation, Data curation.

\section*{Other disclosures}
This work is a substantially extended version of the following non-archival publication:
Katie Seaborn, Tatsuya Itagaki, Mizuki Watanabe, Yijia Wang, Ping Geng, Takao Fujii, Yuto Mandai, Miu Kojima, and Suzuka Yoshida. 2024. Deceptive, Disruptive, No Big Deal: Japanese People React to Simulated Dark Commercial Patterns. In Extended Abstracts of the CHI Conference on Human Factors in Computing Systems (CHI EA ’24). Association for Computing Machinery, New York, NY, USA, Article 95, 1--8. \url{https://doi.org/10.1145/3613905.3651099}.

\section*{Declaration of generative AI and AI-assisted technologies in the writing process}
During the preparation of this work, the authors used DeepL and ChatGPT to translate and back-translate terms. After using these tools, the authors reviewed and edited the content as needed and take full responsibility for the content of the publication.

\section*{Funding}
This work was supported by a JST PRESTO grant (\#JPMJPR24I6) and partially funded by NHK.

\section*{Declaration of competing interest}
The authors declare the following financial interests/personal relationships which may be considered potential competing interests:

Katie Seaborn reports that financial support and administrative support were provided by NHK. Katie Seaborn reports that financial support was provided by Japan Science and Technology Agency (JST). Katie Seaborn reports a relationship with NHK that includes non-financial support. If there are other authors, they declare that they have no known competing financial interests or personal relationships that could have appeared to influence the work reported in this paper.

\section*{Acknowledgements}
Our sincere gratitude to NHK, especially Tatsuro Imono and Yuya Higashi, for funding part of
this research (development and participant fees) and handling the recruitment of participants.
We sincerely thank Paul Riesch and Émilie Fabre for development assistance. We thank Tokyo
Tech Innovation and Dai Senoo for supporting this research in various ways. We thank Peter
Pennefather for reviewing this manuscript before submission.

\section*{Supplementary data}\label{isupdata1}
Supplementary data for this article can be found online at doi:\href{https://doi.org/10.1016/j.ijhcs.2026.103903}{10.\allowbreak{}1016/\allowbreak{}j.\allowbreak{}ijhcs.\allowbreak{}2026.\allowbreak{}103903}.

\section*{Data availability}
The questionnaire data are available at \href{https://osf.io/agbw6}{https:/\allowbreak{}/\allowbreak{}osf.\allowbreak{}io/\allowbreak{}agbw6}.

\newpage

\bibliography{fileName}

\begin{thebibliography}{}
\renewcommand{\doi}[1]{\url{https://doi.org/#1}}
\bibcommenthead

\bibitem [\protect \citeauthoryear {%
Abeler%
\ \BBA {} Nosenzo%
}{%
Abeler%
\ \BBA {} Nosenzo%
}{%
{\protect \APACyear {2014}}%
}]{%
Abeler2014}
\APACinsertmetastar {%
Abeler2014}%
\begin{APACrefauthors}%
Abeler, J.%
\BCBT {}\ \BBA {} Nosenzo, D.%
\end{APACrefauthors}%
\unskip\
\newblock
\APACrefYearMonthDay{2014}{{\APACmonth{03}}}{}.
\newblock
{\BBOQ}\APACrefatitle {Self-selection into laboratory experiments: Pro-social motives versus monetary incentives} {Self-selection into laboratory experiments: Pro-social motives versus monetary incentives}.{\BBCQ}
\newblock
\APACjournalVolNumPages{Experimental Economics}{18}{2}{195–214,}
\newblock
\begin{APACrefDOI} \doi{10.1007/s10683-014-9397-9} \end{APACrefDOI}
\newblock
\begin{APACrefURL} {http://dx.doi.org/10.1007/s10683-014-9397-9} \end{APACrefURL}
\newblock

\newblock

\PrintBackRefs{\CurrentBib}

\bibitem [\protect \citeauthoryear {%
Adelman%
\ \protect \BOthers {.}}{%
Adelman%
\ \protect \BOthers {.}}{%
{\protect \APACyear {2025}}%
}]{%
network2025technoskepticism}
\APACinsertmetastar {%
network2025technoskepticism}%
\begin{APACrefauthors}%
Adelman, D.%
, Brock, A.%
, Dial, A.%
, Dinkins, S.%
, Fouché, R.%
, He, H.%
\BDBL {}Zeitlin-Wu, L.%
\end{APACrefauthors}%
\unskip\
\newblock
\APACrefYearMonthDay{2025}{}{}.
\newblock
{\BBOQ}\APACrefatitle {Technoskepticism: Between Possibility and Refusal} {Technoskepticism: Between possibility and refusal}.{\BBCQ}
\newblock
 \APACrefbtitle {Technoskepticism.} {Technoskepticism.}
\newblock
\APACaddressPublisher{Redwood City, CA, USA}{Stanford University Press}.
\PrintBackRefs{\CurrentBib}

\bibitem [\protect \citeauthoryear {%
Akritas%
, Arnold%
\BCBL {}\ \BBA {} Brunner%
}{%
Akritas%
\ \protect \BOthers {.}}{%
{\protect \APACyear {1997}}%
}]{%
Akritas1997mannwhit}
\APACinsertmetastar {%
Akritas1997mannwhit}%
\begin{APACrefauthors}%
Akritas, M.G.%
, Arnold, S.F.%
\BCBL {} Brunner, E.%
\end{APACrefauthors}%
\unskip\
\newblock
\APACrefYearMonthDay{1997}{{\APACmonth{03}}}{}.
\newblock
{\BBOQ}\APACrefatitle {Nonparametric hypotheses and rank statistics for unbalanced factorial designs} {Nonparametric hypotheses and rank statistics for unbalanced factorial designs}.{\BBCQ}
\newblock
\APACjournalVolNumPages{Journal of the American Statistical Association}{92}{437}{258,}
\newblock
\begin{APACrefDOI} \doi{10.2307/2291470} \end{APACrefDOI}
\newblock
\begin{APACrefURL} {http://dx.doi.org/10.2307/2291470} \end{APACrefURL}
\newblock

\newblock

\PrintBackRefs{\CurrentBib}

\bibitem [\protect \citeauthoryear {%
Alshammari%
, Alhadreti%
\BCBL {}\ \BBA {} Mayhew%
}{%
Alshammari%
\ \protect \BOthers {.}}{%
{\protect \APACyear {2015}}%
}]{%
alshammari2015ask}
\APACinsertmetastar {%
alshammari2015ask}%
\begin{APACrefauthors}%
Alshammari, T.%
, Alhadreti, O.%
\BCBL {} Mayhew, P.%
\end{APACrefauthors}%
\unskip\
\newblock
\APACrefYearMonthDay{2015}{}{}.
\newblock
{\BBOQ}\APACrefatitle {When to ask participants to think aloud: A comparative study of concurrent and retrospective think-aloud methods} {When to ask participants to think aloud: A comparative study of concurrent and retrospective think-aloud methods}.{\BBCQ}
\newblock
\APACjournalVolNumPages{International Journal of Human Computer Interaction}{6}{3}{48--64,}
\newblock

\newblock

\PrintBackRefs{\CurrentBib}

\bibitem [\protect \citeauthoryear {%
Bangor%
, Kortum%
\BCBL {}\ \BBA {} Miller%
}{%
Bangor%
\ \protect \BOthers {.}}{%
{\protect \APACyear {2008}}%
}]{%
bangor2008empirical}
\APACinsertmetastar {%
bangor2008empirical}%
\begin{APACrefauthors}%
Bangor, A.%
, Kortum, P.T.%
\BCBL {} Miller, J.T.%
\end{APACrefauthors}%
\unskip\
\newblock
\APACrefYearMonthDay{2008}{}{}.
\newblock
{\BBOQ}\APACrefatitle {An empirical evaluation of the System Usability Scale} {An empirical evaluation of the system usability scale}.{\BBCQ}
\newblock
\APACjournalVolNumPages{International Journal of Human--Computer Interaction}{24}{6}{574--594,}
\newblock
\begin{APACrefURL} {https://doi.org/10.1080/10447310802205776} \end{APACrefURL}
\newblock

\newblock

\PrintBackRefs{\CurrentBib}

\bibitem [\protect \citeauthoryear {%
Benjamin%
}{%
Benjamin%
}{%
{\protect \APACyear {2019}}%
}]{%
benjamin2019race}
\APACinsertmetastar {%
benjamin2019race}%
\begin{APACrefauthors}%
Benjamin, R.%
\end{APACrefauthors}%
\unskip\
\newblock
\APACrefYear{2019}.
\newblock
\APACrefbtitle {Race after Technology: Abolitionist Tools for the {New Jim Code}} {Race after technology: Abolitionist tools for the {New Jim Code}}.
\newblock
\APACaddressPublisher{Cambridge, UK}{Polity Press}.
\PrintBackRefs{\CurrentBib}

\bibitem [\protect \citeauthoryear {%
Berens%
, Dietmann%
, Krisam%
, Kulyk%
\BCBL {}\ \BBA {} Volkamer%
}{%
Berens%
\ \protect \BOthers {.}}{%
{\protect \APACyear {2022}}%
}]{%
berens2022cookie}
\APACinsertmetastar {%
berens2022cookie}%
\begin{APACrefauthors}%
Berens, B.M.%
, Dietmann, H.%
, Krisam, C.%
, Kulyk, O.%
\BCBL {} Volkamer, M.%
\end{APACrefauthors}%
\unskip\
\newblock
\APACrefYearMonthDay{2022}{}{}.
\newblock
{\BBOQ}\APACrefatitle {Cookie Disclaimers: Impact of Design and Users’ Attitude} {Cookie disclaimers: Impact of design and users’ attitude}.{\BBCQ}
\newblock
 \APACrefbtitle {Proceedings of the 17th International Conference on Availability, Reliability and Security.} {Proceedings of the 17th international conference on availability, reliability and security.}
\newblock
\APACaddressPublisher{New York, NY, USA}{ACM}.
\newblock
\begin{APACrefURL} {https://doi.org/10.1145/3538969.3539008} \end{APACrefURL}
\PrintBackRefs{\CurrentBib}

\bibitem [\protect \citeauthoryear {%
Bermejo~Fernandez%
, Chatzopoulos%
, Papadopoulos%
\BCBL {}\ \BBA {} Hui%
}{%
Bermejo~Fernandez%
\ \protect \BOthers {.}}{%
{\protect \APACyear {2021}}%
}]{%
bermejo2021cookie}
\APACinsertmetastar {%
bermejo2021cookie}%
\begin{APACrefauthors}%
Bermejo~Fernandez, C.%
, Chatzopoulos, D.%
, Papadopoulos, D.%
\BCBL {} Hui, P.%
\end{APACrefauthors}%
\unskip\
\newblock
\APACrefYearMonthDay{2021}{oct}{}.
\newblock
{\BBOQ}\APACrefatitle {This Website Uses Nudging: {MTurk} Workers' Behaviour on Cookie Consent Notices} {This website uses nudging: {MTurk} workers' behaviour on cookie consent notices}.{\BBCQ}
\newblock
\APACjournalVolNumPages{{Proceedings of the ACM on Human-Computer Interaction}}{5}{CSCW2}{,}
\newblock
\begin{APACrefDOI} \doi{10.1145/3476087} \end{APACrefDOI}
\newblock
\begin{APACrefURL} {https://doi.org/10.1145/3476087} \end{APACrefURL}
\newblock

\newblock

\PrintBackRefs{\CurrentBib}

\bibitem [\protect \citeauthoryear {%
Bhoot%
, Shinde%
\BCBL {}\ \BBA {} Mishra%
}{%
Bhoot%
\ \protect \BOthers {.}}{%
{\protect \APACyear {2021}}%
}]{%
bhoot2021enduser}
\APACinsertmetastar {%
bhoot2021enduser}%
\begin{APACrefauthors}%
Bhoot, A.M.%
, Shinde, M.A.%
\BCBL {} Mishra, W.P.%
\end{APACrefauthors}%
\unskip\
\newblock
\APACrefYearMonthDay{2021}{}{}.
\newblock
{\BBOQ}\APACrefatitle {Towards the Identification of Dark Patterns: {An} Analysis Based on End-User Reactions} {Towards the identification of dark patterns: {An} analysis based on end-user reactions}.{\BBCQ}
\newblock
 \APACrefbtitle {{Proceedings of the 11th Indian Conference on Human-Computer Interaction}} {{Proceedings of the 11th Indian Conference on Human-Computer Interaction}}\ (\BPG~24–33).
\newblock
\APACaddressPublisher{New York, NY, USA}{ACM}.
\newblock
\begin{APACrefURL} {https://doi.org/10.1145/3429290.3429293} \end{APACrefURL}
\PrintBackRefs{\CurrentBib}

\bibitem [\protect \citeauthoryear {%
Blandford%
, Furniss%
\BCBL {}\ \BBA {} Makri%
}{%
Blandford%
\ \protect \BOthers {.}}{%
{\protect \APACyear {2016}}%
}]{%
blandford2016qualitative}
\APACinsertmetastar {%
blandford2016qualitative}%
\begin{APACrefauthors}%
Blandford, A.%
, Furniss, D.%
\BCBL {} Makri, S.%
\end{APACrefauthors}%
\unskip\
\newblock
\APACrefYear{2016}.
\newblock
\APACrefbtitle {Qualitative {HCI} Research: Going Behind the Scenes} {Qualitative {HCI} research: Going behind the scenes}.
\newblock
\APACaddressPublisher{Kentfield, CA, USA}{Morgan \& Claypool Publishers}.
\PrintBackRefs{\CurrentBib}

\bibitem [\protect \citeauthoryear {%
Bocchi%
, De~Cicco%
\BCBL {}\ \BBA {} Rossi%
}{%
Bocchi%
\ \protect \BOthers {.}}{%
{\protect \APACyear {2016}}%
}]{%
bocchi2016above}
\APACinsertmetastar {%
bocchi2016above}%
\begin{APACrefauthors}%
Bocchi, E.%
, De~Cicco, L.%
\BCBL {} Rossi, D.%
\end{APACrefauthors}%
\unskip\
\newblock
\APACrefYearMonthDay{2016}{dec}{}.
\newblock
{\BBOQ}\APACrefatitle {Measuring the Quality of Experience of Web users} {Measuring the quality of experience of web users}.{\BBCQ}
\newblock
\APACjournalVolNumPages{SIGCOMM Comput. Commun. Rev.}{46}{4}{8–13,}
\newblock
\begin{APACrefDOI} \doi{10.1145/3027947.3027949} \end{APACrefDOI}
\newblock
\begin{APACrefURL} {https://doi.org/10.1145/3027947.3027949} \end{APACrefURL}
\newblock

\newblock

\PrintBackRefs{\CurrentBib}

\bibitem [\protect \citeauthoryear {%
Bond%
\ \protect \BOthers {.}}{%
Bond%
\ \protect \BOthers {.}}{%
{\protect \APACyear {2012}}%
}]{%
bond201261}
\APACinsertmetastar {%
bond201261}%
\begin{APACrefauthors}%
Bond, R.M.%
, Fariss, C.J.%
, Jones, J.J.%
, Kramer, A.D.%
, Marlow, C.%
, Settle, J.E.%
\BCBL {} Fowler, J.H.%
\end{APACrefauthors}%
\unskip\
\newblock
\APACrefYearMonthDay{2012}{}{}.
\newblock
{\BBOQ}\APACrefatitle {A 61-million-person experiment in social influence and political mobilization} {A 61-million-person experiment in social influence and political mobilization}.{\BBCQ}
\newblock
\APACjournalVolNumPages{Nature}{489}{7415}{295--298,}
\newblock
\begin{APACrefURL} {https://www.nature.com/articles/nature11421} \end{APACrefURL}
\newblock

\newblock

\PrintBackRefs{\CurrentBib}

\bibitem [\protect \citeauthoryear {%
Bongard-Blanchy%
\ \protect \BOthers {.}}{%
Bongard-Blanchy%
\ \protect \BOthers {.}}{%
{\protect \APACyear {2021}}%
}]{%
BongardBlanchy2021}
\APACinsertmetastar {%
BongardBlanchy2021}%
\begin{APACrefauthors}%
Bongard-Blanchy, K.%
, Rossi, A.%
, Rivas, S.%
, Doublet, S.%
, Koenig, V.%
\BCBL {} Lenzini, G.%
\end{APACrefauthors}%
\unskip\
\newblock
\APACrefYearMonthDay{2021}{{\APACmonth{06}}}{}.
\newblock
{\BBOQ}\APACrefatitle {``{I} am Definitely Manipulated, Even When {I} am Aware of it. It's Ridiculous!'' - {D}ark Patterns from the End-User Perspective} {``{I} am definitely manipulated, even when {I} am aware of it. it's ridiculous!'' - {D}ark patterns from the end-user perspective}.{\BBCQ}
\newblock
 \APACrefbtitle {{Designing Interactive Systems Conference 2021}} {{Designing Interactive Systems Conference 2021}}\ (\BPG~763–776).
\newblock
\APACaddressPublisher{}{ACM}.
\newblock
\begin{APACrefURL} {http://dx.doi.org/10.1145/3461778.3462086} \end{APACrefURL}
\PrintBackRefs{\CurrentBib}

\bibitem [\protect \citeauthoryear {%
Bongard-Blanchy%
\ \protect \BOthers {.}}{%
Bongard-Blanchy%
\ \protect \BOthers {.}}{%
{\protect \APACyear {2023}}%
}]{%
bongardblanchy2023}
\APACinsertmetastar {%
bongardblanchy2023}%
\begin{APACrefauthors}%
Bongard-Blanchy, K.%
, Sterckx, J\BHBI L.%
, Rossi, A.%
, Sergeeva, A.%
, Koenig, V.%
, Rivas, S.%
\BCBL {} Distler, V.%
\end{APACrefauthors}%
\unskip\
\newblock
\APACrefYearMonthDay{2023}{}{}.
\newblock
{\BBOQ}\APACrefatitle {Analysing the Influence of Loss-Gain Framing on Data Disclosure Behaviour: A Study on the Use Case of App Permission Requests} {Analysing the influence of loss-gain framing on data disclosure behaviour: A study on the use case of app permission requests}.{\BBCQ}
\newblock
 \APACrefbtitle {{Proceedings of the 2023 European Symposium on Usable Security}} {{Proceedings of the 2023 European Symposium on Usable Security}}\ (\BPG~112–125).
\newblock
\APACaddressPublisher{New York, NY, USA}{ACM}.
\newblock
\begin{APACrefURL} {https://doi.org/10.1145/3617072.3617108} \end{APACrefURL}
\PrintBackRefs{\CurrentBib}

\bibitem [\protect \citeauthoryear {%
Borberg%
, Hougaard%
, Rafnsson%
\BCBL {}\ \BBA {} Kulyk%
}{%
Borberg%
\ \protect \BOthers {.}}{%
{\protect \APACyear {2022}}%
}]{%
borberg2022so}
\APACinsertmetastar {%
borberg2022so}%
\begin{APACrefauthors}%
Borberg, I.%
, Hougaard, R.%
, Rafnsson, W.%
\BCBL {} Kulyk, O.%
\end{APACrefauthors}%
\unskip\
\newblock
\APACrefYearMonthDay{2022}{}{}.
\newblock
{\BBOQ}\APACrefatitle {``{S}o {I} sold my soul'': Effects of dark patterns in cookie notices on end-user behavior and perceptions} {``{S}o {I} sold my soul'': Effects of dark patterns in cookie notices on end-user behavior and perceptions}.{\BBCQ}
\newblock
 \APACrefbtitle {{Workshop on Usable Security and Privacy (USEC)}} {{Workshop on Usable Security and Privacy (USEC)}}\ (\BVOL~3).
\newblock
\APACaddressPublisher{Bloomintgton, IN, USA}{HATS}.
\newblock
\begin{APACrefURL} {https://dx.doi.org/10.14722/usec.2022.23026} \end{APACrefURL}
\PrintBackRefs{\CurrentBib}

\bibitem [\protect \citeauthoryear {%
B{\"o}sch%
, Erb%
, Kargl%
, Kopp%
\BCBL {}\ \BBA {} Pfattheicher%
}{%
B{\"o}sch%
\ \protect \BOthers {.}}{%
{\protect \APACyear {2016}}%
}]{%
bosch2016tales}
\APACinsertmetastar {%
bosch2016tales}%
\begin{APACrefauthors}%
B{\"o}sch, C.%
, Erb, B.%
, Kargl, F.%
, Kopp, H.%
\BCBL {} Pfattheicher, S.%
\end{APACrefauthors}%
\unskip\
\newblock
\APACrefYearMonthDay{2016}{}{}.
\newblock
{\BBOQ}\APACrefatitle {Tales from the dark side: Privacy dark strategies and privacy dark patterns} {Tales from the dark side: Privacy dark strategies and privacy dark patterns}.{\BBCQ}
\newblock
\APACjournalVolNumPages{Proceedings on Privacy Enhancing Technologies}{2016}{4}{237--254,}
\newblock
\begin{APACrefURL} {https://doi.org/10.1515/popets-2016-0038} \end{APACrefURL}
\newblock

\newblock

\PrintBackRefs{\CurrentBib}

\bibitem [\protect \citeauthoryear {%
Bradley%
\ \BBA {} Lang%
}{%
Bradley%
\ \BBA {} Lang%
}{%
{\protect \APACyear {1994}}%
}]{%
bradley1994measuring}
\APACinsertmetastar {%
bradley1994measuring}%
\begin{APACrefauthors}%
Bradley, M.M.%
\BCBT {}\ \BBA {} Lang, P.J.%
\end{APACrefauthors}%
\unskip\
\newblock
\APACrefYearMonthDay{1994}{}{}.
\newblock
{\BBOQ}\APACrefatitle {Measuring emotion: The {Self-Assessment Manikin} and the semantic differential} {Measuring emotion: The {Self-Assessment Manikin} and the semantic differential}.{\BBCQ}
\newblock
\APACjournalVolNumPages{Journal of Behavior Therapy and Experimental Psychiatry}{25}{1}{49--59,}
\newblock
\begin{APACrefURL} {https://www.sciencedirect.com/science/article/abs/pii/0005791694900639} \end{APACrefURL}
\newblock

\newblock

\PrintBackRefs{\CurrentBib}

\bibitem [\protect \citeauthoryear {%
Brignull%
}{%
Brignull%
}{%
{\protect \APACyear {2023}}%
}]{%
Brignull2023}
\APACinsertmetastar {%
Brignull2023}%
\begin{APACrefauthors}%
Brignull, H.%
\end{APACrefauthors}%
\unskip\
\newblock
\APACrefYear{2023}.
\newblock
\APACrefbtitle {Deceptive Patterns: Exposing the Tricks Tech Companies Use to Control You} {Deceptive patterns: Exposing the tricks tech companies use to control you}.
\newblock
\APACaddressPublisher{London, UK}{Testimonium Ltd}.
\PrintBackRefs{\CurrentBib}

\bibitem [\protect \citeauthoryear {%
Carter%
, Mankoff%
, Klemmer%
\BCBL {}\ \BBA {} Matthews%
}{%
Carter%
\ \protect \BOthers {.}}{%
{\protect \APACyear {2008}}%
}]{%
carter2008exiting}
\APACinsertmetastar {%
carter2008exiting}%
\begin{APACrefauthors}%
Carter, S.%
, Mankoff, J.%
, Klemmer, S.R.%
\BCBL {} Matthews, T.%
\end{APACrefauthors}%
\unskip\
\newblock
\APACrefYearMonthDay{2008}{}{}.
\newblock
{\BBOQ}\APACrefatitle {Exiting the cleanroom: On ecological validity and ubiquitous computing} {Exiting the cleanroom: On ecological validity and ubiquitous computing}.{\BBCQ}
\newblock
\APACjournalVolNumPages{Human--Computer Interaction}{23}{1}{47--99,}
\newblock
\begin{APACrefURL} {https://doi.org/10.1080/07370020701851086} \end{APACrefURL}
\newblock

\newblock

\PrintBackRefs{\CurrentBib}

\bibitem [\protect \citeauthoryear {%
Chang%
, Seaborn%
\BCBL {}\ \BBA {} Adams%
}{%
Chang%
\ \protect \BOthers {.}}{%
{\protect \APACyear {2024}}%
}]{%
chang2024theory}
\APACinsertmetastar {%
chang2024theory}%
\begin{APACrefauthors}%
Chang, W.J.%
, Seaborn, K.%
\BCBL {} Adams, A.A.%
\end{APACrefauthors}%
\unskip\
\newblock
\APACrefYearMonthDay{2024}{}{}.
\newblock
{\BBOQ}\APACrefatitle {Theorizing Deception: A Scoping Review of Theory in Research on Dark Patterns and Deceptive Design} {Theorizing deception: A scoping review of theory in research on dark patterns and deceptive design}.{\BBCQ}
\newblock
 \APACrefbtitle {{Extended Abstracts of the 2024 CHI Conference on Human Factors in Computing Systems}.} {{Extended Abstracts of the 2024 CHI Conference on Human Factors in Computing Systems}.}
\newblock
\APACaddressPublisher{New York, NY, USA}{Association for Computing Machinery}.
\newblock
\begin{APACrefURL} {https://doi.org/10.1145/3613905.3650997} \end{APACrefURL}
\PrintBackRefs{\CurrentBib}

\bibitem [\protect \citeauthoryear {%
Chaudhary%
, Saroha%
, Monteiro%
, Forbes%
\BCBL {}\ \BBA {} Parnami%
}{%
Chaudhary%
\ \protect \BOthers {.}}{%
{\protect \APACyear {2022}}%
}]{%
chaudhary2022videostream}
\APACinsertmetastar {%
chaudhary2022videostream}%
\begin{APACrefauthors}%
Chaudhary, A.%
, Saroha, J.%
, Monteiro, K.%
, Forbes, A.G.%
\BCBL {} Parnami, A.%
\end{APACrefauthors}%
\unskip\
\newblock
\APACrefYearMonthDay{2022}{}{}.
\newblock
{\BBOQ}\APACrefatitle {``{A}re You Still Watching?'': Exploring Unintended User Behaviors and Dark Patterns on Video Streaming Platforms} {``{A}re you still watching?'': Exploring unintended user behaviors and dark patterns on video streaming platforms}.{\BBCQ}
\newblock
 \APACrefbtitle {{Proceedings of the 2022 ACM Designing Interactive Systems Conference}} {{Proceedings of the 2022 ACM Designing Interactive Systems Conference}}\ (\BPG~776–791).
\newblock
\APACaddressPublisher{New York, NY, USA}{ACM}.
\newblock
\begin{APACrefURL} {https://doi.org/10.1145/3532106.3533562} \end{APACrefURL}
\PrintBackRefs{\CurrentBib}

\bibitem [\protect \citeauthoryear {%
Cialdini%
}{%
Cialdini%
}{%
{\protect \APACyear {2001}}%
}]{%
cialdini2001science}
\APACinsertmetastar {%
cialdini2001science}%
\begin{APACrefauthors}%
Cialdini, R.B.%
\end{APACrefauthors}%
\unskip\
\newblock
\APACrefYearMonthDay{2001}{}{}.
\newblock
{\BBOQ}\APACrefatitle {The science of persuasion} {The science of persuasion}.{\BBCQ}
\newblock
\APACjournalVolNumPages{Scientific American}{284}{2}{76--81,}
\newblock

\newblock

\PrintBackRefs{\CurrentBib}

\bibitem [\protect \citeauthoryear {%
Cialdini%
, Wosinska%
, Barrett%
, Butner%
\BCBL {}\ \BBA {} Gornik-Durose%
}{%
Cialdini%
\ \protect \BOthers {.}}{%
{\protect \APACyear {1999}}%
}]{%
Cialdini1999}
\APACinsertmetastar {%
Cialdini1999}%
\begin{APACrefauthors}%
Cialdini, R.B.%
, Wosinska, W.%
, Barrett, D.W.%
, Butner, J.%
\BCBL {} Gornik-Durose, M.%
\end{APACrefauthors}%
\unskip\
\newblock
\APACrefYearMonthDay{1999}{{\APACmonth{10}}}{}.
\newblock
{\BBOQ}\APACrefatitle {Compliance with a Request in Two Cultures: The Differential Influence of Social Proof and Commitment/Consistency on Collectivists and Individualists} {Compliance with a request in two cultures: The differential influence of social proof and commitment/consistency on collectivists and individualists}.{\BBCQ}
\newblock
\APACjournalVolNumPages{Personality and Social Psychology Bulletin}{25}{10}{1242–1253,}
\newblock
\begin{APACrefDOI} \doi{10.1177/0146167299258006} \end{APACrefDOI}
\newblock
\begin{APACrefURL} {http://dx.doi.org/10.1177/0146167299258006} \end{APACrefURL}
\newblock

\newblock

\PrintBackRefs{\CurrentBib}

\bibitem [\protect \citeauthoryear {%
Clark-Gordon%
, Bowman%
, Goodboy%
\BCBL {}\ \BBA {} Wright%
}{%
Clark-Gordon%
\ \protect \BOthers {.}}{%
{\protect \APACyear {2019}}%
}]{%
ClarkGordon2019}
\APACinsertmetastar {%
ClarkGordon2019}%
\begin{APACrefauthors}%
Clark-Gordon, C.V.%
, Bowman, N.D.%
, Goodboy, A.K.%
\BCBL {} Wright, A.%
\end{APACrefauthors}%
\unskip\
\newblock
\APACrefYearMonthDay{2019}{{\APACmonth{05}}}{}.
\newblock
{\BBOQ}\APACrefatitle {Anonymity and Online Self-Disclosure: A Meta-Analysis} {Anonymity and online self-disclosure: A meta-analysis}.{\BBCQ}
\newblock
\APACjournalVolNumPages{Communication Reports}{32}{2}{98–111,}
\newblock
\begin{APACrefDOI} \doi{10.1080/08934215.2019.1607516} \end{APACrefDOI}
\newblock
\begin{APACrefURL} {http://dx.doi.org/10.1080/08934215.2019.1607516} \end{APACrefURL}
\newblock

\newblock

\PrintBackRefs{\CurrentBib}

\bibitem [\protect \citeauthoryear {%
Conti%
\ \BBA {} Sobiesk%
}{%
Conti%
\ \BBA {} Sobiesk%
}{%
{\protect \APACyear {2010}}%
}]{%
conti2010malicious}
\APACinsertmetastar {%
conti2010malicious}%
\begin{APACrefauthors}%
Conti, G.%
\BCBT {}\ \BBA {} Sobiesk, E.%
\end{APACrefauthors}%
\unskip\
\newblock
\APACrefYearMonthDay{2010}{}{}.
\newblock
{\BBOQ}\APACrefatitle {Malicious Interface Design: Exploiting the User} {Malicious interface design: Exploiting the user}.{\BBCQ}
\newblock
 \APACrefbtitle {{Proceedings of the 19th International Conference on World Wide Web}} {{Proceedings of the 19th International Conference on World Wide Web}}\ (\BPG~271–280).
\newblock
\APACaddressPublisher{New York, NY, USA}{ACM}.
\newblock
\begin{APACrefURL} {https://doi.org/10.1145/1772690.1772719} \end{APACrefURL}
\PrintBackRefs{\CurrentBib}

\bibitem [\protect \citeauthoryear {%
Cranor%
}{%
Cranor%
}{%
{\protect \APACyear {2022}}%
}]{%
cranor2022cookie}
\APACinsertmetastar {%
cranor2022cookie}%
\begin{APACrefauthors}%
Cranor, L.F.%
\end{APACrefauthors}%
\unskip\
\newblock
\APACrefYearMonthDay{2022}{June}{}.
\newblock
{\BBOQ}\APACrefatitle {Cookie Monster} {Cookie monster}.{\BBCQ}
\newblock
\APACjournalVolNumPages{Commun. ACM}{65}{7}{30–32,}
\newblock
\begin{APACrefDOI} \doi{10.1145/3538639} \end{APACrefDOI}
\newblock
\begin{APACrefURL} {https://doi.org/10.1145/3538639} \end{APACrefURL}
\newblock

\newblock

\PrintBackRefs{\CurrentBib}

\bibitem [\protect \citeauthoryear {%
Di~Geronimo%
, Braz%
, Fregnan%
, Palomba%
\BCBL {}\ \BBA {} Bacchelli%
}{%
Di~Geronimo%
\ \protect \BOthers {.}}{%
{\protect \APACyear {2020}}%
}]{%
digeronimo2020}
\APACinsertmetastar {%
digeronimo2020}%
\begin{APACrefauthors}%
Di~Geronimo, L.%
, Braz, L.%
, Fregnan, E.%
, Palomba, F.%
\BCBL {} Bacchelli, A.%
\end{APACrefauthors}%
\unskip\
\newblock
\APACrefYearMonthDay{2020}{}{}.
\newblock
{\BBOQ}\APACrefatitle {{UI} dark datterns and where to find them: A study on mobile applications and user perception} {{UI} dark datterns and where to find them: A study on mobile applications and user perception}.{\BBCQ}
\newblock
 \APACrefbtitle {{Proceedings of the 2020 CHI Conference on Human Factors in Computing Systems}} {{Proceedings of the 2020 CHI Conference on Human Factors in Computing Systems}}\ (\BPG~1–14).
\newblock
\APACaddressPublisher{New York, NY, USA}{ACM}.
\newblock
\begin{APACrefURL} {https://doi.org/10.1145/3313831.3376600} \end{APACrefURL}
\PrintBackRefs{\CurrentBib}

\bibitem [\protect \citeauthoryear {%
Fansher%
, Chivukula%
\BCBL {}\ \BBA {} Gray%
}{%
Fansher%
\ \protect \BOthers {.}}{%
{\protect \APACyear {2018}}%
}]{%
fansher2018hashtag}
\APACinsertmetastar {%
fansher2018hashtag}%
\begin{APACrefauthors}%
Fansher, M.%
, Chivukula, S.S.%
\BCBL {} Gray, C.M.%
\end{APACrefauthors}%
\unskip\
\newblock
\APACrefYearMonthDay{2018}{}{}.
\newblock
{\BBOQ}\APACrefatitle {\#darkpatterns: {UX} Practitioner Conversations About Ethical Design} {\#darkpatterns: {UX} practitioner conversations about ethical design}.{\BBCQ}
\newblock
 \APACrefbtitle {{Extended Abstracts of the 2018 CHI Conference on Human Factors in Computing Systems}} {{Extended Abstracts of the 2018 CHI Conference on Human Factors in Computing Systems}}\ (\BPG~1–6).
\newblock
\APACaddressPublisher{New York, NY, USA}{ACM}.
\newblock
\begin{APACrefURL} {https://doi.org/10.1145/3170427.3188553} \end{APACrefURL}
\PrintBackRefs{\CurrentBib}

\bibitem [\protect \citeauthoryear {%
Feng%
\ \protect \BOthers {.}}{%
Feng%
\ \protect \BOthers {.}}{%
{\protect \APACyear {2023}}%
}]{%
feng2023analysis}
\APACinsertmetastar {%
feng2023analysis}%
\begin{APACrefauthors}%
Feng, J.%
, Mo, F.%
, Yada, Y.%
, Matsumoto, T.%
, Fukushima, N.%
, Kido, F.%
\BCBL {} Yamana, H.%
\end{APACrefauthors}%
\unskip\
\newblock
\APACrefYearMonthDay{2023}{}{}.
\newblock
{\BBOQ}\APACrefatitle {Analysis of Dark Pattern-related Tweets from 2010} {Analysis of dark pattern-related tweets from 2010}.{\BBCQ}
\newblock
 \APACrefbtitle {{2023 IEEE 8th International Conference on Big Data Analytics (ICBDA)}} {{2023 IEEE 8th International Conference on Big Data Analytics (ICBDA)}}\ (\BPGS\ 100--106).
\newblock
\APACaddressPublisher{Piscataway, NJ, USA}{IEEE}.
\newblock
\begin{APACrefURL} {https://doi.org/10.1109/ICBDA57405.2023.10104855} \end{APACrefURL}
\PrintBackRefs{\CurrentBib}

\bibitem [\protect \citeauthoryear {%
Fogg%
}{%
Fogg%
}{%
{\protect \APACyear {2002}}%
}]{%
Fogg2002}
\APACinsertmetastar {%
Fogg2002}%
\begin{APACrefauthors}%
Fogg, B.J.%
\end{APACrefauthors}%
\unskip\
\newblock
\APACrefYearMonthDay{2002}{{\APACmonth{12}}}{}.
\newblock
{\BBOQ}\APACrefatitle {Persuasive technology: Using computers to change what we think and do} {Persuasive technology: Using computers to change what we think and do}.{\BBCQ}
\newblock
\APACjournalVolNumPages{Ubiquity}{2002}{}{2,}
\newblock
\begin{APACrefDOI} \doi{10.1145/764008.763957} \end{APACrefDOI}
\newblock
\begin{APACrefURL} {http://dx.doi.org/10.1145/764008.763957} \end{APACrefURL}
\newblock

\newblock

\PrintBackRefs{\CurrentBib}

\bibitem [\protect \citeauthoryear {%
Fujihara%
\ \BBA {} Kurokawa%
}{%
Fujihara%
\ \BBA {} Kurokawa%
}{%
{\protect \APACyear {1981}}%
}]{%
fujihara1981}
\APACinsertmetastar {%
fujihara1981}%
\begin{APACrefauthors}%
Fujihara, T.%
\BCBT {}\ \BBA {} Kurokawa, M.%
\end{APACrefauthors}%
\unskip\
\newblock
\APACrefYearMonthDay{1981}{}{}.
\newblock
{\BBOQ}\APACrefatitle {An empirical study of amae (dependence) in interpersonal relations} {An empirical study of amae (dependence) in interpersonal relations}.{\BBCQ}
\newblock
\APACjournalVolNumPages{The Japanese Journal of Experimental Social Psychology}{21}{1}{53–62,}
\newblock
\begin{APACrefDOI} \doi{10.2130/jjesp.21.53} \end{APACrefDOI}
\newblock
\begin{APACrefURL} {http://dx.doi.org/10.2130/jjesp.21.53} \end{APACrefURL}
\newblock

\newblock

\PrintBackRefs{\CurrentBib}

\bibitem [\protect \citeauthoryear {%
Fujimura%
}{%
Fujimura%
}{%
{\protect \APACyear {1999}}%
}]{%
fujimura1999}
\APACinsertmetastar {%
fujimura1999}%
\begin{APACrefauthors}%
Fujimura, K.%
\end{APACrefauthors}%
\unskip\
\newblock
\APACrefYearMonthDay{1999}{Jun}{}.
\newblock
{\BBOQ}\APACrefatitle {The characteristics of consumer satisfaction/dissatisfaction formation in service consumption of {J}apanese people} {The characteristics of consumer satisfaction/dissatisfaction formation in service consumption of {J}apanese people}.{\BBCQ}
\newblock
\APACjournalVolNumPages{Kagawa University Economic Review}{72}{}{215--240,}
\newblock

\newblock

\PrintBackRefs{\CurrentBib}

\bibitem [\protect \citeauthoryear {%
Gra{\ss}l%
, Schraffenberger%
, Zuiderveen~Borgesius%
\BCBL {}\ \BBA {} Buijzen%
}{%
Gra{\ss}l%
\ \protect \BOthers {.}}{%
{\protect \APACyear {2021}}%
}]{%
grassl2021dark}
\APACinsertmetastar {%
grassl2021dark}%
\begin{APACrefauthors}%
Gra{\ss}l, P.%
, Schraffenberger, H.%
, Zuiderveen~Borgesius, F.%
\BCBL {} Buijzen, M.%
\end{APACrefauthors}%
\unskip\
\newblock
\APACrefYearMonthDay{2021}{}{}.
\newblock
{\BBOQ}\APACrefatitle {Dark and bright patterns in cookie consent requests} {Dark and bright patterns in cookie consent requests}.{\BBCQ}
\newblock
\APACjournalVolNumPages{Journal of Digital Social Research}{3}{1}{1--38,}
\newblock
\begin{APACrefDOI} \doi{10.33621/jdsr.v3i1.54} \end{APACrefDOI}
\newblock
\begin{APACrefURL} {https://doi.org/10.33621/jdsr.v3i1.54} \end{APACrefURL}
\newblock

\newblock

\PrintBackRefs{\CurrentBib}

\bibitem [\protect \citeauthoryear {%
Gray%
, Chen%
, Chivukula%
\BCBL {}\ \BBA {} Qu%
}{%
Gray%
, Chen%
\BCBL {}\ \protect \BOthers {.}}{%
{\protect \APACyear {2021}}%
}]{%
Gray2021felt}
\APACinsertmetastar {%
Gray2021felt}%
\begin{APACrefauthors}%
Gray, C.M.%
, Chen, J.%
, Chivukula, S.S.%
\BCBL {} Qu, L.%
\end{APACrefauthors}%
\unskip\
\newblock
\APACrefYearMonthDay{2021}{{\APACmonth{10}}}{}.
\newblock
{\BBOQ}\APACrefatitle {End User Accounts of Dark Patterns as Felt Manipulation} {End user accounts of dark patterns as felt manipulation}.{\BBCQ}
\newblock
\APACjournalVolNumPages{Proceedings of the ACM on Human-Computer Interaction}{5}{CSCW2}{1–25,}
\newblock
\begin{APACrefDOI} \doi{10.1145/3479516} \end{APACrefDOI}
\newblock
\begin{APACrefURL} {http://dx.doi.org/10.1145/3479516} \end{APACrefURL}
\newblock

\newblock

\PrintBackRefs{\CurrentBib}

\bibitem [\protect \citeauthoryear {%
Gray%
, Kou%
, Battles%
, Hoggatt%
\BCBL {}\ \BBA {} Toombs%
}{%
Gray%
\ \protect \BOthers {.}}{%
{\protect \APACyear {2018}}%
}]{%
gray2018darkside}
\APACinsertmetastar {%
gray2018darkside}%
\begin{APACrefauthors}%
Gray, C.M.%
, Kou, Y.%
, Battles, B.%
, Hoggatt, J.%
\BCBL {} Toombs, A.L.%
\end{APACrefauthors}%
\unskip\
\newblock
\APACrefYearMonthDay{2018}{}{}.
\newblock
{\BBOQ}\APACrefatitle {The Dark (Patterns) Side of {UX} Design} {The dark (patterns) side of {UX} design}.{\BBCQ}
\newblock
 \APACrefbtitle {{Proceedings of the 2018 CHI Conference on Human Factors in Computing Systems}} {{Proceedings of the 2018 CHI Conference on Human Factors in Computing Systems}}\ (\BPG~1–14).
\newblock
\APACaddressPublisher{New York, NY, USA}{ACM}.
\newblock
\begin{APACrefURL} {https://doi.org/10.1145/3173574.3174108} \end{APACrefURL}
\PrintBackRefs{\CurrentBib}

\bibitem [\protect \citeauthoryear {%
Gray%
, Mildner%
\BCBL {}\ \BBA {} Gairola%
}{%
Gray%
\ \protect \BOthers {.}}{%
{\protect \APACyear {2025}}%
}]{%
Gray2025time}
\APACinsertmetastar {%
Gray2025time}%
\begin{APACrefauthors}%
Gray, C.M.%
, Mildner, T.%
\BCBL {} Gairola, R.%
\end{APACrefauthors}%
\unskip\
\newblock
\APACrefYearMonthDay{2025}{{\APACmonth{04}}}{}.
\newblock
{\BBOQ}\APACrefatitle {Getting Trapped in Amazon’s “Iliad Flow”: A Foundation for the Temporal Analysis of Dark Patterns} {Getting trapped in amazon’s “iliad flow”: A foundation for the temporal analysis of dark patterns}.{\BBCQ}
\newblock
 \APACrefbtitle {Proceedings of the 2025 CHI Conference on Human Factors in Computing Systems} {Proceedings of the 2025 chi conference on human factors in computing systems}\ (\BPG~1–10).
\newblock
\APACaddressPublisher{}{ACM}.
\newblock
\begin{APACrefURL} {http://dx.doi.org/10.1145/3706598.3713828} \end{APACrefURL}
\PrintBackRefs{\CurrentBib}

\bibitem [\protect \citeauthoryear {%
Gray%
, Sanchez~Chamorro%
, Obi%
\BCBL {}\ \BBA {} Duane%
}{%
Gray%
\ \protect \BOthers {.}}{%
{\protect \APACyear {2023}}%
}]{%
gray2023dpsysreview}
\APACinsertmetastar {%
gray2023dpsysreview}%
\begin{APACrefauthors}%
Gray, C.M.%
, Sanchez~Chamorro, L.%
, Obi, I.%
\BCBL {} Duane, J\BHBI N.%
\end{APACrefauthors}%
\unskip\
\newblock
\APACrefYearMonthDay{2023}{}{}.
\newblock
{\BBOQ}\APACrefatitle {Mapping the Landscape of Dark Patterns Scholarship: A Systematic Literature Review} {Mapping the landscape of dark patterns scholarship: A systematic literature review}.{\BBCQ}
\newblock
 \APACrefbtitle {{Companion Publication of the 2023 ACM Designing Interactive Systems Conference}} {{Companion Publication of the 2023 ACM Designing Interactive Systems Conference}}\ (\BPG~188–193).
\newblock
\APACaddressPublisher{New York, NY, USA}{ACM}.
\newblock
\begin{APACrefURL} {https://doi.org/10.1145/3563703.3596635} \end{APACrefURL}
\PrintBackRefs{\CurrentBib}

\bibitem [\protect \citeauthoryear {%
Gray%
, Santos%
, Bielova%
, Toth%
\BCBL {}\ \BBA {} Clifford%
}{%
Gray%
, Santos%
\BCBL {}\ \protect \BOthers {.}}{%
{\protect \APACyear {2021}}%
}]{%
gray2021legal}
\APACinsertmetastar {%
gray2021legal}%
\begin{APACrefauthors}%
Gray, C.M.%
, Santos, C.%
, Bielova, N.%
, Toth, M.%
\BCBL {} Clifford, D.%
\end{APACrefauthors}%
\unskip\
\newblock
\APACrefYearMonthDay{2021}{}{}.
\newblock
{\BBOQ}\APACrefatitle {Dark Patterns and the Legal Requirements of Consent Banners: An Interaction Criticism Perspective} {Dark patterns and the legal requirements of consent banners: An interaction criticism perspective}.{\BBCQ}
\newblock
 \APACrefbtitle {{Proceedings of the 2021 CHI Conference on Human Factors in Computing Systems}.} {{Proceedings of the 2021 CHI Conference on Human Factors in Computing Systems}.}
\newblock
\APACaddressPublisher{New York, NY, USA}{ACM}.
\newblock
\begin{APACrefURL} {https://doi.org/10.1145/3411764.3445779} \end{APACrefURL}
\PrintBackRefs{\CurrentBib}

\bibitem [\protect \citeauthoryear {%
Gray%
, Santos%
, Bielova%
\BCBL {}\ \BBA {} Mildner%
}{%
Gray%
\ \protect \BOthers {.}}{%
{\protect \APACyear {2024}}%
}]{%
gray2024ontology}
\APACinsertmetastar {%
gray2024ontology}%
\begin{APACrefauthors}%
Gray, C.M.%
, Santos, C.T.%
, Bielova, N.%
\BCBL {} Mildner, T.%
\end{APACrefauthors}%
\unskip\
\newblock
\APACrefYearMonthDay{2024}{}{}.
\newblock
{\BBOQ}\APACrefatitle {An Ontology of Dark Patterns Knowledge: Foundations, Definitions, and a Pathway for Shared Knowledge-Building} {An ontology of dark patterns knowledge: Foundations, definitions, and a pathway for shared knowledge-building}.{\BBCQ}
\newblock
 \APACrefbtitle {{Proceedings of the CHI Conference on Human Factors in Computing Systems}.} {{Proceedings of the CHI Conference on Human Factors in Computing Systems}.}
\newblock
\APACaddressPublisher{New York, NY, USA}{Association for Computing Machinery}.
\newblock
\begin{APACrefURL} {https://doi.org/10.1145/3613904.3642436} \end{APACrefURL}
\PrintBackRefs{\CurrentBib}

\bibitem [\protect \citeauthoryear {%
Greenwood%
\ \BBA {} Nikulin%
}{%
Greenwood%
\ \BBA {} Nikulin%
}{%
{\protect \APACyear {1996}}%
}]{%
greenwood1996guide}
\APACinsertmetastar {%
greenwood1996guide}%
\begin{APACrefauthors}%
Greenwood, P.E.%
\BCBT {}\ \BBA {} Nikulin, M.S.%
\end{APACrefauthors}%
\unskip\
\newblock
\APACrefYear{1996}.
\newblock
\APACrefbtitle {A guide to {Chi-squared} testing} {A guide to {Chi-squared} testing}\ (\BVOL~280).
\newblock
\APACaddressPublisher{New York, NY, USA}{John Wiley \& Sons}.
\PrintBackRefs{\CurrentBib}

\bibitem [\protect \citeauthoryear {%
Gunawan%
, Hartzog%
, Richards%
, Choffnes%
\BCBL {}\ \BBA {} Wilson%
}{%
Gunawan%
\ \protect \BOthers {.}}{%
{\protect \APACyear {2024}}%
}]{%
gunawan2024dark}
\APACinsertmetastar {%
gunawan2024dark}%
\begin{APACrefauthors}%
Gunawan, J.%
, Hartzog, W.%
, Richards, N.%
, Choffnes, D.%
\BCBL {} Wilson, C.%
\end{APACrefauthors}%
\unskip\
\newblock
\APACrefYearMonthDay{2024}{}{}.
\newblock
{\BBOQ}\APACrefatitle {Dark Patterns as Disloyal Design} {Dark patterns as disloyal design}.{\BBCQ}
\newblock
\APACjournalVolNumPages{Indiana Law Journal}{100}{}{1389,}
\newblock
\begin{APACrefURL} {https://scholarship.law.bu.edu/faculty\_scholarship/4107/} \end{APACrefURL}
\newblock

\newblock

\PrintBackRefs{\CurrentBib}

\bibitem [\protect \citeauthoryear {%
Gunawan%
, Pradeep%
, Choffnes%
, Hartzog%
\BCBL {}\ \BBA {} Wilson%
}{%
Gunawan%
\ \protect \BOthers {.}}{%
{\protect \APACyear {2021}}%
}]{%
gunawan2021webmobile}
\APACinsertmetastar {%
gunawan2021webmobile}%
\begin{APACrefauthors}%
Gunawan, J.%
, Pradeep, A.%
, Choffnes, D.%
, Hartzog, W.%
\BCBL {} Wilson, C.%
\end{APACrefauthors}%
\unskip\
\newblock
\APACrefYearMonthDay{2021}{oct}{}.
\newblock
{\BBOQ}\APACrefatitle {A Comparative Study of Dark Patterns Across Web and Mobile Modalities} {A comparative study of dark patterns across web and mobile modalities}.{\BBCQ}
\newblock
\APACjournalVolNumPages{Proceedings of the ACM on Human-Computer Interaction}{5}{CSCW2}{,}
\newblock
\begin{APACrefDOI} \doi{10.1145/3479521} \end{APACrefDOI}
\newblock
\begin{APACrefURL} {https://doi.org/10.1145/3479521} \end{APACrefURL}
\newblock

\newblock

\PrintBackRefs{\CurrentBib}

\bibitem [\protect \citeauthoryear {%
Habib%
, Li%
, Young%
\BCBL {}\ \BBA {} Cranor%
}{%
Habib%
\ \protect \BOthers {.}}{%
{\protect \APACyear {2022}}%
}]{%
habib2022cookie}
\APACinsertmetastar {%
habib2022cookie}%
\begin{APACrefauthors}%
Habib, H.%
, Li, M.%
, Young, E.%
\BCBL {} Cranor, L.%
\end{APACrefauthors}%
\unskip\
\newblock
\APACrefYearMonthDay{2022}{}{}.
\newblock
{\BBOQ}\APACrefatitle {``{O}kay, whatever''': An evaluation of cookie consent interfaces} {``{O}kay, whatever''': An evaluation of cookie consent interfaces}.{\BBCQ}
\newblock
 \APACrefbtitle {{Proceedings of the 2022 CHI Conference on Human Factors in Computing Systems}.} {{Proceedings of the 2022 CHI Conference on Human Factors in Computing Systems}.}
\newblock
\APACaddressPublisher{New York, NY, USA}{ACM}.
\newblock
\begin{APACrefURL} {https://doi.org/10.1145/3491102.3501985} \end{APACrefURL}
\PrintBackRefs{\CurrentBib}

\bibitem [\protect \citeauthoryear {%
Hanjalic%
\ \BBA {} Xu%
}{%
Hanjalic%
\ \BBA {} Xu%
}{%
{\protect \APACyear {2005}}%
}]{%
Hanjalic2005}
\APACinsertmetastar {%
Hanjalic2005}%
\begin{APACrefauthors}%
Hanjalic, A.%
\BCBT {}\ \BBA {} Xu, L\BHBI Q.%
\end{APACrefauthors}%
\unskip\
\newblock
\APACrefYearMonthDay{2005}{{\APACmonth{02}}}{}.
\newblock
{\BBOQ}\APACrefatitle {Affective video content representation and modeling} {Affective video content representation and modeling}.{\BBCQ}
\newblock
\APACjournalVolNumPages{IEEE Transactions on Multimedia}{7}{1}{143–154,}
\newblock
\begin{APACrefDOI} \doi{10.1109/tmm.2004.840618} \end{APACrefDOI}
\newblock
\begin{APACrefURL} {http://dx.doi.org/10.1109/TMM.2004.840618} \end{APACrefURL}
\newblock

\newblock

\PrintBackRefs{\CurrentBib}

\bibitem [\protect \citeauthoryear {%
Head%
, Griffin%
, Bateman%
, Lohman%
\BCBL {}\ \BBA {} Yates%
}{%
Head%
\ \protect \BOthers {.}}{%
{\protect \APACyear {1988}}%
}]{%
head1988priming}
\APACinsertmetastar {%
head1988priming}%
\begin{APACrefauthors}%
Head, T.C.%
, Griffin, R.W.%
, Bateman, T.S.%
, Lohman, L.%
\BCBL {} Yates, V.L.%
\end{APACrefauthors}%
\unskip\
\newblock
\APACrefYearMonthDay{1988}{}{}.
\newblock
{\BBOQ}\APACrefatitle {The priming effect in task design research} {The priming effect in task design research}.{\BBCQ}
\newblock
\APACjournalVolNumPages{Journal of Management}{14}{1}{33--39,}
\newblock
\begin{APACrefURL} {https://doi.org/10.1177/014920638801400104} \end{APACrefURL}
\newblock

\newblock

\PrintBackRefs{\CurrentBib}

\bibitem [\protect \citeauthoryear {%
Heider%
}{%
Heider%
}{%
{\protect \APACyear {2013}}%
}]{%
heider2013psychology}
\APACinsertmetastar {%
heider2013psychology}%
\begin{APACrefauthors}%
Heider, F.%
\end{APACrefauthors}%
\unskip\
\newblock
\APACrefYear{2013}.
\newblock
\APACrefbtitle {The Psychology of Interpersonal Relations} {The psychology of interpersonal relations}.
\newblock
\APACaddressPublisher{London, UK}{Psychology Press}.
\PrintBackRefs{\CurrentBib}

\bibitem [\protect \citeauthoryear {%
Herbig%
\ \BBA {} Kramer%
}{%
Herbig%
\ \BBA {} Kramer%
}{%
{\protect \APACyear {1994}}%
}]{%
Herbig1994}
\APACinsertmetastar {%
Herbig1994}%
\begin{APACrefauthors}%
Herbig, P.A.%
\BCBT {}\ \BBA {} Kramer, H.%
\end{APACrefauthors}%
\unskip\
\newblock
\APACrefYearMonthDay{1994}{{\APACmonth{06}}}{}.
\newblock
{\BBOQ}\APACrefatitle {The Effect of Information Overload on the Innovation Choice Process: Innovation Overload} {The effect of information overload on the innovation choice process: Innovation overload}.{\BBCQ}
\newblock
\APACjournalVolNumPages{Journal of Consumer Marketing}{11}{2}{45–54,}
\newblock
\begin{APACrefDOI} \doi{10.1108/07363769410058920} \end{APACrefDOI}
\newblock
\begin{APACrefURL} {http://dx.doi.org/10.1108/07363769410058920} \end{APACrefURL}
\newblock

\newblock

\PrintBackRefs{\CurrentBib}

\bibitem [\protect \citeauthoryear {%
Hidaka%
, Kobuki%
, Watanabe%
\BCBL {}\ \BBA {} Seaborn%
}{%
Hidaka%
\ \protect \BOthers {.}}{%
{\protect \APACyear {2023}}%
}]{%
hidaka2023linguistic}
\APACinsertmetastar {%
hidaka2023linguistic}%
\begin{APACrefauthors}%
Hidaka, S.%
, Kobuki, S.%
, Watanabe, M.%
\BCBL {} Seaborn, K.%
\end{APACrefauthors}%
\unskip\
\newblock
\APACrefYearMonthDay{2023}{}{}.
\newblock
{\BBOQ}\APACrefatitle {{Linguistic Dead-Ends and Alphabet Soup: Finding dark patterns in Japanese apps}} {{Linguistic Dead-Ends and Alphabet Soup: Finding dark patterns in Japanese apps}}.{\BBCQ}
\newblock
 \APACrefbtitle {{Proceedings of the 2023 CHI Conference on Human Factors in Computing Systems}.} {{Proceedings of the 2023 CHI Conference on Human Factors in Computing Systems}.}
\newblock
\APACaddressPublisher{New York, NY, USA}{ACM}.
\newblock
\begin{APACrefURL} {https://doi.org/10.1145/3544548.3580942} \end{APACrefURL}
\PrintBackRefs{\CurrentBib}

\bibitem [\protect \citeauthoryear {%
Hogan%
, Barry%
\BCBL {}\ \BBA {} Lang%
}{%
Hogan%
\ \protect \BOthers {.}}{%
{\protect \APACyear {2022}}%
}]{%
Hogan2022}
\APACinsertmetastar {%
Hogan2022}%
\begin{APACrefauthors}%
Hogan, M.%
, Barry, C.%
\BCBL {} Lang, M.%
\end{APACrefauthors}%
\unskip\
\newblock
\APACrefYearMonthDay{2022}{{\APACmonth{11}}}{}.
\newblock
{\BBOQ}\APACrefatitle {Dissecting Optional Micro-Decisions in Online Transactions: Perceptions, Deceptions, and Errors} {Dissecting optional micro-decisions in online transactions: Perceptions, deceptions, and errors}.{\BBCQ}
\newblock
\APACjournalVolNumPages{ACM Transactions on Computer-Human Interaction}{29}{6}{1–27,}
\newblock
\begin{APACrefDOI} \doi{10.1145/3531005} \end{APACrefDOI}
\newblock
\begin{APACrefURL} {http://dx.doi.org/10.1145/3531005} \end{APACrefURL}
\newblock

\newblock

\PrintBackRefs{\CurrentBib}

\bibitem [\protect \citeauthoryear {%
Iturregui-Gallardo%
\ \BBA {} M{\'e}ndez-Ulrich%
}{%
Iturregui-Gallardo%
\ \BBA {} M{\'e}ndez-Ulrich%
}{%
{\protect \APACyear {2020}}%
}]{%
iturregui2020towards}
\APACinsertmetastar {%
iturregui2020towards}%
\begin{APACrefauthors}%
Iturregui-Gallardo, G.%
\BCBT {}\ \BBA {} M{\'e}ndez-Ulrich, J.L.%
\end{APACrefauthors}%
\unskip\
\newblock
\APACrefYearMonthDay{2020}{}{}.
\newblock
{\BBOQ}\APACrefatitle {Towards the creation of a tactile version of the {Self-Assessment Manikin (T-SAM)} for the emotional assessment of visually impaired people} {Towards the creation of a tactile version of the {Self-Assessment Manikin (T-SAM)} for the emotional assessment of visually impaired people}.{\BBCQ}
\newblock
\APACjournalVolNumPages{International Journal of Disability, Development and Education}{67}{6}{657--674,}
\newblock
\begin{APACrefURL} {https://doi.org/10.1080/1034912X.2019.1626007} \end{APACrefURL}
\newblock

\newblock

\PrintBackRefs{\CurrentBib}

\bibitem [\protect \citeauthoryear {%
Kalton%
\ \BBA {} Schuman%
}{%
Kalton%
\ \BBA {} Schuman%
}{%
{\protect \APACyear {1982}}%
}]{%
Kalton1982}
\APACinsertmetastar {%
Kalton1982}%
\begin{APACrefauthors}%
Kalton, G.%
\BCBT {}\ \BBA {} Schuman, H.%
\end{APACrefauthors}%
\unskip\
\newblock
\APACrefYearMonthDay{1982}{}{}.
\newblock
{\BBOQ}\APACrefatitle {The effect of the question on survey responses: A review} {The effect of the question on survey responses: A review}.{\BBCQ}
\newblock
\APACjournalVolNumPages{Journal of the Royal Statistical Society. Series A (General)}{145}{1}{42,}
\newblock
\begin{APACrefDOI} \doi{10.2307/2981421} \end{APACrefDOI}
\newblock
\begin{APACrefURL} {http://dx.doi.org/10.2307/2981421} \end{APACrefURL}
\newblock

\newblock

\PrintBackRefs{\CurrentBib}

\bibitem [\protect \citeauthoryear {%
Koh%
\ \BBA {} Seah%
}{%
Koh%
\ \BBA {} Seah%
}{%
{\protect \APACyear {2023}}%
}]{%
Koh2023yu}
\APACinsertmetastar {%
Koh2023yu}%
\begin{APACrefauthors}%
Koh, W.C.%
\BCBT {}\ \BBA {} Seah, Y.Z.%
\end{APACrefauthors}%
\unskip\
\newblock
\APACrefYearMonthDay{2023}{{\APACmonth{12}}}{}.
\newblock
{\BBOQ}\APACrefatitle {Unintended consumption: {The} effects of four e-commerce dark patterns} {Unintended consumption: {The} effects of four e-commerce dark patterns}.{\BBCQ}
\newblock
\APACjournalVolNumPages{Cleaner and Responsible Consumption}{11}{100145}{100145,}
\newblock

\newblock

\PrintBackRefs{\CurrentBib}

\bibitem [\protect \citeauthoryear {%
Kuppens%
, Tuerlinckx%
, Russell%
\BCBL {}\ \BBA {} Barrett%
}{%
Kuppens%
\ \protect \BOthers {.}}{%
{\protect \APACyear {2013}}%
}]{%
Kuppens2013}
\APACinsertmetastar {%
Kuppens2013}%
\begin{APACrefauthors}%
Kuppens, P.%
, Tuerlinckx, F.%
, Russell, J.A.%
\BCBL {} Barrett, L.F.%
\end{APACrefauthors}%
\unskip\
\newblock
\APACrefYearMonthDay{2013}{{\APACmonth{07}}}{}.
\newblock
{\BBOQ}\APACrefatitle {The relation between valence and arousal in subjective experience} {The relation between valence and arousal in subjective experience}.{\BBCQ}
\newblock
\APACjournalVolNumPages{Psychological Bulletin}{139}{4}{917–940,}
\newblock
\begin{APACrefDOI} \doi{10.1037/a0030811} \end{APACrefDOI}
\newblock
\begin{APACrefURL} {http://dx.doi.org/10.1037/a0030811} \end{APACrefURL}
\newblock

\newblock

\PrintBackRefs{\CurrentBib}

\bibitem [\protect \citeauthoryear {%
Kyi%
\ \protect \BOthers {.}}{%
Kyi%
\ \protect \BOthers {.}}{%
{\protect \APACyear {2023}}%
}]{%
kyi2023gdpr}
\APACinsertmetastar {%
kyi2023gdpr}%
\begin{APACrefauthors}%
Kyi, L.%
, Ammanaghatta~Shivakumar, S.%
, Santos, C.T.%
, Roesner, F.%
, Zufall, F.%
\BCBL {} Biega, A.J.%
\end{APACrefauthors}%
\unskip\
\newblock
\APACrefYearMonthDay{2023}{}{}.
\newblock
{\BBOQ}\APACrefatitle {Investigating Deceptive Design in {GDPR's} Legitimate Interest} {Investigating deceptive design in {GDPR's} legitimate interest}.{\BBCQ}
\newblock
 \APACrefbtitle {{Proceedings of the 2023 CHI Conference on Human Factors in Computing Systems}.} {{Proceedings of the 2023 CHI Conference on Human Factors in Computing Systems}.}
\newblock
\APACaddressPublisher{New York, NY, USA}{ACM}.
\newblock
\begin{APACrefURL} {https://doi.org/10.1145/3544548.3580637} \end{APACrefURL}
\PrintBackRefs{\CurrentBib}

\bibitem [\protect \citeauthoryear {%
Lewis%
}{%
Lewis%
}{%
{\protect \APACyear {2018}}%
}]{%
lewis2018system}
\APACinsertmetastar {%
lewis2018system}%
\begin{APACrefauthors}%
Lewis, J.R.%
\end{APACrefauthors}%
\unskip\
\newblock
\APACrefYearMonthDay{2018}{}{}.
\newblock
{\BBOQ}\APACrefatitle {The {System Usability Scale}: Past, present, and future} {The {System Usability Scale}: Past, present, and future}.{\BBCQ}
\newblock
\APACjournalVolNumPages{International Journal of Human--Computer Interaction}{34}{7}{577--590,}
\newblock

\newblock

\PrintBackRefs{\CurrentBib}

\bibitem [\protect \citeauthoryear {%
Linxen%
\ \protect \BOthers {.}}{%
Linxen%
\ \protect \BOthers {.}}{%
{\protect \APACyear {2021}}%
}]{%
Linxen2021}
\APACinsertmetastar {%
Linxen2021}%
\begin{APACrefauthors}%
Linxen, S.%
, Sturm, C.%
, Br\"{u}hlmann, F.%
, Cassau, V.%
, Opwis, K.%
\BCBL {} Reinecke, K.%
\end{APACrefauthors}%
\unskip\
\newblock
\APACrefYearMonthDay{2021}{{\APACmonth{05}}}{}.
\newblock
{\BBOQ}\APACrefatitle {How {WEIRD} is {CHI}?} {How {WEIRD} is {CHI}?}{\BBCQ}
\newblock
 \APACrefbtitle {{Proceedings of the 2021 CHI Conference on Human Factors in Computing Systems}} {{Proceedings of the 2021 CHI Conference on Human Factors in Computing Systems}}\ (\BPGS\ 1--14).
\newblock
\APACaddressPublisher{New York, NY, USA}{ACM}.
\newblock
\begin{APACrefURL} {http://dx.doi.org/10.1145/3411764.3445488} \end{APACrefURL}
\PrintBackRefs{\CurrentBib}

\bibitem [\protect \citeauthoryear {%
Loewenstein%
\ \BBA {} Prelec%
}{%
Loewenstein%
\ \BBA {} Prelec%
}{%
{\protect \APACyear {1991}}%
}]{%
loewenstein1991negative}
\APACinsertmetastar {%
loewenstein1991negative}%
\begin{APACrefauthors}%
Loewenstein, G.%
\BCBT {}\ \BBA {} Prelec, D.%
\end{APACrefauthors}%
\unskip\
\newblock
\APACrefYearMonthDay{1991}{}{}.
\newblock
{\BBOQ}\APACrefatitle {Negative time preference} {Negative time preference}.{\BBCQ}
\newblock
\APACjournalVolNumPages{The American Economic Review}{81}{2}{347--352,}
\newblock
\begin{APACrefURL} {https://www.jstor.org/stable/2006883} \end{APACrefURL}
\newblock

\newblock

\PrintBackRefs{\CurrentBib}

\bibitem [\protect \citeauthoryear {%
Luguri%
\ \BBA {} Strahilevitz%
}{%
Luguri%
\ \BBA {} Strahilevitz%
}{%
{\protect \APACyear {2021}}%
}]{%
luguri2021shining}
\APACinsertmetastar {%
luguri2021shining}%
\begin{APACrefauthors}%
Luguri, J.%
\BCBT {}\ \BBA {} Strahilevitz, L.J.%
\end{APACrefauthors}%
\unskip\
\newblock
\APACrefYearMonthDay{2021}{}{}.
\newblock
{\BBOQ}\APACrefatitle {Shining a light on dark patterns} {Shining a light on dark patterns}.{\BBCQ}
\newblock
\APACjournalVolNumPages{Journal of Legal Analysis}{13}{1}{43--109,}
\newblock
\begin{APACrefURL} {https://doi.org/10.1093/jla/laaa006} \end{APACrefURL}
\newblock

\newblock

\PrintBackRefs{\CurrentBib}

\bibitem [\protect \citeauthoryear {%
Machuletz%
\ \BBA {} B\"{o}hme%
}{%
Machuletz%
\ \BBA {} B\"{o}hme%
}{%
{\protect \APACyear {2020}}%
}]{%
Machuletz2020}
\APACinsertmetastar {%
Machuletz2020}%
\begin{APACrefauthors}%
Machuletz, D.%
\BCBT {}\ \BBA {} B\"{o}hme, R.%
\end{APACrefauthors}%
\unskip\
\newblock
\APACrefYearMonthDay{2020}{{\APACmonth{04}}}{}.
\newblock
{\BBOQ}\APACrefatitle {Multiple Purposes, Multiple Problems: A User Study of Consent Dialogs after {GDPR}} {Multiple purposes, multiple problems: A user study of consent dialogs after {GDPR}}.{\BBCQ}
\newblock
\APACjournalVolNumPages{Proceedings on Privacy Enhancing Technologies}{2020}{2}{481–498,}
\newblock
\begin{APACrefDOI} \doi{10.2478/popets-2020-0037} \end{APACrefDOI}
\newblock
\begin{APACrefURL} {http://dx.doi.org/10.2478/popets-2020-0037} \end{APACrefURL}
\newblock

\newblock

\PrintBackRefs{\CurrentBib}

\bibitem [\protect \citeauthoryear {%
Martin%
}{%
Martin%
}{%
{\protect \APACyear {2003}}%
}]{%
martin2003agile}
\APACinsertmetastar {%
martin2003agile}%
\begin{APACrefauthors}%
Martin, R.C.%
\end{APACrefauthors}%
\unskip\
\newblock
\APACrefYear{2003}.
\newblock
\APACrefbtitle {Agile Software Development: Principles, Patterns, and Practices} {Agile software development: Principles, patterns, and practices}.
\newblock
\APACaddressPublisher{Hoboken, NJ, USA}{Prentice Hall}.
\PrintBackRefs{\CurrentBib}

\bibitem [\protect \citeauthoryear {%
Mathur%
\ \protect \BOthers {.}}{%
Mathur%
\ \protect \BOthers {.}}{%
{\protect \APACyear {2019}}%
}]{%
mathur2019atscale}
\APACinsertmetastar {%
mathur2019atscale}%
\begin{APACrefauthors}%
Mathur, A.%
, Acar, G.%
, Friedman, M.J.%
, Lucherini, E.%
, Mayer, J.%
, Chetty, M.%
\BCBL {} Narayanan, A.%
\end{APACrefauthors}%
\unskip\
\newblock
\APACrefYearMonthDay{2019}{nov}{}.
\newblock
{\BBOQ}\APACrefatitle {Dark Patterns at Scale: Findings from a Crawl of 11K Shopping Websites} {Dark patterns at scale: Findings from a crawl of 11k shopping websites}.{\BBCQ}
\newblock
\APACjournalVolNumPages{Proceedings of the ACM on Human--Computer Interaction}{3}{CSCW}{,}
\newblock
\begin{APACrefDOI} \doi{10.1145/3359183} \end{APACrefDOI}
\newblock
\begin{APACrefURL} {https://doi.org/10.1145/3359183} \end{APACrefURL}
\newblock

\newblock

\PrintBackRefs{\CurrentBib}

\bibitem [\protect \citeauthoryear {%
Mathur%
, Kshirsagar%
\BCBL {}\ \BBA {} Mayer%
}{%
Mathur%
\ \protect \BOthers {.}}{%
{\protect \APACyear {2021}}%
}]{%
mathur2021whatdark}
\APACinsertmetastar {%
mathur2021whatdark}%
\begin{APACrefauthors}%
Mathur, A.%
, Kshirsagar, M.%
\BCBL {} Mayer, J.%
\end{APACrefauthors}%
\unskip\
\newblock
\APACrefYearMonthDay{2021}{}{}.
\newblock
{\BBOQ}\APACrefatitle {What Makes a Dark Pattern... Dark? {D}esign Attributes, Normative Considerations, and Measurement Methods} {What makes a dark pattern... dark? {D}esign attributes, normative considerations, and measurement methods}.{\BBCQ}
\newblock
 \APACrefbtitle {{Proceedings of the 2021 CHI Conference on Human Factors in Computing Systems}.} {{Proceedings of the 2021 CHI Conference on Human Factors in Computing Systems}.}
\newblock
\APACaddressPublisher{New York, NY, USA}{ACM}.
\newblock
\begin{APACrefURL} {https://doi.org/10.1145/3411764.3445610} \end{APACrefURL}
\PrintBackRefs{\CurrentBib}

\bibitem [\protect \citeauthoryear {%
Mauss%
\ \BBA {} Robinson%
}{%
Mauss%
\ \BBA {} Robinson%
}{%
{\protect \APACyear {2009}}%
}]{%
Mauss2009}
\APACinsertmetastar {%
Mauss2009}%
\begin{APACrefauthors}%
Mauss, I.B.%
\BCBT {}\ \BBA {} Robinson, M.D.%
\end{APACrefauthors}%
\unskip\
\newblock
\APACrefYearMonthDay{2009}{{\APACmonth{02}}}{}.
\newblock
{\BBOQ}\APACrefatitle {Measures of emotion: A review} {Measures of emotion: A review}.{\BBCQ}
\newblock
\APACjournalVolNumPages{Cognition \& Emotion}{23}{2}{209–237,}
\newblock
\begin{APACrefDOI} \doi{10.1080/02699930802204677} \end{APACrefDOI}
\newblock
\begin{APACrefURL} {http://dx.doi.org/10.1080/02699930802204677} \end{APACrefURL}
\newblock

\newblock

\PrintBackRefs{\CurrentBib}

\bibitem [\protect \citeauthoryear {%
Meckem%
\ \BBA {} Carlson%
}{%
Meckem%
\ \BBA {} Carlson%
}{%
{\protect \APACyear {2010}}%
}]{%
soni2010rapid}
\APACinsertmetastar {%
soni2010rapid}%
\begin{APACrefauthors}%
Meckem, S.%
\BCBT {}\ \BBA {} Carlson, J.L.%
\end{APACrefauthors}%
\unskip\
\newblock
\APACrefYearMonthDay{2010}{}{}.
\newblock
{\BBOQ}\APACrefatitle {Using ``rapid experimentation'' to inform customer service experience design} {Using ``rapid experimentation'' to inform customer service experience design}.{\BBCQ}
\newblock
 \APACrefbtitle {{CHI `10 Extended Abstracts on Human Factors in Computing Systems}} {{CHI `10 Extended Abstracts on Human Factors in Computing Systems}}\ (\BPG~4553–4566).
\newblock
\APACaddressPublisher{New York, NY, USA}{Association for Computing Machinery}.
\newblock
\begin{APACrefURL} {https://doi.org/10.1145/1753846.1754193} \end{APACrefURL}
\PrintBackRefs{\CurrentBib}

\bibitem [\protect \citeauthoryear {%
Mildner%
\ \protect \BOthers {.}}{%
Mildner%
\ \protect \BOthers {.}}{%
{\protect \APACyear {2023}}%
}]{%
mildner2023}
\APACinsertmetastar {%
mildner2023}%
\begin{APACrefauthors}%
Mildner, T.%
, Freye, M.%
, Savino, G\BHBI L.%
, Doyle, P.R.%
, Cowan, B.R.%
\BCBL {} Malaka, R.%
\end{APACrefauthors}%
\unskip\
\newblock
\APACrefYearMonthDay{2023}{}{}.
\newblock
{\BBOQ}\APACrefatitle {Defending Against the Dark Arts: Recognising Dark Patterns in Social Media} {Defending against the dark arts: Recognising dark patterns in social media}.{\BBCQ}
\newblock
 \APACrefbtitle {{Proceedings of the 2023 ACM Designing Interactive Systems Conference}} {{Proceedings of the 2023 ACM Designing Interactive Systems Conference}}\ (\BPG~2362–2374).
\newblock
\APACaddressPublisher{New York, NY, USA}{Association for Computing Machinery}.
\newblock
\begin{APACrefURL} {https://doi.org/10.1145/3563657.3595964} \end{APACrefURL}
\PrintBackRefs{\CurrentBib}

\bibitem [\protect \citeauthoryear {%
Miller%
\ \BBA {} Ross%
}{%
Miller%
\ \BBA {} Ross%
}{%
{\protect \APACyear {1975}}%
}]{%
miller1975self}
\APACinsertmetastar {%
miller1975self}%
\begin{APACrefauthors}%
Miller, D.T.%
\BCBT {}\ \BBA {} Ross, M.%
\end{APACrefauthors}%
\unskip\
\newblock
\APACrefYearMonthDay{1975}{}{}.
\newblock
{\BBOQ}\APACrefatitle {Self-serving biases in the attribution of causality: Fact or fiction?} {Self-serving biases in the attribution of causality: Fact or fiction?}{\BBCQ}
\newblock
\APACjournalVolNumPages{Psychological Bulletin}{82}{2}{213,}
\newblock

\newblock

\PrintBackRefs{\CurrentBib}

\bibitem [\protect \citeauthoryear {%
Mittone%
\ \BBA {} Savadori%
}{%
Mittone%
\ \BBA {} Savadori%
}{%
{\protect \APACyear {2009}}%
}]{%
Mittone2009}
\APACinsertmetastar {%
Mittone2009}%
\begin{APACrefauthors}%
Mittone, L.%
\BCBT {}\ \BBA {} Savadori, L.%
\end{APACrefauthors}%
\unskip\
\newblock
\APACrefYearMonthDay{2009}{{\APACmonth{06}}}{}.
\newblock
{\BBOQ}\APACrefatitle {The scarcity bias} {The scarcity bias}.{\BBCQ}
\newblock
\APACjournalVolNumPages{Applied Psychology}{58}{3}{453–468,}
\newblock
\begin{APACrefDOI} \doi{10.1111/j.1464-0597.2009.00401.x} \end{APACrefDOI}
\newblock

\newblock

\PrintBackRefs{\CurrentBib}

\bibitem [\protect \citeauthoryear {%
Miura%
}{%
Miura%
}{%
{\protect \APACyear {2021}}%
}]{%
miura2021norms}
\APACinsertmetastar {%
miura2021norms}%
\begin{APACrefauthors}%
Miura, T.%
\end{APACrefauthors}%
\unskip\
\newblock
\APACrefYear{2021}.
\unskip\
\newblock
\APACrefbtitle {Characteristics and Historical Origins of {Japanese} Consumer Behavior and {Japanese} Corporate Behavior: Individual Strength and Organizational Weakness} {Characteristics and historical origins of {Japanese} consumer behavior and {Japanese} corporate behavior: Individual strength and organizational weakness}\ \APACtypeAddressSchool {\BUPhD}{}{}.
\unskip\
\newblock
\APACaddressSchool {}{Chuo University}.
\PrintBackRefs{\CurrentBib}

\bibitem [\protect \citeauthoryear {%
Monge~Roffarello%
\ \BBA {} De~Russis%
}{%
Monge~Roffarello%
\ \BBA {} De~Russis%
}{%
{\protect \APACyear {2022}}%
}]{%
roffarello2022steal}
\APACinsertmetastar {%
roffarello2022steal}%
\begin{APACrefauthors}%
Monge~Roffarello, A.%
\BCBT {}\ \BBA {} De~Russis, L.%
\end{APACrefauthors}%
\unskip\
\newblock
\APACrefYearMonthDay{2022}{}{}.
\newblock
{\BBOQ}\APACrefatitle {Towards Understanding the Dark Patterns That Steal Our Attention} {Towards understanding the dark patterns that steal our attention}.{\BBCQ}
\newblock
 \APACrefbtitle {{Extended Abstracts of the 2022 CHI Conference on Human Factors in Computing Systems}.} {{Extended Abstracts of the 2022 CHI Conference on Human Factors in Computing Systems}.}
\newblock
\APACaddressPublisher{New York, NY, USA}{ACM}.
\newblock
\begin{APACrefURL} {https://doi.org/10.1145/3491101.3519829} \end{APACrefURL}
\PrintBackRefs{\CurrentBib}

\bibitem [\protect \citeauthoryear {%
Nakano%
}{%
Nakano%
}{%
{\protect \APACyear {2022}}%
}]{%
nakano2022zadak}
\APACinsertmetastar {%
nakano2022zadak}%
\begin{APACrefauthors}%
Nakano, Y.%
\end{APACrefauthors}%
\unskip\
\newblock
\APACrefYear{2022}.
\newblock
\APACrefbtitle {ザ・ダークパターン : ユーザーの心や行動をあざむくデザイン ({The Dark Pattern})} {ザ・ダークパターン : ユーザーの心や行動をあざむくデザイン ({The Dark Pattern})}.
\newblock
\APACaddressPublisher{Tokyo, Japan}{Shoei Publishing Co.}
\PrintBackRefs{\CurrentBib}

\bibitem [\protect \citeauthoryear {%
Narayanan%
, Mathur%
, Chetty%
\BCBL {}\ \BBA {} Kshirsagar%
}{%
Narayanan%
\ \protect \BOthers {.}}{%
{\protect \APACyear {2020}}%
}]{%
Narayanan2020}
\APACinsertmetastar {%
Narayanan2020}%
\begin{APACrefauthors}%
Narayanan, A.%
, Mathur, A.%
, Chetty, M.%
\BCBL {} Kshirsagar, M.%
\end{APACrefauthors}%
\unskip\
\newblock
\APACrefYearMonthDay{2020}{{\APACmonth{04}}}{}.
\newblock
{\BBOQ}\APACrefatitle {Dark Patterns: Past, Present, and Future: The evolution of tricky user interfaces} {Dark patterns: Past, present, and future: The evolution of tricky user interfaces}.{\BBCQ}
\newblock
\APACjournalVolNumPages{Queue}{18}{2}{67–92,}
\newblock
\begin{APACrefDOI} \doi{10.1145/3400899.3400901} \end{APACrefDOI}
\newblock

\newblock

\PrintBackRefs{\CurrentBib}

\bibitem [\protect \citeauthoryear {%
Nazarov%
\ \BBA {} Baimukhambetov%
}{%
Nazarov%
\ \BBA {} Baimukhambetov%
}{%
{\protect \APACyear {2022}}%
}]{%
nazarov2022clustering}
\APACinsertmetastar {%
nazarov2022clustering}%
\begin{APACrefauthors}%
Nazarov, D.%
\BCBT {}\ \BBA {} Baimukhambetov, Y.%
\end{APACrefauthors}%
\unskip\
\newblock
\APACrefYearMonthDay{2022}{}{}.
\newblock
{\BBOQ}\APACrefatitle {Clustering of Dark Patterns in the User Interfaces of Websites and Online Trading Portals (E-Commerce)} {Clustering of dark patterns in the user interfaces of websites and online trading portals (e-commerce)}.{\BBCQ}
\newblock
\APACjournalVolNumPages{Mathematics}{10}{18}{3219,}
\newblock
\begin{APACrefURL} {https://doi.org/10.3390/math10183219} \end{APACrefURL}
\newblock

\newblock

\PrintBackRefs{\CurrentBib}

\bibitem [\protect \citeauthoryear {%
Nik~Ahmad%
\ \BBA {} Megat~Sazali%
}{%
Nik~Ahmad%
\ \BBA {} Megat~Sazali%
}{%
{\protect \APACyear {2021}}%
}]{%
NikAhmad2021}
\APACinsertmetastar {%
NikAhmad2021}%
\begin{APACrefauthors}%
Nik~Ahmad, N.A.%
\BCBT {}\ \BBA {} Megat~Sazali, P.N.N.%
\end{APACrefauthors}%
\unskip\
\newblock
\APACrefYearMonthDay{2021}{{\APACmonth{08}}}{}.
\newblock
{\BBOQ}\APACrefatitle {Performing User Acceptance Test with {System Usability Scale} for Graduation Application} {Performing user acceptance test with {System Usability Scale} for graduation application}.{\BBCQ}
\newblock
 \APACrefbtitle {2021 International Conference on Software Engineering \& Computer Systems and 4th International Conference on Computational Science and Information Management (ICSECS-ICOCSIM)} {2021 international conference on software engineering \& computer systems and 4th international conference on computational science and information management (icsecs-icocsim)}\ (\BPG~86–91).
\newblock
\APACaddressPublisher{}{IEEE}.
\newblock
\begin{APACrefURL} {http://dx.doi.org/10.1109/ICSECS52883.2021.00023} \end{APACrefURL}
\PrintBackRefs{\CurrentBib}

\bibitem [\protect \citeauthoryear {%
Nimkoompai%
}{%
Nimkoompai%
}{%
{\protect \APACyear {2022}}%
}]{%
Nimkoompai2022}
\APACinsertmetastar {%
Nimkoompai2022}%
\begin{APACrefauthors}%
Nimkoompai, A.%
\end{APACrefauthors}%
\unskip\
\newblock
\APACrefYearMonthDay{2022}{November}{}.
\newblock
{\BBOQ}\APACrefatitle {Risk Analysis of Encountering Dark Patterns of {UX} E-commerce Applications Affecting Personal Data} {Risk analysis of encountering dark patterns of {UX} e-commerce applications affecting personal data}.{\BBCQ}
\newblock
 \APACrefbtitle {{2022 6th International Conference on Information Technology (InCIT)}} {{2022 6th International Conference on Information Technology (InCIT)}}\ (\BPGS\ 115--119).
\newblock
\APACaddressPublisher{New York, NY, USA}{IEEE}.
\newblock
\begin{APACrefURL} {http://dx.doi.org/10.1109/InCIT56086.2022.10067640} \end{APACrefURL}
\PrintBackRefs{\CurrentBib}

\bibitem [\protect \citeauthoryear {%
OECD%
}{%
OECD%
}{%
{\protect \APACyear {2022}}%
}]{%
oecd2022}
\APACinsertmetastar {%
oecd2022}%
\begin{APACrefauthors}%
OECD%
\end{APACrefauthors}%
\unskip\
\newblock
\APACrefYear{2022}.
\newblock
\APACrefbtitle {Dark commercial patterns} {Dark commercial patterns}\ (\BNUM~336).
\newblock
\APACaddressPublisher{Paris, France}{OECD Publishing}.
\PrintBackRefs{\CurrentBib}

\bibitem [\protect \citeauthoryear {%
OECD%
}{%
OECD%
}{%
{\protect \APACyear {2023}}%
}]{%
oecd2023}
\APACinsertmetastar {%
oecd2023}%
\begin{APACrefauthors}%
OECD%
\end{APACrefauthors}%
\unskip\
\newblock
\APACrefYear{2023}.
\newblock
\APACrefbtitle {Consumer Vulnerability in the Digital Age} {Consumer vulnerability in the digital age}\ (\BNUM~355).
\newblock
\APACaddressPublisher{Paris, France}{OECD Publishing}.
\PrintBackRefs{\CurrentBib}

\bibitem [\protect \citeauthoryear {%
Owens%
\ \protect \BOthers {.}}{%
Owens%
\ \protect \BOthers {.}}{%
{\protect \APACyear {2022}}%
}]{%
Owens2022}
\APACinsertmetastar {%
Owens2022}%
\begin{APACrefauthors}%
Owens, K.%
, Gunawan, J.%
, Choffnes, D.%
, Emami-Naeini, P.%
, Kohno, T.%
\BCBL {} Roesner, F.%
\end{APACrefauthors}%
\unskip\
\newblock
\APACrefYearMonthDay{2022}{{\APACmonth{09}}}{}.
\newblock
{\BBOQ}\APACrefatitle {Exploring Deceptive Design Patterns in Voice Interfaces} {Exploring deceptive design patterns in voice interfaces}.{\BBCQ}
\newblock
 \APACrefbtitle {{Proceedings of the 2022 European Symposium on Usable Security}} {{Proceedings of the 2022 European Symposium on Usable Security}}\ (\BPGS\ 64--78).
\newblock
\APACaddressPublisher{New York, NY, USA}{ACM}.
\newblock
\begin{APACrefURL} {http://dx.doi.org/10.1145/3549015.3554213} \end{APACrefURL}
\PrintBackRefs{\CurrentBib}

\bibitem [\protect \citeauthoryear {%
Richards%
\ \BBA {} Hartzog%
}{%
Richards%
\ \BBA {} Hartzog%
}{%
{\protect \APACyear {2021}}%
}]{%
richards2021duty}
\APACinsertmetastar {%
richards2021duty}%
\begin{APACrefauthors}%
Richards, N.%
\BCBT {}\ \BBA {} Hartzog, W.%
\end{APACrefauthors}%
\unskip\
\newblock
\APACrefYearMonthDay{2021}{}{}.
\newblock
{\BBOQ}\APACrefatitle {A duty of loyalty for privacy law} {A duty of loyalty for privacy law}.{\BBCQ}
\newblock
\APACjournalVolNumPages{Wash. UL Rev.}{99}{}{961,}
\newblock

\newblock

\PrintBackRefs{\CurrentBib}

\bibitem [\protect \citeauthoryear {%
Ruxton%
}{%
Ruxton%
}{%
{\protect \APACyear {2006}}%
}]{%
Ruxton2006}
\APACinsertmetastar {%
Ruxton2006}%
\begin{APACrefauthors}%
Ruxton, G.D.%
\end{APACrefauthors}%
\unskip\
\newblock
\APACrefYearMonthDay{2006}{{\APACmonth{05}}}{}.
\newblock
{\BBOQ}\APACrefatitle {{The unequal variance t-test is an underused alternative to Student's t-test and the Mann–Whitney U test}} {{The unequal variance t-test is an underused alternative to Student's t-test and the Mann–Whitney U test}}.{\BBCQ}
\newblock
\APACjournalVolNumPages{Behavioral Ecology}{17}{4}{688–690,}
\newblock
\begin{APACrefDOI} \doi{10.1093/beheco/ark016} \end{APACrefDOI}
\newblock
\begin{APACrefURL} {http://dx.doi.org/10.1093/beheco/ark016} \end{APACrefURL}
\newblock

\newblock

\PrintBackRefs{\CurrentBib}

\bibitem [\protect \citeauthoryear {%
Sakamoto%
, Murozono%
\BCBL {}\ \BBA {} Ota%
}{%
Sakamoto%
\ \protect \BOthers {.}}{%
{\protect \APACyear {2020}}%
}]{%
sakamotoinvestigation2020}
\APACinsertmetastar {%
sakamotoinvestigation2020}%
\begin{APACrefauthors}%
Sakamoto, K.%
, Murozono, T.%
\BCBL {} Ota, Y.%
\end{APACrefauthors}%
\unskip\
\newblock
\APACrefYearMonthDay{2020}{{\APACmonth{05}}}{}.
\newblock
{\BBOQ}\APACrefatitle {An {Investigation} of {Consent} {Management} {Platforms} in {Japan}} {An {Investigation} of {Consent} {Management} {Platforms} in {Japan}}.{\BBCQ}
\newblock
\APACjournalVolNumPages{Research Report Security Psychology and Trust (SPT)}{2020-SPT-37}{3}{1--8,}
\newblock

\newblock

\PrintBackRefs{\CurrentBib}

\bibitem [\protect \citeauthoryear {%
Sch\"{a}fer%
, Preuschoff%
\BCBL {}\ \BBA {} Borchers%
}{%
Sch\"{a}fer%
\ \protect \BOthers {.}}{%
{\protect \APACyear {2023}}%
}]{%
schafer2023countermeasures}
\APACinsertmetastar {%
schafer2023countermeasures}%
\begin{APACrefauthors}%
Sch\"{a}fer, R.%
, Preuschoff, P.M.%
\BCBL {} Borchers, J.%
\end{APACrefauthors}%
\unskip\
\newblock
\APACrefYearMonthDay{2023}{}{}.
\newblock
{\BBOQ}\APACrefatitle {Investigating Visual Countermeasures Against Dark Patterns in User Interfaces} {Investigating visual countermeasures against dark patterns in user interfaces}.{\BBCQ}
\newblock
 \APACrefbtitle {{Proceedings of Mensch Und Computer 2023}} {{Proceedings of Mensch Und Computer 2023}}\ (\BPG~161–172).
\newblock
\APACaddressPublisher{New York, NY, USA}{ACM}.
\newblock
\begin{APACrefURL} {https://doi.org/10.1145/3603555.3603563} \end{APACrefURL}
\PrintBackRefs{\CurrentBib}

\bibitem [\protect \citeauthoryear {%
Seaborn%
\ \BBA {} Chang%
}{%
Seaborn%
\ \BBA {} Chang%
}{%
{\protect \APACyear {2024}}%
}]{%
seabornanother2024}
\APACinsertmetastar {%
seabornanother2024}%
\begin{APACrefauthors}%
Seaborn, K.%
\BCBT {}\ \BBA {} Chang, W.J.%
\end{APACrefauthors}%
\unskip\
\newblock
\APACrefYearMonthDay{2024}{{\APACmonth{05}}}{}.
\newblock
{\BBOQ}\APACrefatitle {Another subtle pattern: {Examining} demographic biases in dark patterns and deceptive design research} {Another subtle pattern: {Examining} demographic biases in dark patterns and deceptive design research}.{\BBCQ}
\newblock
 \APACrefbtitle {Mobilizing {Research} and {Regulatory} {Action} on {Dark} {Patterns} and {Deceptive} {Design} {Practices} {Workshop} at {CHI} {Conference} on {Human} {Factors} in {Computing} {Systems}.} {Mobilizing {Research} and {Regulatory} {Action} on {Dark} {Patterns} and {Deceptive} {Design} {Practices} {Workshop} at {CHI} {Conference} on {Human} {Factors} in {Computing} {Systems}.}
\newblock
\APACaddressPublisher{online}{CEUR-WS.org}.
\newblock
\begin{APACrefURL} {https://ceur-ws.org/Vol-3720/} \end{APACrefURL}
\PrintBackRefs{\CurrentBib}

\bibitem [\protect \citeauthoryear {%
Seaborn%
\ \protect \BOthers {.}}{%
Seaborn%
\ \protect \BOthers {.}}{%
{\protect \APACyear {2024}}%
}]{%
Seaborn2024lbw}
\APACinsertmetastar {%
Seaborn2024lbw}%
\begin{APACrefauthors}%
Seaborn, K.%
, Itagaki, T.%
, Watanabe, M.%
, Wang, Y.%
, Geng, P.%
, Fujii, T.%
\BDBL {}Yoshida, S.%
\end{APACrefauthors}%
\unskip\
\newblock
\APACrefYearMonthDay{2024}{}{}.
\newblock
{\BBOQ}\APACrefatitle {Deceptive, Disruptive, No Big Deal: Japanese People React to Simulated Dark Commercial Patterns} {Deceptive, disruptive, no big deal: Japanese people react to simulated dark commercial patterns}.{\BBCQ}
\newblock
 \APACrefbtitle {{Extended Abstracts of the CHI Conference on Human Factors in Computing Systems}.} {{Extended Abstracts of the CHI Conference on Human Factors in Computing Systems}.}
\newblock
\APACaddressPublisher{New York, NY, USA}{ACM}.
\newblock
\begin{APACrefURL} {http://dx.doi.org/10.1145/3613905.3651099} \end{APACrefURL}
\PrintBackRefs{\CurrentBib}

\bibitem [\protect \citeauthoryear {%
Seaborn%
\ \BBA {} Nakamura%
}{%
Seaborn%
\ \BBA {} Nakamura%
}{%
{\protect \APACyear {2025}}%
}]{%
seaborn2025ycs}
\APACinsertmetastar {%
seaborn2025ycs}%
\begin{APACrefauthors}%
Seaborn, K.%
\BCBT {}\ \BBA {} Nakamura, S.%
\end{APACrefauthors}%
\unskip\
\newblock
\APACrefYearMonthDay{2025}{{\APACmonth{08}}}{}.
\newblock
{\BBOQ}\APACrefatitle {Quality and representativeness of research online with Yahoo! Crowdsourcing} {Quality and representativeness of research online with yahoo! crowdsourcing}.{\BBCQ}
\newblock
\APACjournalVolNumPages{Frontiers in Psychology}{16}{}{,}
\newblock
\begin{APACrefDOI} \doi{10.3389/fpsyg.2025.1588579} \end{APACrefDOI}
\newblock
\begin{APACrefURL} {http://dx.doi.org/10.3389/fpsyg.2025.1588579} \end{APACrefURL}
\newblock

\newblock

\PrintBackRefs{\CurrentBib}

\bibitem [\protect \citeauthoryear {%
Shahin~Sharifi%
\ \BBA {} Rahim~Esfidani%
}{%
Shahin~Sharifi%
\ \BBA {} Rahim~Esfidani%
}{%
{\protect \APACyear {2014}}%
}]{%
ShahinSharifi2014}
\APACinsertmetastar {%
ShahinSharifi2014}%
\begin{APACrefauthors}%
Shahin~Sharifi, S.%
\BCBT {}\ \BBA {} Rahim~Esfidani, M.%
\end{APACrefauthors}%
\unskip\
\newblock
\APACrefYearMonthDay{2014}{{\APACmonth{06}}}{}.
\newblock
{\BBOQ}\APACrefatitle {The impacts of relationship marketing on cognitive dissonance, satisfaction, and loyalty: The mediating role of trust and cognitive dissonance} {The impacts of relationship marketing on cognitive dissonance, satisfaction, and loyalty: The mediating role of trust and cognitive dissonance}.{\BBCQ}
\newblock
\APACjournalVolNumPages{International Journal of Retail \& Distribution Management}{42}{6}{553–575,}
\newblock
\begin{APACrefDOI} \doi{10.1108/ijrdm-05-2013-0109} \end{APACrefDOI}
\newblock

\newblock

\PrintBackRefs{\CurrentBib}

\bibitem [\protect \citeauthoryear {%
Soe%
, Nordberg%
, Guribye%
\BCBL {}\ \BBA {} Slavkovik%
}{%
Soe%
\ \protect \BOthers {.}}{%
{\protect \APACyear {2020}}%
}]{%
soe2020norway}
\APACinsertmetastar {%
soe2020norway}%
\begin{APACrefauthors}%
Soe, T.H.%
, Nordberg, O.E.%
, Guribye, F.%
\BCBL {} Slavkovik, M.%
\end{APACrefauthors}%
\unskip\
\newblock
\APACrefYearMonthDay{2020}{}{}.
\newblock
{\BBOQ}\APACrefatitle {Circumvention by design - {D}ark patterns in cookie consent for online news outlets} {Circumvention by design - {D}ark patterns in cookie consent for online news outlets}.{\BBCQ}
\newblock
 \APACrefbtitle {{Proceedings of the 11th Nordic Conference on Human-Computer Interaction: Shaping Experiences, Shaping Society}.} {{Proceedings of the 11th Nordic Conference on Human-Computer Interaction: Shaping Experiences, Shaping Society}.}
\newblock
\APACaddressPublisher{New York, NY, USA}{ACM}.
\newblock
\begin{APACrefURL} {https://doi.org/10.1145/3419249.3420132} \end{APACrefURL}
\PrintBackRefs{\CurrentBib}

\bibitem [\protect \citeauthoryear {%
Stanton%
\ \protect \BOthers {.}}{%
Stanton%
\ \protect \BOthers {.}}{%
{\protect \APACyear {2017}}%
}]{%
stanton2017human}
\APACinsertmetastar {%
stanton2017human}%
\begin{APACrefauthors}%
Stanton, N.A.%
, Salmon, P.M.%
, Rafferty, L.A.%
, Walker, G.H.%
, Baber, C.%
\BCBL {} Jenkins, D.P.%
\end{APACrefauthors}%
\unskip\
\newblock
\APACrefYear{2017}.
\newblock
\APACrefbtitle {Human Factors Methods: A Practical Guide for Engineering and Design} {Human factors methods: A practical guide for engineering and design}.
\newblock
\APACaddressPublisher{Boca Raton, Florida, USA}{CRC Press}.
\PrintBackRefs{\CurrentBib}

\bibitem [\protect \citeauthoryear {%
Swain%
}{%
Swain%
}{%
{\protect \APACyear {2018}}%
}]{%
Swain2018}
\APACinsertmetastar {%
Swain2018}%
\begin{APACrefauthors}%
Swain, J.%
\end{APACrefauthors}%
\unskip\
\newblock
\APACrefYearMonthDay{2018}{}{}.
\newblock
\APACrefbtitle {A Hybrid Approach to Thematic Analysis in Qualitative Research: Using a Practical Example.} {A hybrid approach to thematic analysis in qualitative research: Using a practical example.}
\newblock
\APACaddressPublisher{}{SAGE Publications Ltd}.
\newblock
\begin{APACrefURL} {http://dx.doi.org/10.4135/9781526435477} \end{APACrefURL}
\PrintBackRefs{\CurrentBib}

\bibitem [\protect \citeauthoryear {%
Synodinos%
}{%
Synodinos%
}{%
{\protect \APACyear {2001}}%
}]{%
Synodinos2001}
\APACinsertmetastar {%
Synodinos2001}%
\begin{APACrefauthors}%
Synodinos, N.E.%
\end{APACrefauthors}%
\unskip\
\newblock
\APACrefYearMonthDay{2001}{{\APACmonth{11}}}{}.
\newblock
{\BBOQ}\APACrefatitle {Understanding Japanese consumers: Some important underlying factors} {Understanding japanese consumers: Some important underlying factors}.{\BBCQ}
\newblock
\APACjournalVolNumPages{Japanese Psychological Research}{43}{4}{235–248,}
\newblock
\begin{APACrefDOI} \doi{10.1111/1468-5884.00181} \end{APACrefDOI}
\newblock
\begin{APACrefURL} {http://dx.doi.org/10.1111/1468-5884.00181} \end{APACrefURL}
\newblock

\newblock

\PrintBackRefs{\CurrentBib}

\bibitem [\protect \citeauthoryear {%
Takahashi%
}{%
Takahashi%
}{%
{\protect \APACyear {2010}}%
}]{%
takashi2010seijitsu}
\APACinsertmetastar {%
takashi2010seijitsu}%
\begin{APACrefauthors}%
Takahashi, H.%
\end{APACrefauthors}%
\unskip\
\newblock
\APACrefYearMonthDay{2010}{}{}.
\newblock
{\BBOQ}\APACrefatitle {Consumer behaviour and brand theory (2): Organizing the evolution and positioning of brand theory} {Consumer behaviour and brand theory (2): Organizing the evolution and positioning of brand theory}.{\BBCQ}
\newblock
\APACjournalVolNumPages{Kansei Gakuin Business Studies}{}{62}{17--49,}
\newblock

\newblock

\PrintBackRefs{\CurrentBib}

\bibitem [\protect \citeauthoryear {%
Takano%
\ \BBA {} Osaka%
}{%
Takano%
\ \BBA {} Osaka%
}{%
{\protect \APACyear {1997}}%
}]{%
Takano1997}
\APACinsertmetastar {%
Takano1997}%
\begin{APACrefauthors}%
Takano, Y.%
\BCBT {}\ \BBA {} Osaka, E.%
\end{APACrefauthors}%
\unskip\
\newblock
\APACrefYearMonthDay{1997}{}{}.
\newblock
{\BBOQ}\APACrefatitle {{``Japanese collectivism'' and ``American individualism'': Reexamining the dominant view}} {{``Japanese collectivism'' and ``American individualism'': Reexamining the dominant view}}.{\BBCQ}
\newblock
\APACjournalVolNumPages{The Japanese Journal of Psychology}{68}{4}{312–327,}
\newblock
\begin{APACrefDOI} \doi{10.4992/jjpsy.68.312} \end{APACrefDOI}
\newblock

\newblock

\PrintBackRefs{\CurrentBib}

\bibitem [\protect \citeauthoryear {%
Thaler%
\ \BBA {} Sunstein%
}{%
Thaler%
\ \BBA {} Sunstein%
}{%
{\protect \APACyear {2021}}%
}]{%
thaler2021nudge}
\APACinsertmetastar {%
thaler2021nudge}%
\begin{APACrefauthors}%
Thaler, R.H.%
\BCBT {}\ \BBA {} Sunstein, C.R.%
\end{APACrefauthors}%
\unskip\
\newblock
\APACrefYear{2021}.
\newblock
\APACrefbtitle {Nudge: The final edition} {Nudge: The final edition}.
\newblock
\APACaddressPublisher{}{Penguin}.
\PrintBackRefs{\CurrentBib}

\bibitem [\protect \citeauthoryear {%
Tokuhara%
, Yuichiro%
, Takaku%
, Komatsubara%
\BCBL {}\ \BBA {} Nakamura%
}{%
Tokuhara%
\ \protect \BOthers {.}}{%
{\protect \APACyear {2023}}%
}]{%
tokuhara2023choicedelay}
\APACinsertmetastar {%
tokuhara2023choicedelay}%
\begin{APACrefauthors}%
Tokuhara, M.%
, Yuichiro, K.%
, Takaku, T.%
, Komatsubara, T.%
\BCBL {} Nakamura, S.%
\end{APACrefauthors}%
\unskip\
\newblock
\APACrefYearMonthDay{2023}{{\APACmonth{09}}}{}.
\newblock
{\BBOQ}\APACrefatitle {Effect of delay in sequential display of choices on selection} {Effect of delay in sequential display of choices on selection}.{\BBCQ}
\newblock
\APACjournalVolNumPages{IEICE Technical Report}{123}{HCS-188, HIP-189}{65--70,}
\newblock

\newblock

\PrintBackRefs{\CurrentBib}

\bibitem [\protect \citeauthoryear {%
Tripepi%
, Jager%
, Dekker%
\BCBL {}\ \BBA {} Zoccali%
}{%
Tripepi%
\ \protect \BOthers {.}}{%
{\protect \APACyear {2010}}%
}]{%
Tripepi2010}
\APACinsertmetastar {%
Tripepi2010}%
\begin{APACrefauthors}%
Tripepi, G.%
, Jager, K.J.%
, Dekker, F.W.%
\BCBL {} Zoccali, C.%
\end{APACrefauthors}%
\unskip\
\newblock
\APACrefYearMonthDay{2010}{{\APACmonth{04}}}{}.
\newblock
{\BBOQ}\APACrefatitle {Selection Bias and Information Bias in Clinical Research} {Selection bias and information bias in clinical research}.{\BBCQ}
\newblock
\APACjournalVolNumPages{Nephron Clinical Practice}{115}{2}{c94–c99,}
\newblock
\begin{APACrefDOI} \doi{10.1159/000312871} \end{APACrefDOI}
\newblock
\begin{APACrefURL} {http://dx.doi.org/10.1159/000312871} \end{APACrefURL}
\newblock

\newblock

\PrintBackRefs{\CurrentBib}

\bibitem [\protect \citeauthoryear {%
Utz%
, Degeling%
, Fahl%
, Schaub%
\BCBL {}\ \BBA {} Holz%
}{%
Utz%
\ \protect \BOthers {.}}{%
{\protect \APACyear {2019}}%
}]{%
utz20219gdprconsent}
\APACinsertmetastar {%
utz20219gdprconsent}%
\begin{APACrefauthors}%
Utz, C.%
, Degeling, M.%
, Fahl, S.%
, Schaub, F.%
\BCBL {} Holz, T.%
\end{APACrefauthors}%
\unskip\
\newblock
\APACrefYearMonthDay{2019}{}{}.
\newblock
{\BBOQ}\APACrefatitle {{(Un)}Informed Consent: Studying {GDPR} Consent Notices in the Field} {{(Un)}informed consent: Studying {GDPR} consent notices in the field}.{\BBCQ}
\newblock
 \APACrefbtitle {{Proceedings of the 2019 ACM SIGSAC Conference on Computer and Communications Security}} {{Proceedings of the 2019 ACM SIGSAC Conference on Computer and Communications Security}}\ (\BPG~973–990).
\newblock
\APACaddressPublisher{New York, NY, USA}{ACM}.
\newblock
\begin{APACrefURL} {https://doi.org/10.1145/3319535.3354212} \end{APACrefURL}
\PrintBackRefs{\CurrentBib}

\bibitem [\protect \citeauthoryear {%
van~den Haak%
, de Jong%
\BCBL {}\ \BBA {} Jan~Schellens%
}{%
van~den Haak%
\ \protect \BOthers {.}}{%
{\protect \APACyear {2003}}%
}]{%
van2003retrospective}
\APACinsertmetastar {%
van2003retrospective}%
\begin{APACrefauthors}%
van~den Haak, M.%
, de Jong, M.%
\BCBL {} Jan~Schellens, P.%
\end{APACrefauthors}%
\unskip\
\newblock
\APACrefYearMonthDay{2003}{}{}.
\newblock
{\BBOQ}\APACrefatitle {Retrospective vs. concurrent think-aloud protocols: Testing the usability of an online library catalogue} {Retrospective vs. concurrent think-aloud protocols: Testing the usability of an online library catalogue}.{\BBCQ}
\newblock
\APACjournalVolNumPages{Behaviour \& Information Technology}{22}{5}{339--351,}
\newblock
\begin{APACrefURL} {https://doi.org/10.1080/0044929031000} \end{APACrefURL}
\newblock

\newblock

\PrintBackRefs{\CurrentBib}

\bibitem [\protect \citeauthoryear {%
van~der Ham%
, Faber%
, Venselaar%
, van Kreveld%
\BCBL {}\ \BBA {} L{\"o}ffler%
}{%
van~der Ham%
\ \protect \BOthers {.}}{%
{\protect \APACyear {2015}}%
}]{%
vanderHam2015}
\APACinsertmetastar {%
vanderHam2015}%
\begin{APACrefauthors}%
van~der Ham, I.J.M.%
, Faber, A.M.E.%
, Venselaar, M.%
, van Kreveld, M.J.%
\BCBL {} L{\"o}ffler, M.%
\end{APACrefauthors}%
\unskip\
\newblock
\APACrefYearMonthDay{2015}{{\APACmonth{05}}}{}.
\newblock
{\BBOQ}\APACrefatitle {Ecological validity of virtual environments to assess human navigation ability} {Ecological validity of virtual environments to assess human navigation ability}.{\BBCQ}
\newblock
\APACjournalVolNumPages{Frontiers in Psychology}{6}{}{,}
\newblock
\begin{APACrefDOI} \doi{10.3389/fpsyg.2015.00637} \end{APACrefDOI}
\newblock
\begin{APACrefURL} {http://dx.doi.org/10.3389/fpsyg.2015.00637} \end{APACrefURL}
\newblock

\newblock

\PrintBackRefs{\CurrentBib}

\bibitem [\protect \citeauthoryear {%
van Nimwegen%
\ \BBA {} de Wit%
}{%
van Nimwegen%
\ \BBA {} de Wit%
}{%
{\protect \APACyear {2022}}%
}]{%
van2022shopping}
\APACinsertmetastar {%
van2022shopping}%
\begin{APACrefauthors}%
van Nimwegen, C.%
\BCBT {}\ \BBA {} de Wit, J.%
\end{APACrefauthors}%
\unskip\
\newblock
\APACrefYearMonthDay{2022}{}{}.
\newblock
{\BBOQ}\APACrefatitle {Shopping in the dark: Effects of platform choice on dark pattern recognition} {Shopping in the dark: Effects of platform choice on dark pattern recognition}.{\BBCQ}
\newblock
 \APACrefbtitle {International Conference on Human-Computer Interaction} {International conference on human-computer interaction}\ (\BPGS\ 462--475).
\newblock
\APACaddressPublisher{London, UK}{Springer}.
\newblock
\begin{APACrefURL} {https://doi.org/10.1007/978-3-031-05412-9\_32} \end{APACrefURL}
\PrintBackRefs{\CurrentBib}

\bibitem [\protect \citeauthoryear {%
Voigt%
, Schl{\"o}gl%
\BCBL {}\ \BBA {} Groth%
}{%
Voigt%
\ \protect \BOthers {.}}{%
{\protect \APACyear {2021}}%
}]{%
voigt2021dark}
\APACinsertmetastar {%
voigt2021dark}%
\begin{APACrefauthors}%
Voigt, C.%
, Schl{\"o}gl, S.%
\BCBL {} Groth, A.%
\end{APACrefauthors}%
\unskip\
\newblock
\APACrefYearMonthDay{2021}{}{}.
\newblock
{\BBOQ}\APACrefatitle {Dark patterns in online shopping: Of sneaky tricks, perceived annoyance and respective brand trust} {Dark patterns in online shopping: Of sneaky tricks, perceived annoyance and respective brand trust}.{\BBCQ}
\newblock
 \APACrefbtitle {{International Conference on Human-Computer Interaction}} {{International Conference on Human-Computer Interaction}}\ (\BPGS\ 143--155).
\newblock
\APACaddressPublisher{London, UK}{Springer}.
\newblock
\begin{APACrefURL} {https://link.springer.com/chapter/10.1007/978-3-030-77750-0\_10} \end{APACrefURL}
\PrintBackRefs{\CurrentBib}

\bibitem [\protect \citeauthoryear {%
Wilson%
}{%
Wilson%
}{%
{\protect \APACyear {2014}}%
}]{%
wilson2013interview}
\APACinsertmetastar {%
wilson2013interview}%
\begin{APACrefauthors}%
Wilson, C.%
\end{APACrefauthors}%
\unskip\
\newblock
\APACrefYear{2014}.
\newblock
\APACrefbtitle {Interview Techniques for {UX} Practitioners: A User-Centered Design Method} {Interview techniques for {UX} practitioners: A user-centered design method}.
\newblock
\APACaddressPublisher{Waltham, MA, USA}{Morgan Kaufmann}.
\PrintBackRefs{\CurrentBib}

\bibitem [\protect \citeauthoryear {%
Yada%
\ \protect \BOthers {.}}{%
Yada%
\ \protect \BOthers {.}}{%
{\protect \APACyear {2022}}%
}]{%
yada2022dark}
\APACinsertmetastar {%
yada2022dark}%
\begin{APACrefauthors}%
Yada, Y.%
, Feng, J.%
, Matsumoto, T.%
, Fukushima, N.%
, Kido, F.%
\BCBL {} Yamana, H.%
\end{APACrefauthors}%
\unskip\
\newblock
\APACrefYearMonthDay{2022}{}{}.
\newblock
{\BBOQ}\APACrefatitle {Dark patterns in e-commerce: A dataset and its baseline evaluations} {Dark patterns in e-commerce: A dataset and its baseline evaluations}.{\BBCQ}
\newblock
 \APACrefbtitle {{2022 IEEE International Conference on Big Data (Big Data)}} {{2022 IEEE International Conference on Big Data (Big Data)}}\ (\BPGS\ 3015--3022).
\newblock
\APACaddressPublisher{Piscataway, NJ, USA}{IEEE}.
\newblock
\begin{APACrefURL} {https://ieeexplore.ieee.org/document/10020800} \end{APACrefURL}
\PrintBackRefs{\CurrentBib}

\bibitem [\protect \citeauthoryear {%
Yamanouchi%
}{%
Yamanouchi%
}{%
{\protect \APACyear {2015}}%
}]{%
yamano2015jpsus}
\APACinsertmetastar {%
yamano2015jpsus}%
\begin{APACrefauthors}%
Yamanouchi, S.%
\end{APACrefauthors}%
\unskip\
\newblock
\APACrefYearMonthDay{2015}{}{}.
\newblock
{\BBOQ}\APACrefatitle {エンジニアのための人を対象とする研究計画入門 ({I}ntroduction to Human Subjects Research Planning for Engineers)} {エンジニアのための人を対象とする研究計画入門 ({I}ntroduction to human subjects research planning for engineers)}.{\BBCQ}
\newblock
\BIn{} (\BPGS\ 74--118).
\newblock
\APACaddressPublisher{Tokyo, Japan}{Maruzen}.
\PrintBackRefs{\CurrentBib}

\end{thebibliography}

\end{CJK}
\end{document}